\documentclass[11pt]{article}

\usepackage{mydef2col}
\usepackage{eufrak}
\usepackage{enumitem}
\usepackage{bbm}
\usepackage{bm}
\usepackage{autobreak}
\usepackage{kantlipsum,widetext}
\usepackage[T1]{fontenc}
\usepackage[autostyle=true]{csquotes}
\usepackage{todonotes}

\allowdisplaybreaks[4]

\graphicspath{{./Graphs/}}
\usepackage{tikz}
\usetikzlibrary{calc,decorations.markings}
\usetikzlibrary{arrows,patterns,positioning}

\usepackage[round]{natbib}

\newcommand{\Ind}{\mathbbm{1}}
\newcommand{\iu}{\mathrm{i}\mkern1mu}

\newcommand{\ThetaVec}{\boldsymbol{\Theta}}
\newcommand{\thetaVec}{\boldsymbol{\theta}}
\newcommand{\sigmaVec}{\boldsymbol{\sigma}}
\newcommand{\IndVec}{\boldsymbol{\Ind}}
\newcommand{\PhiVec}{\boldsymbol{\Phi}}
\newcommand{\phiVec}{\boldsymbol{\phi}}
\newcommand{\OmegaVec}{\boldsymbol{\Omega}}
\newcommand{\omegaVec}{\boldsymbol{\omega}}

\newcommand{\RotM}{\mathcal{R}}
\newcommand{\calK}{\mathcal{K}}
\newcommand{\HM}{\mathcal{H}}
\newcommand{\StrM}{\mathcal{S}}
\newcommand{\FM}{\mathcal{F}}
\newcommand{\BM}{\mathcal{B}}
\newcommand{\PM}{\mathcal{P}}
\newcommand{\GM}{\mathcal{G}}

\newcommand{\YM}{\mathcal{Y}}
\newcommand{\calT}{\mathcal{T}}
\newcommand{\LambdaM}{\boldsymbol{\Lambda}}

\newcommand{\etaM}{\boldsymbol{\eta}}
\newcommand{\by}{\mathbf{y}}
\newcommand{\bl}{\bar{\lambda}}

\newcommand{\tl}{\tilde{\lambda}}
\DeclareMathOperator\arctanh{arctanh}

\title{Multilayer heat equations and their solutions via oscillating integral transforms.}
\def\thetitle1{Multilayer heat equations and their solutions via oscillating integral transforms.}
\author{
\authorstyle{
Andrey Itkin{}
\textsuperscript{1}
Alexander Lipton{}
\textsuperscript{2}
and Dmitry Muravey
\textsuperscript{3}
}
\newline\newline
\textsuperscript{1}
\institution{Tandon School of Engineering, New York University, New York, USA} \\
\textsuperscript{2}
\institution{The Jerusalem School of Business Administration, The Hebrew University of Jerusalem, Jerusalem, Israel;} \\
\textsuperscript{\ \ }
\institution{Connection Science and Engineering, Massachusetts Institute of Technology, Cambridge, MA, USA} \\
\textsuperscript{3}
\institution{Moscow State University, Moscow, Russia}
}

\date{\today}
\begin{document}

\maketitle

\lettrineabstract{By expanding the Dirac delta function in terms of the eigenfunctions of the corresponding Sturm-Liouville problem, we construct some new (oscillating) integral transforms. These transforms are then used to solve various finance, physics, and mathematics problems, which could be characterized by the existence of a multilayer spatial structure and moving (time-dependent) boundaries (internal interfaces) between the layers. Thus, constructed solutions are semi-analytical and extend the authors' previous work (Itkin, Lipton, Muravey, Multilayer heat equations: application to finance, FMF, 1, 2021). However, our new method doesn't duplicate the previous one but provides alternative representations of the solution which have different properties and serve other purposes.
}

\vspace{0.5in}

\section*{Introduction} \label{Introduction}

Undoubtedly, integral transforms are one of the most beautiful developments of classical mathematics. They build a robust and unified basis for solving various linear differential and integral equations of mathematical physics, applied mathematics, and engineering and can be utilized for both the initial and initial-boundary problems, usually with constant coefficients. Moreover, with ever greater modern demand for mathematical methods in science and engineering, the utility of and interest in the integral transforms increase with time. Thus, even though integral transforms already have many mathematical and physical applications, they constantly find new applications in advanced study and research.

As mentioned in \citep{Stetza2010}, the idea that beauty is "right" and "true" has been around since humankind developed abstract thought. The Latin phrase {\it Pulchritudo splendor varitatis} - "beauty is the splendor of truth" - is thousands of years old and suggests that beauty and truth are interrelated. Indeed, it seems appropriate that something beautiful would be true, but it is more realistic to think that something hideous could also be true.

This thought immediately comes to mind when looking at some modern problems in physics, biology, finance, etc. Among other essential problems, let us mention just a few:
(a) the transdermal drug release from an iontophoretic system, \citep{Pontrelli2016};
(b) the transmembrane diffusion in cells;
(c)  the growth of diffusive brain tumors, which considers the brain tissue's heterogeneity, \citep{Asvestas2014};
(d) distribution of organisms in rivers - an advection-diffusion problem, \citep{Ramirez2013};
(e)  pricing and calibration  with piecewise constant, linear and quadratic volatility, \citep{LiptonSepp2011, ItkinLipton2017, Lipton2018a, LiptonGal2014};
(f) the oscillating/skew Brownian Motion, the local time diffusion, \citep{Lejay2006, Lejay2018};
(g) the occupation times, optimal stopping/first passage time problems for the processes with broken drift, \citep{Salminen2021, Mordecki2019, ankirchner2021};
(h) parabolic equations and diffusion processes on graphs, \citep{Lejay2006}.

Solving these and similar problems, characterized by multiple spatial layers (with different physical properties) and moving boundaries between these layers, is very important, even though it is impossible to find their solutions by using the standard operational calculus of classical integral transforms. Therefore, to proceed, we need to sacrifice the beauty of elegant and straightforward classical transforms and instead build a new class of integral transforms. Of course, these transforms have to be adapted to the specific structure at hand; see, e.g., \citep{ItkinLiptonMuraveyBook} where various time-dependent problems of mathematical finance and physics with moving boundaries are solved by doing precisely that. Usually, those transforms are constructed by using some basis in eigenfunctions found by solving the corresponding Sturm-Liouville (SL) problem, \citep{Antimirov}.

In this paper, we extend the approach of \citep{ItkinLiptonMuraveyBook} by applying its main idea to solving various multilayer problems for the one-dimensional heat equation. Let us introduce the time $t \in [0,\infty)$, the space coordinate $x \in \mathbb{R}$, and a function $u(t,x): \mathbb{R}^+\times \mathbb{R} \mapsto \mathbb{R}^+, \ u \in {\cal C}^2$. Suppose that the whole domain $\Omega = \mathbb{R}$ could be split into $N+2$ non-overlapping layers $-\infty < y_0(t) < ... < y_{N}(t) < \infty$, $\forall t \geq 0$, i.e. $\Omega = \bigcup_{i=0}^N \Omega_i$, where each layer is a curvilinear strip
\begin{align} \label{Omega_i_def}
\Omega_i &= [y_{i-1}(t), y_{i}(t)] \times \mathbb{R}^{+}, \quad y_{i}(t) < y_{i+1}(t), \quad \forall t \ge 0, \quad \forall i = 1,\dots,N,
\end{align}
\noindent and $y_i(t) \in {\cal C}^1, \ i = 0,\dots,N$. Also, let us consider a diffusion (thermal conductivity)  process in $\mathbb{R}$ such that the diffusion coefficient $\sigma$ is a piecewise constant function of $x$, i.e.
\begin{align}
\sigma(x) &=
\begin{cases}
 \sigma_i, & \forall x \in [y_{i-1}(t),y_i(t)], \quad i = 1,\dots,N, \\
\sigma_-,  & x \in (-\infty,y_0(t)], \\
\sigma_+, &  x \in [y_N(t),\infty),
\end{cases}
\end{align}
\noindent where $\sigma_i > 0, \sigma_- > 0, \sigma_+ > 0$ are constants. This assumption can be relaxed, so that $\sigma_\pm = \sigma_pm(t), \sigma_i = \sigma_i(t)$, however, for simplicity we assume them to be constant. This assumption doesn't impact the main idea of the proposed approach and the construction of oscillating integral transforms but makes many formulae look lighter.

Suppose the evolution of this diffusion process is described by the heat equation
\begin{equation} \label{Intro:ML_ex}
u_t = \sigma^2(x)  u_{xx},
\end{equation}
\noindent which should be solved subject to the initial condition
\begin{equation} \label{ic}
u(0, x) = \delta(x - x_0),		
\end{equation}
\noindent where $\delta(x - x_0)$ is the Dirac delta function, and also the matching conditions (see the corresponding discussion in \citep{ItkinLiptonMuraveyMulti} and references therein)
\begin{align} \label{Intro:ML_ex_matching_cond}
\lim_{x \to y_i(t) - 0}u(t, x) &= \lim_{x \to y_i(t) + 0}u(t, x), 	 	\\
\lim_{x \to y_i(t) - 0} \sigma_{i}^2 u_x(t, x) &= \lim_{x \to y_i(t) + 0} \sigma_{i+1}^2 u(t, x), \qquad i = 0,\dots,N. \nonumber
\end{align}
These conditions mean the continuity of the function $u$ and its flux over each boundary in the multilayer media. Note that such conditions could vary depending on the problem; see \citep{CarrMarch2018}.

The known methods of solving \eqref{Intro:ML_ex}, \eqref{ic}, \eqref{Intro:ML_ex_matching_cond} can be conventionally classified as follows
\begin{enumerate}[label=(\alph*)]
\item  eigenfunction expansions, \citep{CarrMarch2018};
\item the Laplace Transform, \citep{LiptonSepp2011,ItkinLipton2017};
\item the method of heat potentials (HP) and generalized integral transform (GIT),  \citep{ItkinLiptonMuraveyMulti, ItkinLiptonMuraveyBook};
\item the Fokas method, \citep{Asvestas2014, Deconinck2014, Deconinck2016};
\item probabilistic methods, \citep{Lejay2006}.
\end{enumerate}
The methods (a), (b) (d) can be used only for problems with straight boundaries; method (c) works even if the boundaries are curvilinear. Therefore, we want to treat multilayer problems by using a flavor of the framework (c).

To illustrate the main idea, we recall a recent paper \citep{ItkinLiptonMuraveyMulti} where the multilayer method has been introduced in detail. Let us consider the $i$-th layer in \eqref{Intro:ML_ex} and the corresponding integral transform
\begin{equation*}
\bar{u}(t,\lambda) = \int_{y_{i-1}(t)}^{y_i(t)} \sin \left(\lambda (x - y_i(t)) \right) u(t, x) dx.
\end{equation*}
Applying this transform to \eqref{Intro:ML_ex} for the $i$-th layer and then constructing the inverse transform allows  a closed form representation of the solution $u(t,x), \ x \in  [y_{i-1}(t), y_i(t)]$ which, however, also depends on four yet unknown functions
\begin{alignat}{2}
\psi_i^-(t) &= \lim_{x \to y_{i}(t)-0}u(t, x),     & \quad \psi_i^+(t) &= \lim_{x \to y_{i}(t)+0}u(t, x), \\
\Psi_i^-(t) &= \lim_{x \to y_{i}(t)-0} u_x(t, x), &  \quad \Psi_i^+(t) &= \lim_{x \to y_{i}(t)+0} u_x(t, x). \nonumber
\end{alignat}
It turns out that with allowance for the matching \eqref{Intro:ML_ex_matching_cond} and boundary \eqref{Intro:ML_ex} conditions the unknown functions $\psi^\pm_i(t)$, $\Psi^\pm_i(t)$ solve a system of Volterra equations of the second kind. Once this system is solved (numerically), the solution of the problem \eqref{Intro:ML_ex} is obtained. It is important to emphasize that this procedure relies on the {\it first} step, which consists in using an integral transform to obtain a formal solution at every layer, and the {\it second} step utilizing the matching conditions to derive the Volterra equations.

Although it is shown in \citep{ItkinLiptonMuraveyMulti} that this approach works well, here we slightly modify it by changing the order of the corresponding steps. It means that for each problem under consideration, we want to design a special integral transform with the kernel satisfying the matching conditions \eqref{Intro:ML_ex_matching_cond}. Once this is done, the problem \eqref{Intro:ML_ex_matching_cond} can be solved by using a direct/inverse integral transform in line of \citep{ItkinLiptonMuraveyBook}. Surprisingly, in many cases, the corresponding integral transforms can be explicitly obtained. By analogy to the oscillating Brownian Motion, \citep{Lejay2006}, we call them "oscillating integral transforms" (OIT).

The novelty of this paper is twofold. First, we derive some integral transforms which, to the best of our knowledge,
yet have never been reported in the literature. Therefore, this is a contribution in a purely mathematical context. Second, we develop a new method of solving the multilayer heat equation (MHE) problem in \eqref{Intro:ML_ex}, \eqref{ic}, \eqref{Intro:ML_ex_matching_cond} with curvilinear boundaries. Moreover, since the Green's function of this problem is constructed in a semi-explicit form, several problems, such as determining the first passage time, optimal stopping, etc. for processes associated with \eqref{Intro:ML_ex} with discontinuous coefficients, can be solved by either the HP or GIT methods, \citep{ItkinLiptonMuraveyBook}.

It is worth discussing the second point in more detail. In \citep{ItkinLiptonMuraveyMulti} we developed a multilayer method that makes it possible to find semi-analytical solutions of some one-dimensional PDEs that otherwise cannot be
solved analytically. Our approach works because the corresponding PDEs cannot be reduced to the heat or Bessel equation in the whole spatial domain but are reducible in each layer separately. As a result, the layer boundaries can be chosen arbitrarily (to some extent) based on the argument of accuracy and convenience. In particular, we constructed these boundaries so that the corresponding Volterra equations could be solved by using the Laplace transform. However, suppose we want to use this method to solve physical interlayer (interface) boundaries, which are defined by the essence of the problem and cannot be chosen at will. In this case, we have to solve the Volterra equations the way they are, so no further simplifications are possible.

In this case, the gradients of the solution could potentially misbehave close to each internal boundary in a sense that our series representation of the solution could converge very slowly; see  \citep{ItkinLiptonMuraveyMulti}. Therefore, the numerical solution for the corresponding Volterra equations could be unstable. However, by adding artificial boundaries to make the Volterra integral a convolution, the Volterra equation can be solved using the Laplace transform, thus eliminating potential problems. As a result, the corresponding solution is well-behaved.

To address the issues mentioned above, we develop another flavor of the multilayer method, which gives rise to different are more numerically stable Volterra equations. To proceed, we need the boundary points to be {\it  internal} for an appropriate basis, rather than the {\it boundary} points for another basis. If this choice is possible, the described problem of potential instability in the numerical solution of the Volterra equations, which occurs close to the layer boundaries, disappears. Therefore,
our new method doesn't duplicate that one in  \citep{ItkinLiptonMuraveyMulti} because now we obtain alternative representations of the solution with better properties.

The paper is organized as follows. Section ~\ref{sec:ODE} presents a construction of the OIT by using series expansions of the Dirac delta function and building an orthonormal basis of eigenfunctions for the corresponding Sturm-Liouville problem. We consider various cases characterized by a spectrum of eigenvalues: continuous, discrete, or mixed. In Section~\ref{examples1} some examples of this technique are presented in more detail. We describe a model with two non-zero layers and a discrete spectrum and another model with three layers and a continuous spectrum. Next, section~\ref {MHEcurv} is devoted to problems that give rise to multilayer heat equations with curvilinear boundaries. We discuss in detail the time-dependent oscillating Brownian motion and various applications from finance and physics. For a two-layer problem with moving (time-dependent) boundary (the interlayer interface), we derive a system of linear Volterra integral equations for the values of the solution and its gradient at the interface. By solving this system, we can express the solution of the whole problem in the closed-form. Thus, our method is semi-analytical. To the best of our knowledge, in the existing literature, the problem at hand was solved just numerically (for the time-dependent boundary) or analytically (for the constant boundary). Thus, our approach is new and contributes to the existing literature. In the next section, we demonstrate the power of our method as applied to solving a practical problem - solidification (or melting) of a liquid-solid system at the time-dependent phase interface. Again, so far in the literature, this problem was solved only numerically, e.g., by using finite differences (FD), which brings some issues when constructing an efficient FD scheme at the domain with the interphase moving boundary. Then, we introduce and solve a multilayer problem with the number of layers $N > 2$ and piecewise constant coefficients and time-dependent interfaces in Section~\ref{MHEcurv}. We summarize our conclusions in the final section.

We tried doing our best to simplify the exposition of the method and make it transparent to all readers regardless of their background (finance, physics, or mathematics). For this reason, we moved a significant part of derivations into appendices. Section~\ref{deltaExp} is a bit technical; however, it would be difficult to understand our approach without it. If necessary, it could be skipped at first reading so that after Introduction, the reader can immediately proceed to Section~\ref{examples1}.

\section{Building integral transforms for multilayer problems} \label{sec:ODE}

For the reader's convenience and to eliminate overcomplicated formulae, we begin this section by introducing a special notation used across the whole paper.

\subsection{Notation} \label{sec:notation}

By $\by$ we denote the following vector
\begin{equation} 	\label{BG:ydef}
\by = \left(y_0, y_1,\dots,y_N \right)^\top, \quad y_{i-1} < y_{i}, \ i = 1\dots N. 	
\end{equation}
The vector $\sigmaVec$ is the $N+2$-dimensional vector with positive entries
\begin{equation} \label{BG:sigmadef}
\sigmaVec = \left(\sigma_-, \sigma_1,\dots,\sigma_N, \sigma_+\right)^\top,
\end{equation}
\noindent and by $\sigmaVec^2$ we mean a component-wise square of $\sigmaVec$. We also introduce the vector indicator function $\IndVec_{x | \by}$ associated with $\by$
\begin{equation}
\IndVec_{x|\by} = \left(\Ind_{x < y_0}, \Ind_{y_0<x < y_1},\dots,\Ind_{y_{N-1} < x <y_N}, \Ind_{x > y_N}\right)^\top.
\end{equation}
If $\by$ is a scalar, i.e. $\by = y$, the indicator vector $\IndVec_{x | y}$ is defined as
\begin{equation}
\IndVec_{x | y} = \left(\Ind_{x < y}, \Ind_{x > y}\right)^\top.
\end{equation}

Let the operator $\langle  \mathbf{a},\mathbf{b}\rangle $ denote a scalar product
\begin{equation}
\langle  \mathbf{a},\mathbf{b}\rangle  = \mathbf{a}^\top \mathbf{b} = \mathbf{b}^\top \mathbf{a} =  \sum_{i = 1}^{N} a_i b_i.
\end{equation}

Let us also introduce $\RotM(\varphi)$, $\HM(\varphi)$ and $\StrM(\alpha, \beta)$
\begin{equation} \label{BG:RotStretch_def}
\RotM_\varphi =     	
\begin{bmatrix}
	\cos \varphi & 	 -	\sin \varphi \\
	\sin \varphi & 		\cos \varphi
\end{bmatrix},
\quad
\HM_\varphi =     	
\begin{bmatrix}
	\cosh \varphi & 	 \sinh \varphi \\
	\sinh \varphi & 		\cosh \varphi
\end{bmatrix},
\quad
\StrM_{\alpha, \beta} =
\begin{bmatrix}
	\alpha & 	0 \\
	0 & \beta
\end{bmatrix},
\end{equation}
\noindent which are matrices of rotation, hyperbolic rotation and stretch, respectively. The special case $\StrM_{1, \beta}$ is denoted as $\StrM_\beta = \StrM_{1, \beta}$.

Suppose we are given a set of squared matrices $\mathbf{A}_k, \ k=1,\dots,K$. A product over a set $k \in [a,b]$ is defined as
\begin{equation} 	\label{prodM}
\prod_{k = a}^{b} \mathbf{A}_k = \mathbf{A}_b \mathbf{A}_{b-1}\dots\mathbf{A}_{a+1} \mathbf{A}_a.
\end{equation}

\subsection{Construction of transforms using series expansion of the Dirac delta function} \label{deltaExp}

Our focus in this paper is on solving various multilayer problems by constructing an appropriate integral transform.  In doing so, we begin with an observation that the Dirac delta function can be represented as a series expansion in eigenfunctions of some Sturm-Liouville (SL) problem, \citep{Titchmarsh1962}. We discuss various approaches to obtaining this representation in Section~\ref{ExpDelta}. Then an integral transform associated with a given expansion can be constructed as follows.

Let $\by$ and $\sigmaVec$ be the vectors defined in \eqref{BG:ydef} and \eqref{BG:sigmadef} respectively. Let us introduce a function $u_{\lambda}(x|x_0): \mathbb{R} \mapsto \mathbb{R}^+, \ u \in {\cal C}^2$ parameterized by $\lambda \in \mathbb{R}$ and the initial condition $u(x_0) = \delta(x-x_0)$.  We consider the following  SL problem for the function $u_{\lambda}(x | x_0)$
\begin{align} \label{SL_problem}
\fp{}{x} \left[ \langle \IndVec_{x|\by}, \sigmaVec^2 \rangle  \fp{}{x} u_\lambda(x|x_0) \right] - \lambda u_\lambda(x | x_0) = -\delta(x - x_0),
\end{align}
\noindent  which should be solved subject to the vanishing boundary conditions
\begin{align}  \label{bc}
\lim_{x \to \pm \infty} u_\lambda(x | x_0) &= 0,
\end{align}
\noindent and also the matching conditions
\begin{align} \label{SL_problem:MatchingCond}
\lim_{x \to y_i-0}  u_\lambda(x | x_0)  &=	\lim_{x \to y_i+0}u_\lambda(x | x_0),  \\
\lim_{x \to y_i -0}  \sigma_i^2 u'_\lambda(x | x_0)  &=	\lim_{x \to y_i +0}  \sigma_{i+1}^2 u'_\lambda(x | x_0),
\qquad i = 0,\dots,N, \nonumber \\
\lim_{x \to y_0 -0}  \sigma_-^2 u'_\lambda(x | x_0)  &=	\lim_{x \to y_0 +0}  \sigma_0^2 u'_\lambda(x | x_0),
\nonumber \\
\lim_{x \to y_N -0}  \sigma_N^2 u'_\lambda(x | x_0) &=	\lim_{x \to y_N +0}  \sigma_+^2 u'_\lambda(x | x_0). \nonumber
\end{align}

Once the function $u_\lambda(x | x_0)$ is known, the following representations of the Dirac delta function holds
\begin{equation} \label{SL:Dirac_def}
\delta(x - x_0) = \frac{1}{2\pi \iu} \int_{\gamma - \iu \infty}^{\gamma + \iu \infty} u_\lambda(x | x_0) d\lambda.
\end{equation}	
Indeed, suppose that the function $u(t,x)$ solves the heat equation at $(t,x) \in \mathbb{R}^+ \times \mathbb{R}$ with the initial condition $u(0,x) = \delta(x-x_0)$. Taking the Laplace transform, one can see that the image $u_\lambda(x|x_0)$
solves \eqref{SL_problem}. Therefore, given $u_\lambda(x|x_0)$, $u(t,x)$ can be restored via the inverse Laplace transform (the Bromwich integral). Then setting $t=0$ we obtain \eqref{SL:Dirac_def}.

In \eqref{SL:Dirac_def} $\gamma$ is a real-valued constant chosen in such a way that all poles of the function under the integral are located in a complex plane to the left of a straight line parallel to the imaginary axis $x \in (\gamma, -\iu \infty),  (\gamma, \iu \infty)$. As per \citep{Titchmarsh1962}, this gives rise to the delta function expansion in non-normalized eigenfunctions of the SL problem of the form
\begin{equation} \label{deltaForm}
\delta(x - x_0) = \int_{-\infty}^{+\infty} F_\omega(x) B_\omega(x_0) d\omega,	
\end{equation}
\noindent where $F_\omega(x)$ is some function of $x$ and $B_\omega(x_0)$ - some function of $x_0$, so the integrand in \eqref{deltaForm} factorizes into a product. Then, using the filtering property of delta function
\begin{equation} \label{deltaFilt}
f(x) = \int_{-\infty}^{\infty} \delta(x - x_0)  f(x_0)dx_0,
\end{equation}
\noindent we arrive at the following construction
\begin{align*}
\bar f (\omega) = \int_{-\infty}^{+\infty} F_\omega(x) f(x) dx, 	\qquad
f(x) = \int_{-\infty}^{+\infty}  B_\omega(x) \bar f (\omega)  d\omega,
\end{align*}
\noindent which is an integral transform built based on the eigenfunctions of the corresponding SL problem. Thus, constructing the necessary integral transform is almost straightforward once the corresponding expansion of the Dirac delta function is derived. In the next section, we consider the latter problem in more detail.

\subsection{Expansion of the Dirac delta function into series of eigenfunctions} \label{ExpDelta}

Since solutions to this problem could have different properties depending on the explicit form of $\by,\sigmaVec$, below, we develop a more granular classification.

\paragraph{Problems with continuous spectrum.} This corresponds to the special case $\by = y$, so \eqref{bc}, and \eqref{SL_problem:MatchingCond} take the form
\begin{align} \label{SL_problem:CONT_SPEC}
\lim_{x \to \pm \infty} u_\lambda(x | x_0) = 0, \quad \lim_{x \to y-0}  u_\lambda(x | x_0)  =	\lim_{x \to y +0}u_\lambda(x | x_0), \quad \lim_{x \to y -0}  \sigma_-^2 u'_\lambda(x | x_0) =	\lim_{x \to y +0}  \sigma_+^2 u'_\lambda(x | x_0).
\end{align}

\paragraph{Problems with discrete spectrum.} This occurs when the vector $\sigmaVec$ is of a special form
\begin{equation} \label{sigma_discrete_spectrum}
\sigmaVec = \left(0, \sigma_1, \sigma_2, \dots, \sigma_{N-1}, \sigma_N, 0\right)^\top,
\end{equation}
\noindent or when the absorbing boundary conditions at $y_0$ and $y_N$ are used. Then \eqref{bc}, \eqref{SL_problem:MatchingCond} read
\begin{align} \label{SL_problem:DISC_SPEC}
\lim_{x \to y_0 + 0 } u_\lambda(x | x_0) &= 0, \qquad \lim_{x \to y_N - 0 } u_\lambda(x | x_0) = 0, \\
\lim_{x \to y_i-0}  u_\lambda(x | x_0)  &=	\lim_{x \to y_i+0}u_\lambda(x | x_0), \quad \lim_{x \to y_{i} -0}  \sigma_{i}^2 u'_\lambda(x | x_0) =	\lim_{x \to y_{i} +0}  \sigma_{i+1}^2 u'_\lambda(x | x_0),  \quad i = 1,\dots,N-1. \nonumber
\end{align}

\paragraph{Problems with mixed spectrum.} In the general case the problem \eqref{SL_problem}, \eqref{bc} \eqref{SL_problem:MatchingCond}, as well as the problems corresponding to the vector $\sigmaVec$ with $\sigma_-=0$ or $\sigma_+ = 0$ have mixed spectra .  \hfill $\square$.

\bigskip

Below we analyze each that case in more detail, find an explicit solution for $u_\lambda(x | x_0)$ and obtain the corresponding representation for the Dirac delta function $\delta(x - x_0)$.

\subsubsection{Continuous spectrum. The oscillating Fourier transform} \label{sec:osc_FT}

In this section, we derive a new (to the best of our knowledge) integral transform, which can be viewed as a  generalization of the Fourier transform. We call it the oscillating Fourier transform (OIT).

It is well-known that the classical Fourier transform can be easily derived from the following identity
\begin{equation} \label{CS:Dirac_delta_def}
\delta(x - x_0) = \frac{1}{2\pi} \int_{-\infty}^{+\infty} e^{\iu \omega (x - x_0)} d\omega.
\end{equation}
The expansion \eqref{CS:Dirac_delta_def} is associated with the exponential function
\begin{equation} \label{CS:classical_function_f}
f_\omega(x) = e^{\iu \omega x}, \quad x \in (-\infty, \infty).
\end{equation}
By analogy, for our problem as the basis function we consider a piecewise exponential function
\begin{equation} \label{CS:osc_function_f}
g_\omega(x) =
\begin{cases}
e^{\iu \omega (x-y) / \sigma_-}, \quad x <y, \\
e^{\iu \omega (x-y) / \sigma_+}, \quad x >y. \\
\end{cases}	
\end{equation}
In case $\sigma_-=\sigma_+$ the function $g$ up to a constant multiplier coincides with the function $f$.

Given the vector $\sigmaVec^2 = \left(\sigma_-^2, \sigma_+^2\right)^\top$ let us define the function $u_\lambda (x | x_0)$ as the solution of \eqref{SL_problem}, \eqref{SL_problem:CONT_SPEC}. Our goal is to find an explicit expression for $u_\lambda(x|x_0)$. Once this is done, we use \eqref{SL:Dirac_def} and get the necessary representation of $\delta(x - x_0)$.

The solution of this problem is given in Appendix~\ref{appDiracCont}. The result reads
\begin{align} \label{CS:delta_FB_repr1}
\delta( x - x_0) = \frac{1}{2\pi}\int_{-\infty}^{+\infty}	 \IndVec_{x | y}^\top \FM_{x -y, \sigmaVec} (\omega) \BM_{x_0 - y, \sigmaVec}(\omega) 	\IndVec_{x_0 | y} 	d\omega,
\end{align}
\noindent where the definitions of matrices $\FM_{x -y, \sigmaVec} (\omega), \ \BM_{x_0 - y, \sigmaVec}(\omega)$ are given in \eqref{FBmatrices}.

Since the Forward $\FM$ and Backward $\BM$  matrices in \eqref{CS:delta_FB_repr1} depend only on $x$ and $x_0$, respectively, $\FM$ can be interpreted as a kernel of some integral transform, and $\BM$ - as an inversion kernel.  Indeed, using \eqref{deltaFilt} together with \eqref{CS:delta_FB_repr1} yields
\begin{equation} \label{CS:osc_identity_f}
f(x_0) =\frac{1}{2\pi}\int_{-\infty}^{+\infty}\int_{-\infty}^{+\infty}	
\IndVec_{x | y}^\top \FM_{x -y, \sigmaVec}(\omega)\BM_{x_0 - y, \sigmaVec}(\omega) \IndVec_{x_0 | y} f(x) dx d\omega.
\end{equation}
Splitting \eqref{CS:osc_identity_f} into two integrals we arrive at the following integral transform
\begin{equation} \label{CS:OSC_FT_inv}
f(x)= \frac{1}{2 \pi}\int_{-\infty}^{+\infty}	\boldsymbol{\bar f}(\omega)^\top \BM_{x-y, \sigmaVec}(\omega) \IndVec_{x | y} d\omega.
\end{equation}
The two-component image $\boldsymbol{\bar f}(\omega)$ is defined as follows
\begin{equation} \label{CS:OSC_FT_dir}
\boldsymbol{\bar f}(\omega) =
\begin{bmatrix}
\bar f_-(\omega) \\ \bar f _+ (\omega)
\end{bmatrix}
= \int_{-\infty}^{+\infty} \FM_{x-y, \sigmaVec}(\omega) \IndVec_{x | y} f(x)dx.
\end{equation}
We call it the \textit{Oscillating Fourier transform}. Contrary to the original Fourier transform, the oscillating Fourier transform in \eqref{CS:OSC_FT_dir}, \eqref{CS:OSC_FT_inv} has a two-component image.

\subsubsection{Discrete spectrum. Problem in a strip} \label{discSpect}

In this case it is possible to directly derive an orthogonal basis associated with the solution of the corresponding problem in a strip with matching conditions in \eqref{SL_problem:DISC_SPEC}.

But first, let us consider a slightly modified problem
\begin{align} \label{disEq1}
\fp{}{x}\left( \langle  \sigmaVec^2,\IndVec_{x|\by} \rangle  \fp{}{x}Z(x)\right) &= \lambda^2 Z(x), \qquad
Z(x \leq y_0) = Z(x \geq y_N) = 0.
\end{align}
Here the function $Z(x) = Z_{\lambda,  \by}(x) \in {\cal C}^1 \  \forall x \in [y_0, y_N]$, and so is the flux
\begin{equation}
\sigma_i^2 Z'_x \Big|_{x \to y_{i}-0} =\sigma_{i+1}^2 Z'_x \Big|_{x \to y_{i}+0}, \quad i = 1,\dots,N-1.
\end{equation}

The functions $Z_{\lambda, \by}(x)$ form an orthogonal basis in $[y_0, y_N]$. Indeed, let us introduce two functions $Z_{\lambda,\by}(x)$ and $Y_{\mu,\by}(x)$ associated with two different parameters $\lambda, \mu, \ \lambda \neq \mu$ in \eqref{disEq1}, and consider the following identity
\begin{align} \label{int0}
\int_{y_0}^{y_N} \left[
\fp{}{x}\left( \langle  \sigmaVec^2,\IndVec_{x|\by} \rangle  \fp{}{x}Z(x)\right) Y(x) -
\fp{}{x}\left( \langle  \sigmaVec^2,\IndVec_{x|\by} \rangle  \fp{}{x}Y(x)\right) Z(x)
\right] dx  = (\lambda^2 - \mu^2) \int_{y_0}^{y_N}  Z(x) Y(x) dx.
\end{align}

The integral in the LHS of \eqref{int0} vanishes. Indeed, integrating by parts, we obtain
\begin{align}
\int_{y_0}^{y_N} &\left[\fp{}{x}\left( \langle  \sigmaVec^2,\IndVec_{x|\by} \rangle  \fp{}{x}Z(x)\right) Y(x) -
\fp{}{x}\left( \langle  \sigmaVec^2,\IndVec_{x|\by} \rangle  \fp{}{x}Y(x)\right) Z(x) \right] dx  \\
&= \int_{y_0}^{y_N} \left[\langle \sigmaVec^2,\IndVec_{x | \by}\rangle   \left(Z'(x) Y'(x) - Y'(x)Z'(x) \right) 		\right] dx \nonumber \\
&+ \sum_{i=1}^{N}\langle \sigmaVec^2,\IndVec_{x | \by}\rangle \left[Z'(x) Y(x)   - Y'(x) Z(x)  \right]\bigg|_{x = y_i + 0}	- \sum_{i=0}^{N - 1}\langle \sigmaVec^2,\IndVec_{x | \by}\rangle \left[Z'(x)  Y(x) - Y'(x)Z(x) \right]\bigg|_{x = y_i -0} \nonumber		\\
&= \sum_{i=1}^{N - 1} \left( \sigma_i^2 \left[ Z'(y_i + 0) Y(y_i)   - Y'(y_i+0) Z(y_i)\right] - \sigma_{i-1}^2
\left[ Z'(y_i - 0) Y(y_i)   - Y'(y_i-0) Z(y_i)\right] \right) \nonumber \\	
&= \sum_{i=1}^{N - 1} \left( Y(y_i) \left[ \sigma_i^2 Z'(y_i + 0) -\sigma_{i-1}^2 Z'(y_i - 0)\right] -
Z(y_i) \left[ \sigma_i^2 Y'(y_i + 0) -\sigma_{i-1}^2 Y'(y_i - 0)\right] \right) =0. \nonumber
\end{align}
Since $\lambda \neq \mu$, that means that the functions $Z(x)$ and $Y(x)$ are orthogonal.

We proceed with an explicit construction of the orthogonal basis associated with the problem  \eqref{SL_problem}, \eqref{bc}, \eqref{SL_problem:DISC_SPEC}. Let us define the function $\ThetaVec_{\by, \lambda}(x)$ as follows
\begin{equation} \label{DS:theta_def}
\ThetaVec_{\by, \lambda}(x) = \left[0, \Theta_{\by, \lambda, 1}(x),\dots,\Theta_{\by, \lambda, N}(x), 0\right]^\top.	
\end{equation}
In turn, each component of $\ThetaVec_{\by, \lambda}(x)$ can be represented as
\begin{equation} \label{DS:theta_def_CD}
\Theta_{\by, \lambda, i}(x) = C_i(\lambda, \by) \cos\left(\bl_i x\right) + D_i(\lambda, \by)\sin\left(\bl_i x \right), \qquad \bl_i = \lambda/\sigma_i.
\end{equation}
The unknown coefficients $C_i, D_i, \ i=1,\dots,N$ can be found via recursion
\begin{align} \label{DS:rec_eq1}
C_1(\lambda, \by) = -\sin\left(\bl_i y_0\right), \qquad	 D_1(\lambda, \by) &=  \cos\left(\bl_i y_0\right), \\
\begin{bmatrix}
  \cos\left(\bl_{i+1} y_i\right)  &  \sin\left(\bl_{i+1} y_i\right) \\
- \sin\left(\bl_{i+1} y_i\right)  &  \cos\left(\bl_{i+1} y_i\right)
\end{bmatrix}
\begin{bmatrix}
C_{i+1}(\lambda, \by) \\
D_{i+1}(\lambda, \by)
\end{bmatrix}
&=
\begin{bmatrix}
      \cos\left(\bl_i y_i\right)   &       \sin\left(\bl_i y_i\right) \\
-s_i \sin\left(\bl_i y_i\right)   & 	s_i \cos\left(\bl_i y_i\right)
\end{bmatrix}
\begin{bmatrix}
C_i(\lambda, \by) \\
D_i(\lambda, \by)
\end{bmatrix}
, \ s_i = \frac{\sigma_i}{\sigma_{i+1}}. \nonumber
\end{align}

An explicit solution of \eqref{DS:rec_eq1} can be expressed by using the stretch and rotation matrices introduced in \eqref{BG:RotStretch_def}
\begin{equation}
\begin{bmatrix}
C_{i+1}(\lambda, \by) \\
D_{i+1}(\lambda, \by)
\end{bmatrix}
= \left(\RotM_{-\bl_{i+1} y_i}\right)^{-1} \StrM_{s_i}  \RotM_{-\bl_i y_i}
\begin{bmatrix}
C_{i}(\lambda, \by) \\
D_{i}(\lambda, \by),
\end{bmatrix}
\qquad
\end{equation}

Defining the length $l_i = y_{i} - y_{i-1}$ of each segment and using the property  $\RotM_{\alpha}^{-1} = \RotM_{-\alpha}$, $\RotM_{\alpha + \beta} = \RotM_{\alpha} \RotM_{\beta}$ this recurrent formula can be written as
\begin{equation} \label{DS:rec_eq2}
\begin{bmatrix}
C_{i}(\lambda, \by) \\
D_{i}(\lambda, \by)
\end{bmatrix} =
\RotM_{\bl_i y_{i-1}} \left(\prod_{j=1}^{i-1} \Lambda_j \right) \RotM_{\bl_1 y_{1}}
\begin{bmatrix}
C_{1}(\lambda, \by) \\
D_{1}(\lambda, \by,
\end{bmatrix}
, \quad \Lambda_1 = \StrM_{s_1}, \quad \Lambda_i = \StrM_{s_{i}} \, \RotM_{- \bl_{i-1}l_{i-1}}, \quad i > 1.
\end{equation}

The eigenvalues $\lambda_n$ solve the transcendental equation
\begin{align} \label{DS:eig_eq_aux}
\cos\left( \bl_N y_{N} \right) C_{N}(\lambda, \by)  + \sin\left( \bl_N y_{N}\right) D_{N}(\lambda, \by)  = 0,
\qquad \lambda = \lambda_n.'
\end{align}
Taking into account the following identities
\begin{align*}
\RotM_{-\bl_1 y_1}	
\begin{bmatrix}
C_{1}(\lambda, \by) \\
D_{1}(\lambda, \by)
\end{bmatrix} \
&=
\RotM_{-\bl_1 y_1}	
\begin{bmatrix}
-\sin\left(\bl_1 y_0\right) \\
\cos\left(\bl_1 y_0\right)
\end{bmatrix}
=
\begin{bmatrix}
\sin \left(\bl_1 l_1\right) \\
\cos \left(\bl_1 l_1\right)
\end{bmatrix}, \\
\begin{bmatrix}
\cos\left(\bl_N y_N\right) \\
\sin\left(\bl_N y_N\right)
\end{bmatrix}^\top \RotM_{-\bl_N y_{N-1}}
&=
\begin{bmatrix}
\cos\left(\bl_N l_N\right) \\
\sin\left(\bl_N l_N\right)
\end{bmatrix}^\top,
\end{align*}
\noindent the \eqref{DS:eig_eq_aux} can also be re-written in a more explicit form
\begin{align}  \label{DS:eig_eq}
\begin{bmatrix}
\cos\left(\bl_N l_{N}\right) \\
\sin\left(\bl l_{N}\right)
\end{bmatrix}^\top
\left(\prod_{i=1}^{N-1} \Lambda_i  \right)
\begin{bmatrix}
\sin\left(\bl_1  l_{1}\right) \\
\cos\left(\bl_1 l_{1}\right)
\end{bmatrix}
=0. \qquad \lambda = \lambda_n.
\end{align}

Thus, we arrive at the following series expansion for the delta function
\begin{equation} \label{DS:delta_repr}
\delta(x - x_0) =  \sum_{n = 1}^\infty \frac{
\langle  \ThetaVec_{\lambda_n, \by}(x_0), \IndVec_{x_0 | \by}\rangle
\langle  \ThetaVec_{\lambda_n, \by}(x), \IndVec_{x| \by} \rangle }
{\int_{y_0}^{y_N} \langle  \ThetaVec^2_{\lambda_n, \by}(\xi), \IndVec_{\xi | \by}\rangle  d\xi}.	
\end{equation}

\paragraph{Oscillating Fourier Series.} Using the representation \eqref{DS:delta_repr} we obtain the following \textit{Oscillating Fourier series}
\begin{equation} \label{DS:osc_Fourier_series}
f(x) = \sum_{n = 1}^\infty \frac{\langle  \ThetaVec_{\lambda_n, \by}(x), \IndVec_{x| \by}\rangle  \bar f (\lambda_n)} {\int_{y_0}^{y_N} \langle  \ThetaVec_{\lambda_n, \by}(\xi), \IndVec_{\xi | \by}\rangle ^2 d\xi},
\qquad
\bar f(\lambda) = \int_{y_0}^{y_N} f(\xi) \langle  \ThetaVec_{\lambda, \by}(\xi), \IndVec_{\xi | \by}\rangle  d\xi.
\end{equation}
The eigenvalues $\lambda_n$ are defined as an increasing sequence of the solutions of \eqref{DS:eig_eq}.

The function $\ThetaVec_{\lambda_n, \by}(x)$ can be considered as a generalization of the sine function $\sin\left(\frac{\pi n x}{l}\right)$. Indeed, if $\sigmaVec$ has the form
\begin{equation*}
\sigmaVec = \left[0, \sigma, \sigma, \dots, \sigma, 0\right]^\top,
\end{equation*}
\noindent all stretch matrices $\StrM$ in \eqref{DS:rec_eq2} and \eqref{DS:eig_eq} are equal to the identity matrix. Using a semi-group property of the rotation matrices yields the following equation for the eigenvalues $\lambda_n$
\begin{align}
\begin{bmatrix}
\cos\left( \frac{\lambda_n }{\sigma} l_{N}\right) \\
\sin\left( \frac{\lambda_n }{\sigma} l_{N}\right)
\end{bmatrix}^\top
\RotM_{ -\sum_{k = 2}^{N-1} \frac{\lambda_n l_{k}}{\sigma}}
\begin{bmatrix}
\sin\left( \frac{\lambda_n }{\sigma} l_{1}\right) \\
\cos\left( \frac{\lambda_n }{\sigma} l_{1}\right)
\end{bmatrix}
&= 0,
\end{align}
\noindent from which it follows that $\sin\left( \frac{l \lambda_n}{\sigma}\right) = 0$ and $\lambda_n = \frac{\pi n \sigma}{l}, \ l = \sum_{k =1}^{N-1} l_k = y_N - y_0$. Computing the constants $C_i$ and $D_i$ and substituting the values of $\lambda_n$ gives rise to the following formula for $\Theta_{y, \lambda_n}(x)$
\begin{equation}
\Theta_{\by, \lambda_n}(x) = \left[ 0, \sin\left(\pi n\frac{x - y_0}{l}\right),  \sin\left(\pi n\frac{x - y_0}{l}\right), \dots , \sin\left(\pi n\frac{x - y_0}{l}\right), 0 \right]^\top.	
\end{equation}
Hence, when $\sigma_i = \sigma= const, \ \forall i=1,\dots,N$ the eigenfunctions coincide with the sine functions.

\subsubsection{Mixed spectrum}

As compared with the previous cases this is the most complicated problem. 	Let us slightly change the notation to $\sigma_- = \sigma_0, \ \sigma_+ = \sigma_{N+1}, \ l_i = y_{i} - y_{i-1}, \ l_0 = 0$. Let us look for the solution $u_\lambda(x|x_0)$ in the form
\begin{equation} \label{uLmbdaMixed}
u_\lambda(x | x_0) = \langle  \mathbf{u}, \IndVec_{x | \by}\rangle , \quad \mathbf{u} = \left(u_0, u_1,\dots, u_N, u_{N+1}\right),
\end{equation}
\noindent where
\begin{align} \label{MS:u_def}
u_{i} = (1-\delta_{i,N+1}) C_i \cosh\left(\tl_i x\right)  &+ (1-\delta_{i,0})D_i \sinh\left(\tl_i x\right) + \frac{1}{2\sqrt{\lambda} \sigma_i} e^{-\tl_i |x-x_0|},  \\
\tl_i &= \frac{\sqrt{\lambda}}{\sigma_i}, \quad i=0,\dots,N+1, \nonumber
\end{align}
\noindent and $\delta_{i,b}$ is the Kronecker delta. Let us define the auxiliary matrix $\GM_{\lambda} (\alpha, \beta | z)$
\begin{equation} \label{MS:GM_def}
\GM_\lambda (\alpha, \beta | z) =
\begin{bmatrix}
e_\lambda(z,\beta, \alpha) - e_\lambda(z,\beta, \beta)  & e_\lambda(z, \alpha, \alpha) - e_\lambda(z,\beta, \beta) \\
e_\lambda(z,\alpha, \beta) - e_\lambda(z,\beta, \beta)  & e_\lambda(z, \alpha, \beta) - e_\lambda(z,\beta, \beta)
\end{bmatrix}
, \qquad e_\lambda(z,a,b) = \frac{1}{2\sqrt{\lambda} b}e^{-\frac{|z|\sqrt{\lambda}}{a}}.
\end{equation}
Using the matching conditions at $x = y_0$ for the functions $u_0$ and $u_1$, we obtain the equation
\begin{equation} \label{MS:C-_def}
\begin{bmatrix} C_1 \\ D_1 \end{bmatrix} =
\HM_{-\tl_1 y_0} \StrM_{s_0}
\begin{bmatrix} 1 \\ 1 \end{bmatrix}
C_-  + \HM_{-\tl_1 y_0} \GM_\lambda(\sigma_0, \sigma_1 | x_0 - y_0) \IndVec_{x_0 | y_0}, \qquad
s_- = \frac{\sigma_-}{\sigma_1}.
\end{equation}	
By analogy with Section~\ref{discSpect}, it is possible to derive recurrent relationships between the coefficients $C_i$, $D_i$ and $C_{i + 1}$, $D_{i +1}$ which read
\begin{equation}  \label{MS:Cn_def}
\begin{bmatrix} C_{i+1}  \\ D_{i+1} \end{bmatrix}
= \HM_{-\tl_{i+1} y_i} \StrM_{s_i} \HM_{\tl_i y_i}
\begin{bmatrix} 	C_{i}  \\ D_{i} 	\end{bmatrix} 	
+ 	\HM_{-\tl_{i+1} y_i} \GM_\lambda(\sigma_i, \sigma_{i + 1} | x_0 - y_i) \IndVec_{x_0 | y_i}.
\end{equation}
And at the last point $x = y_N$ the matching conditions yield the equation
\begin{equation} \label{MS:D+_def}
\begin{bmatrix} 1 \\ -1 \end{bmatrix}
D_+ = \StrM_{s_N} \HM_{\tl_N y_N}
\begin{bmatrix} 	C_n \\ D_n \end{bmatrix}
+ \GM_\lambda(\sigma_N, \sigma_{N+1} | x_0 - y_N) \IndVec_{x_0 | y_N}.
\end{equation}

Combining \eqref{MS:C-_def}, \eqref{MS:Cn_def}, \eqref{MS:D+_def} together we arrive at the system of equations to determine $C_-$ and $D_+$
\begin{equation}\label{DS:CD_system_raw}
\begin{bmatrix} 	1 \\ -1 \end{bmatrix}
D_+ = \prod_{j = 0}^N \StrM_{s_j} \HM_{\tl_j l_j}	
\begin{bmatrix} 	1 \\ 1 \end{bmatrix} C_-
+ \sum_{k =0}^N \left( \prod_{p = k + 1}^N \StrM_{s_p} \HM_{\tl_p l_p}\right) \GM_\lambda(\sigma_k, \sigma_{k+1} | x_0 - y_k) \IndVec_{x_0 |y_k}.
\end{equation}
This system can also be re-written in the matrix form if we define the matrix $\LambdaM(\lambda)$
\begin{equation}
\LambdaM(\lambda) = \prod_{j = 0}^N \StrM_{s_j} \HM_{\tl_j l_j},
\end{equation}
\noindent with elements $\Lambda_{i,j}, \ i,j=1,\dots,N+1$, so
\begin{equation}
\begin{bmatrix}
	\LambdaM_{11}(\lambda)  + \LambdaM_{12}(\lambda)  & -1 \\
	\LambdaM_{21}(\lambda)  + \LambdaM_{22}(\lambda)  &  1 \\	
\end{bmatrix}
\begin{bmatrix} 	C_- \\ D_+ 	\end{bmatrix}
= -\sum_{k =0}^N \left( \prod_{p = k + 1}^N \StrM_{s_p} \HM_{\tl_p l_p} \right) \GM_\lambda(\sigma_k, \sigma_{k+1} | x_0 - y_k) \IndVec_{x_0 |y_k}.
\end{equation}
Solving for the coefficients $C_-$ and $D_+$ yields
\begin{equation} \label{MS:CD_eq}
\begin{bmatrix}
C_- \\ D_+
\end{bmatrix} =
- \frac{\LambdaM^*(\lambda) }{\det \LambdaM^*(\lambda) } \sum_{k =0}^N
\left(\prod_{p = k + 1}^N \StrM_{s_p} \HM_{\tl_p l_p} \right) \GM_\lambda(\sigma_k, \sigma_{k+1} | x_0 - y_k) \IndVec_{x_0 |y_k},
\end{equation}
\noindent where $\LambdaM^*$ is an adjacent matrix
\begin{equation}
\LambdaM^*(\lambda)  =
\begin{bmatrix}
1 & 1 \\
-\LambdaM_{21}(\lambda)  - \LambdaM_{22}(\lambda)  &
\LambdaM_{11}(\lambda)  + \LambdaM_{12}(\lambda).
\end{bmatrix}.
\end{equation}

In a similar manner let us define a sequence of matrices
\begin{equation} \label{MS:LambdaSeq}
\LambdaM_{k}^{p}(\lambda)  = \prod_{j=k}^{p-1} \StrM_{s_j} \HM_{\tl_j l_j}, \qquad
\LambdaM(\lambda) = \LambdaM_{0}^{N+1}(\lambda), \qquad \LambdaM_{p}^{p}(\lambda) =
\begin{bmatrix} 1 & 0 \\ 0 & 1 \end{bmatrix}.
\end{equation}
Re-writing the recursion in \eqref{MS:Cn_def} by using these matrices and solving for the coefficients $C_i$ and $D_i$ yields
\begin{equation} \label{solMixed}
\begin{bmatrix} C_i \\ D_i \end{bmatrix}
= C_- \HM_{- \tl_i  y_{i-1}} \LambdaM_{0}^{i}(\lambda)
\begin{bmatrix} 1 \\ 1 \end{bmatrix}
+ \HM_{-\tl_i y_{i-1}} \sum_{k = 0}^i \LambdaM_{k+1}^{i+1}(\lambda)
\GM_\lambda (\sigma_k, \sigma_{k+1} | x_0 - y_k) \IndVec_{x_0 | y_k}.
\end{equation}
Therefore, the $i$-th component of $\mathbf{u}$ can be expressed as
\begin{equation} \label{DS:ui_def}
u_i(x) =
\begin{bmatrix}
\cosh\left(\tl_i (x - y_{i-1}\right) \\
\sinh\left( \tl_i (x - y_{i-1})\right) \\ 	
\end{bmatrix}^\top
\left( C_- \LambdaM_{0}^{i}(\lambda)
\begin{bmatrix} 1\\ 1\end{bmatrix}
+ \sum_{k = 0}^i \LambdaM_{k+1}^{i+1}(\lambda) \GM_\lambda(\sigma_k, \sigma_{k+1} | x_0 - y_k) \IndVec_{x_0 | y_k} \right) + \frac{e^{- \tl_i |x-x_0|}}{2\sqrt{\lambda} \sigma_i}.
\end{equation}

Having \eqref{solMixed} we can substitute it into \eqref{uLmbdaMixed} and then into the Bromwich integral in \eqref{SL:Dirac_def}. It turns out, however, that the integrands in \eqref{DS:ui_def} have a branching point at $\lambda = 0$.  and poles at  $\lambda = \lambda_n$, $n = 1 \dots \infty$ where $\lambda_n$ are the roots of the equation $\det \LambdaM^*(\lambda) = 0$. By doing contour integration, see Appendix~\ref{appContour}, we obtain the necessary representation of the Dirac delta function
\begin{equation} \label{MS:delta_repr}
\delta(x - x_0) = \left\langle \frac{1}{\pi \iu} \int_{0}^{\infty} \left[\mathbf{u}_{e^{-\iu \pi} \omega^2}(x | x_0) - \mathbf{u}_{e^{\iu \pi} \omega^2}(x | x_0)\right] \omega\, d\omega
+\sum_{k=1}^\infty \underset{\lambda=\lambda_k}{\operatorname{Res}}\left(\mathbf{u}_\lambda (x| x_0)\right), \IndVec_{x | \by} \right\rangle.
\end{equation}

\section{Representative examples} \label{examples1}

As mentioned already, the series expansion of the Dirac delta function in the basis of eigenfunctions of the corresponding SL problem provides a mechanism of building integral transforms with necessary properties suitable for solving a given multilayer problem. This section offers examples showing how this relatively abstract concept can be implemented in two and three-dimensional cases.

\subsection{Model with two non-zero layers}

Let us consider the problem \eqref{SL_problem},\eqref{SL_problem:DISC_SPEC} with $N = 2$, so the corresponding vectors $\by$ and $\sigmaVec$ take the form
\begin{equation} \label{DS:ex:y_and_sigma_specs}
\by = \left( y_0, y_1, y_2\right)^\top, \quad \sigmaVec = \left(0, \sigma_1, \sigma_2, 0\right)^\top,
\end{equation}
\noindent and the basis function $\Theta_{\by, \lambda}(x)$ reads
\begin{equation}
\Theta_{\by, \lambda}(x) =
\begin{cases}
0, & x \in (-\infty, y_0], \\
C_1(\lambda, \by) \cos\left(\bl_1 x \right) + 	D_1(\lambda, \by) \sin\left(\bl_1 x \right), & x\in [y_0, y_1],\\
C_2(\lambda, \by) \cos\left(\bl_2 x \right) + D_2(\lambda, \by) \sin\left(\bl_2 x \right),	 & x\in [y_1, y_2],\\
0, & x \in [y_2, \infty).
\end{cases}
\end{equation}
The coefficients $C_1(\lambda, \by)$ and $D_1(\lambda, \by)$ found by using the corresponding procedure described in Section~\ref{discSpect} read
\begin{equation*}
C_1(\lambda, \by) = -\sin\left(\bl_1 y_{0}\right), \quad D_1(\lambda, \by) = \cos\left(\bl_1 y_{0}\right),
\end{equation*}
\noindent and so $C_2(\lambda, \by)$ and $D_2(\lambda, \by)$ follow as
\begin{equation}
\begin{bmatrix}		C_{2}(\lambda, \by) \\  D_{2}(\lambda, \by) \end{bmatrix} =
\RotM_{\bl_2 y_{1}} \StrM_{s_1} \RotM_{\bl_1 y_{1}}
\begin{bmatrix} C_{1}(\lambda, \by) \\  D_{1}(\lambda, \by) \end{bmatrix}.
\end{equation}
Thus, the function $\ThetaVec_{\lambda, \by}(x)$ in the explicit form reads
\begin{equation} \label{DS:2layer_example:theta_gen}
\Theta_{\by, \lambda}(x) =
\begin{cases}
0, & x \in (-\infty, y_0], \\
\sin\left(\bl_1 (x - y_0)\right), 	& x\in [y_0, y_1],\\
\cos\left(\bl_2(x - y_1)\right) \sin\left(\bl_1 l_1\right) + s_1 \sin\left(\bl_2(x - y_1)\right) \cos\left(\bl_1 l_1\right),	& x\in [y_1, y_2],\\
0, & x \in [y_2, \infty).
\end{cases}
\end{equation}

In turn, the eigenvalues $\left\{ \lambda_n\right\}_{n = 1}^\infty$ solve the equation
\begin{align}
\begin{bmatrix}
\cos\left(\bl_2(y_2 - y_1)\right) & \sin\left(\bl_2(y_2 - y_1)\right)
\end{bmatrix}
\begin{bmatrix}
1 & 0\\
0 & s_1
\end{bmatrix}
\begin{bmatrix} \sin\left(\bl_1(y_1 - y_0)\right) \\ \cos\left(\bl_1(y_1 - y_0)\right) \end{bmatrix} = 0,
\end{align}
\noindent or, alternatively
\begin{equation} \label{posRoots}
\left(\sigma_2 - \sigma_1\right) \sin\left(\bl_2 l_2 - \bl_1 l_1 \right) =
\left(\sigma_2 + \sigma_1\right) \sin\left(\bl_2 l_2 + \bl_1 l_1\right).
\end{equation}
Therefore, we  define the eigenvalues $\lambda_1, \lambda_2, \cdots$ as the ordered sequence of positive roots of this equation. With allowance for \eqref{posRoots}, the function in \eqref{DS:2layer_example:theta_gen} can be simplified
\begin{equation}
\ThetaVec_{\by, n}(x) = \ThetaVec_{\by, \lambda_n}(x) =
\begin{cases}
0, & x \in (-\infty, y_0], \\
\sin\left(\bl_1(x - y_0)\right), & x\in [y_0, y_1],\\
\frac{\sin\left(\bl_1 l_1\right)}{\sin\left(\bl_2 l_2\right)} \sin\left(\bl_2 (y_2 - x)\right), & x\in [y_1, y_2],\\
0, & x \in [y_2, \infty),
\end{cases}
\end{equation}
\noindent with $\lambda = \lambda_n$.

\begin{figure}[!htb]
\vspace{-0.1in}
\begin{center}
\includegraphics[totalheight=3.5in]{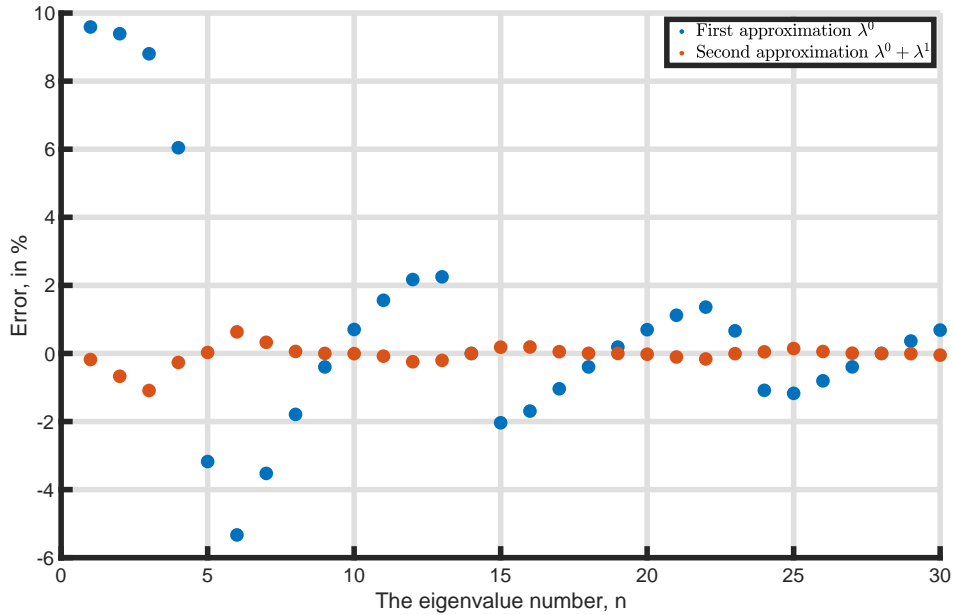}
\caption{The error of zero and first order approximations in solving \eqref{posRoots} as compared with the numerical solution of \eqref{posRoots} when $\sigma_1 = 7$, $\sigma_2 = 0.7$, $l_1 = 1.2$, $l_2 = 1$. }
\label{Fig1}
\end{center}
\end{figure}

\paragraph{Flat $\boldsymbol\sigma$.} If $\sigma_1$ is close to $\sigma_2$, so the function $\sigma$ is almost flat, the following approximation holds
\begin{equation} \label{DS:ex:lambda_approx_zero}
	\lambda_n \approx \lambda_n^0 = \frac{\pi n }{l_1 / \sigma_1 + l_2 / \sigma_2}
\end{equation}
Using the Taylor series expansion, the next approximation $\lambda_n^1$ can also be obtained by solving the equation
\begin{equation*}	
\alpha
\sin\left( \beta \left[\lambda_n^0  + \lambda_n^1\right] \right) =
\sin\left( \left[\lambda_n^0  + \lambda_n^1\right] \left( \frac{l_2}{\sigma_2} + \frac{l_1}{\sigma_1} \right)\right), \qquad
\alpha = \frac{\sigma_2 - \sigma_1}{\sigma_2 + \sigma_1}, \quad
\beta = \frac{l_2}{ \sigma_2} - \frac{l_1}{ \sigma_1},
\end{equation*}
\noindent or
\begin{equation*}
\alpha	\left[	\sin\left(\lambda_n^0 \beta \right)  + \lambda_n^1 \beta \cos\left( \lambda_n^0 \beta \right) \right] \approx (-1)^{n} \lambda_n^1 \left[\frac{l_2}{\sigma_2} + \frac{l_1}{\sigma_1} \right],
\end{equation*}
\noindent which yields
\begin{equation*}
\lambda_n^1 \approx \frac{\alpha \sin\left(\beta \lambda_n^0 \right)}{(-1)^{n}  \left[\frac{l_2}{\sigma_2} + \frac{l_1}{\sigma_1} \right] - \alpha \cos \left(\beta \lambda_n^0 \right)}.
\end{equation*}
Therefore, in the next order of approximation we get $\lambda_n \approx \lambda_n^0  + \lambda_n^1$. Surprisingly, this formula works well even if $|\sigma_1 - \sigma_2|$ is sufficiently large. In Fig.\ref{Fig1} the error in computing the first 30 eigenvalues using the approximations with one and two terms in the Taylor series expansion of $\sigma$ is presented using $\sigma_1 = 7$, $\sigma_2 = 0.7$, $l_1 = 1.2$ and $l_2 = 1$. The figure shows that the zero-order approximation $\lambda_n^0$ has poor quality, and the error in computing the first five eigenvalues is about 5-10 percent. However, the first-order approximation is much more accurate and provides about one percent difference in the worst case (the 3rd and 6th eigenvalues, respectively). In any case, those approximations can be used as a good initial guess when solving  \eqref{posRoots} numerically. Note, that it is possible to derive similar approximations for the general multi-layer case in \eqref{DS:eig_eq}.

\subsection{Model with three layers and continuous spectrum}

Here we consider the problem \eqref{SL_problem}, \eqref{SL_problem:CONT_SPEC} with $N = 2$, so the corresponding vectors $\by$ and $\sigmaVec$ are
\begin{equation} \label{MS:ex:y_and_sigma_specs}
\by = \left(y_0, y_1\right)^\top, \quad
\sigmaVec = \left(\sigma_-, \sigma_1, \sigma_+\right)^\top.
\end{equation}

Using the result in \eqref{MS:u_def}, the function $u_\lambda(x | x_0)$ can be represented as
\begin{align} \label{MS:example:u_final}
u_\lambda(x | x_0) &= \frac{1}{2 \lambda}
\begin{bmatrix}
\tl_- e^{-\tl_- |x - x_0|} \\
\tl_1 e^{-\tl_1 |x - x_0|} \\
\tl_+ e^{-\tl_+ |x - x_0|}
\end{bmatrix}^\top	
\IndVec_{x | \by} + \IndVec_{y_0 < x <y_1}
\begin{bmatrix}
\cosh\left( \tl_1 (x - y_0)\right)		\\
\sinh\left( \tl_1 (x- y_0) \right)
\end{bmatrix} ^\top
\GM_\lambda(\sigma_-, \sigma_1 | x_0 - y_0) \Ind_{x_0 |y_0} \\
&+ \left(
\begin{bmatrix} \Ind_{x < y_0} \\ \Ind_{x > y_1} \end{bmatrix}^\top
\begin{bmatrix} e^{-\tl_- x}  & 0 \\ 0 & e^{-\tl_+ x} \end{bmatrix}
+ \IndVec_{y_0 < x <y_1}
\begin{bmatrix} \cosh\left(\tl_1(x - y_0)\right) \\ \sinh\left(\tl_1 (x- y_0) \right) \end{bmatrix}^\top
\begin{bmatrix} 1 & 0 \\ -s_- & 0  \end{bmatrix}
\right)
\begin{bmatrix} C_- \\ D_+ 	\end{bmatrix}, \nonumber
\end{align}
\noindent while the constants $C_-$ and $D_+$ read
\begin{align} \label{MS:example:CD_eq}
\begin{bmatrix} C_- \\ D_+ \end{bmatrix}
(\lambda) &= -\frac{\LambdaM^*(\lambda) }{\det \LambdaM^*(\lambda) }
\left( \StrM_{s_1} \HM_{\tl_1 l_1} \GM_\lambda(\sigma_-, \sigma_{1} | x_0 - y_0) \IndVec_{x_0 |y_0} +
\GM_\lambda(\sigma_1, \sigma_{+} | x_0 - y_1)\IndVec_{x_0 |y_1} \right), \\
\LambdaM(\lambda) &= \StrM_{s_1} \HM_{\tl_1 l_1} \StrM_{s_-} =
\begin{bmatrix}
\cosh\left(\tl_1  l_1\right) & s_- \sinh\left(\tl_1 l_1\right) \\
s_1 \sinh\left(\tl_1 l_1\right) & \frac{s_-}{s_1} \cosh\left(\tl_1 l_1\right)
\end{bmatrix}. \nonumber
\end{align}
The determinant $\det \LambdaM^*$ can be found explicitly
\begin{align}
\det \LambdaM^*(\lambda) &= \left(s_1  + s_- \right) \sinh\left(\tl_1 l_1\right) + \left(1 + s_-\right) \cosh\left(\tl_1 l_1\right).
\end{align}
Moreover, \eqref{DS:eig_eq_aux} can also be solved in closed form which yields
\begin{equation} \label{MS:example_spectrum_eq}
\tanh\left(\tl_1 l_1\right) = -\frac{\sigma_1(\sigma_- + \sigma_+)}{\sigma_1^2 + \sigma_-\sigma_+}
= -\frac{s_- + 1/s_1}{1 + s_-/s_1}.
\end{equation}
From here we find the explicit representation for the eigenvalues $\lambda_n^\pm$
\begin{align}
\lambda_n^\pm &= \frac{\sigma_1^2}{l_1^2} \left(\chi^2 -\pi^2 n^2  \pm 2\iu \pi \chi\right), \quad
\chi = -\arctanh \left(\frac{s_- + 1/s_1}{1 + s_-/s_1}\right).
\end{align}
Thus, we have managed to explicitly find all the components in \eqref{MS:delta_repr}, which makes it possible to construct an oscillating integral transform in this case. However, since the final result looks bulky, and we don't use it in this paper, it will be published elsewhere.

\section{MHE for problems with curvilinear boundaries} \label{MHEcurv}

The previous two sections cover the construction of a particular type of integral transform - the OIT. Although these sections are a bit technical, we cannot omit them since OIT is new and has never been reported in the literature. However, once these results have been obtained and described, we can concentrate on solving some more practical problems which regularly occur in various areas of science and engineering. In particular, in this section, we solve MHE defined at the space domain with curvilinear (time-dependent) boundaries by using the corresponding OIT, constructed in \eqref{CS:osc_identity_f} and \eqref{DS:osc_Fourier_series}. A similar OIT can also be derived for problems with mixed spectrum by using a general representation \eqref{MS:delta_repr}, but we don't consider it in this paper.

\subsection{Time-dependent oscillating Brownian motion}

The oscillating Brownian motion (OBM) is described in detail in \citep{KeilsonWellner1978}. Let $\sigma(x) = \sigma_+, \ x \ge 0, \ \sigma(x) = \sigma_-, \ x < 0$ and let $m(dx) = 2\sigma^{-2}(x) dx$ be speed measure. In \citep{ItoMcKean1965} the authors construct a diffusion process $Y_t$ with an arbitrary speed measure $m$ (which has support at an interval $Q \in \mathbb{R}$) from Brownian motion and show that $Y_t$ is a strong Markov, conservative diffusion process on $Q$. They call it  the OBM with the desired speed measure $m$. The process $Y_t$ "oscillates" in a sense that it behaves like a Brownian motion which changes variance parameter each time it crosses zero. In the well-known special case $\sigma_- = \sigma_+ = \sigma$, $Y_t$ is an ordinary Brownian motion with variance $\sigma^2$. Another known case is $\sigma_+ = \sigma$, $\sigma_- = \infty$, and then $Y_t$ is a reflecting Brownian motion.

Applications of the OBM in finance could be found in \citep{LejayPigato2019,Gairat2016} who proposed a threshold local volatility model with piecewise constant volatility and drift (the geometric oscillating Brownian motion (GOBM))
\footnote{Note that in \citep{Schoutens2006} the authors study a class of tractable diffusions suitable for the model's primitives of interest rates. They consider scalar diffusions with scale $s(x)$ and speed $m(x)$ densities discontinuous at the level $x^*$, which they call a family of Self Exciting Threshold diffusions. Those processes can also be considered as examples of the OBM.}. This model is an instance of the tiled volatility model considered in \citep{LiptonSepp2011} and later extended in \citep{ItkinLipton2017}. A detailed description of the mathematical aspects of this problem and solutions can also be found in \citep{Lipton2018a}.

In the GOBM model, a fixed threshold separates two regimes for the prices. The volatility and the drift parameter can assume two possible values, according to the stock price position, above or below the threshold. Thus, $\sigma_-$ is
the volatility below the threshold, $\sigma_+$ - the volatility above the threshold, and similarly $b_-, b_+$ for the drift. It turns out that such model accounts for the leverage effect when $\sigma_- > \sigma_+$. In this case, when prices are low, volatility increases, consistently with what is observed on empirical financial data. As mentioned in \citep{LejayPigato2019}, a motivation for considering such price dynamics coming from a different viewpoint is given in \citep{Ankirchner2017}. It is shown in \citep{Ankirchner2017} that the GOBM describes the price dynamics corresponding to the optimal strategy for a manager who can control, in a stylized setting, the volatility of the value of a firm, getting bonus payments when the value process performs better than a reference index.

In terms of financial mathematics, the GOBM model is a local volatility model of the form
\begin{eqnarray}
	d S_t = \mu(S_t) S_t dt + \sigma(S_t) S_t dW_t,
\end{eqnarray}
\noindent where $S_t$ is the stock price, $T$ is the time, $\mu(S_t)$ is the drift, $\sigma(S_t)$ is the local volatility function, $W_t$ is the standard Brownian motion, and the drift and local volatility are defined as
\begin{equation}
\sigma(x) =
\begin{cases}
	\sigma_-, & x < m, \\
	\sigma_+, & x \ge m.
\end{cases}
\quad \textrm{and} \quad
\mu(x) =
\begin{cases}
	\mu_-, & x < m, \\
	\mu_+, & x \ge m.
\end{cases}
\end{equation}
In the GOBM model, the threshold $m$ is constant, yet the time-dependent version of the model has not been considered in the literature.

Another step in this direction has been done in \citep{FrizPigatoSeibel2020} who proposed the Step Stochastic Volatility Model (SSVM).  The authors were looking for a (possible) simple (without adding jumps or non-Markovian rough fractional volatility dynamics) modification of a class of stochastic volatility models to be capable of producing extreme short-dated implied volatility skew\footnote{As the authors mentioned, much of the recent success of rough SVMs is due to the (desired) implied skew blow up at rate $T^{H-1/2}$, $T$ is the time to maturity, and the Hurst parameter $H \in (0, 1/2]$ quantifies the roughness of the volatility process. The blowup can be at most of the order $T^{-1/2}$ which is a model-free consequence of no-arbitrage, \citep{Lee2002}.}. They found that introducing a leverage effect by making volatility discontinuous at the money (by multiplying the (backbone) stochastic volatility with distinct factors, say $\sigma_-$ and $\sigma_+$ depending on when the considered option is out-of-the-money or in-the-money) achieves this goal, and the implied skew generated by such model explodes as $T^{-1/2}$, \citep{Pigato2019}.

Going from finance to physics, diffusion in nonuniform media is another example of the OBM or skew Brownian motion playing an essential role in practice; see, e.g., in \citep{Sattin2008, Andreucci2019} and references therein. In the presence of substantial inhomogeneities, even the hydrodynamic Fokker-Planck equation limit may become inaccurate and mask some features of the real solution, as computed from the Master Equation. For example, the author considers an experimental test - a study of tracer diffusion between gelatin solutions with different viscosity or different effective diffusivity. The width of the interface between the two solutions is minimal and can be put to zero. Hence, the whole system may be modeled as two regions with different jumping lengths - the same setting as regions with different variance in the mathematical finance context. Again, the relevant domain is split by a constant interface $x=0$. However, this problem can be further generalized by considering time-dependent boundaries.

For the financial models mentioned above, it seems more plausible that since the boundary (threshold) for the oscillating volatility should be close to the at-the-money (ATM) level, it would be beneficial to consider it a function of time. In this paper, we concentrate on MHEs. However, our method can be further generalized to the problems that can be reduced to the MHE by transformations. In many cases, even the time-dependent models can be reduced to the MHE (or to the Multilayer Bessel equations, \citep{Schoutens20061}). However, the boundaries inherent to those problems and constant in the original variables become time-dependent in the new variables; see various examples in \citep{ItkinLiptonMuraveyBook}. Therefore, our framework covers both problems with time-dependent interfaces, as well as models with time-dependent parameters.

With allowance for all that, we consider a problem with two layers and time-dependent interface $y(\tau)$ where $\tau = T-t$ is the backward time
\begin{alignat}{2} 	\label{PDE:osc_main_PDE}
\frac{\partial u}{\partial \tau} &= \frac{\partial}{\partial x}\left(\langle \Ind_{x | y(\tau)}, \sigmaVec^2\rangle
\frac{\partial u}{\partial x} \right), &\qquad \sigmaVec &= \left(\sigma_-, \sigma_+\right) \\
\lim_{x \to \pm \infty} u(\tau, x) &= 0, &\qquad u(0, x) &= \delta(x-x_0), \nonumber \\
\lim_{x \to y(\tau)-0} u(\tau, x)  &= \lim_{x \to y(\tau) +0} u(\tau, x), &\qquad
\lim_{x \to y(\tau) -0}  \sigma_-^2 u'_x(\tau,x) &=	\lim_{x \to y(\tau) +0}  \sigma_+^2 u'_x(\tau,x). \nonumber
\end{alignat}

The solution of this problem is given in Appendix~\ref{app4} and reads
\begin{align} \label{sol1}
u(\tau, x) &= \left\langle \IndVec_{x_0 | y(0)},  \PM_{\sigmaVec}(x - y(\tau), \tau, | x_0 - y(\tau), 0)
\IndVec_{x | y(\tau)} \right\rangle \\
&+ \int_0^\tau \psi(s) \left\langle
\begin{bsmallmatrix}  1 \\ -1  \end{bsmallmatrix},
\PM_{\sigmaVec}(x - y(\tau), \tau | y(s) - y(\tau), s) 	\IndVec_{x | y(\tau)} ds \right\rangle \nonumber \\
 &+ \int_0^\tau \varphi(s) \left \langle
\begin{bsmallmatrix}	1 \\ -1 \end{bsmallmatrix},
\etaM_{\sigmaVec}(x - y(\tau), \tau | y(s) - y(\tau), s) \IndVec_{x | y(\tau)} \right \rangle ds, \nonumber
\end{align}
\noindent where functions $\PM_{\sigmaVec} (z, \tau | \zeta, s), \ \etaM_{\sigmaVec}(z, \tau | \zeta, s)$ are defined in \eqref{etaM}, and $\psi(\tau) = \Phi(s) + y'(s) \varphi(s)$. The functions $\varphi(\tau)$ and $\Phi(\tau)$ are introduced in \eqref{gradDef} and are yet unknown values of the solution and its spatial gradient at the interface boundary $y(\tau)$. However, differentiating \eqref{sol1} on $x$ and substituting $x \to y(\tau)-0$ \footnote{Alternatively,  due to the last boundary condition in \eqref{PDE:osc_main_PDE} we can use $x \to y(\tau)+0$.} gives rise to a system of linear Volterra integral equations of the second kind for $\varphi(\tau)$ and $\Phi(\tau)$
\begin{align} \label{phiVolterra}
\varphi(\tau) &= \langle \IndVec_{x_0 | y(0)}, \PM_{\sigmaVec}^{1}(0, \tau, | x_0-y(\tau), 0) \rangle
+ \int_0^\tau \psi(s) \left\langle
\begin{bsmallmatrix} 1 \\ -1 \end{bsmallmatrix},
\PM^1_{\sigmaVec}(0, \tau | y(s)-y(\tau), s) \right\rangle ds \\
&+ \int_0^\tau \varphi(s) \left\langle
\begin{bsmallmatrix}	1 \\ -1 \end{bsmallmatrix},
\etaM^1_{\sigmaVec}(0, \tau | y(s)-y(\tau), s) \right\rangle  ds, \nonumber \\
\frac{\Phi(\tau)}{\sigma_-^2} &= \langle \IndVec_{x_0 | y(0)}, \fp{\PM_{\sigmaVec}^{1}}{x}(0, \tau, | x_0-y(\tau), 0) \rangle  + \int_0^\tau \psi(s) \left\langle
\begin{bsmallmatrix} 1 \\ -1 \end{bsmallmatrix}
\fp{\PM^1_{\sigmaVec}}{x}(0, \tau | y(s)-y(\tau), s) \right\rangle ds \nonumber \\
&+ \int_0^\tau \varphi(s) \left\langle
\begin{bsmallmatrix}	1 \\ -1 \end{bsmallmatrix},
\fp{\etaM^1_{\sigmaVec}}{x}(0, \tau | y(s)-y(\tau), s) \right\rangle  ds. \nonumber
\end{align}
\noindent where the superscript in $\PM^i, \ \etaM^i$ denotes a transposed $i$-th column, e.g., $\PM^i =  \begin{bsmallmatrix} \PM_{1i} \\ \PM_{2i} \end{bsmallmatrix}^\top$. This system can be solved numerically; see a detailed discussion in \citep{ItkinLiptonMuraveyBook}. Therefore, the proposed method is semi-analytical in a sense that the solution \eqref{sol1} is explicit but depends on two functions that should be found numerically.

Note, that in the case $y(\tau) = const$, from \eqref{etaM} we have $\etaM^1_{\sigmaVec}(0, \tau | 0, s) = 0$,
\begin{align*}
\PM_{\sigmaVec} (0, \tau | 0, s) &= \frac{1}{2\sqrt{\pi(\tau  -s)}}
\begin{bmatrix} \frac{1}{\sigma_-} & 0 \\ 0 &    \frac{1}{\sigma_+} \end{bmatrix}
\begin{bmatrix} 1 + \Sigma  & 1 + \Sigma \\ 1-\Sigma & 1 - \Sigma  \end{bmatrix},
\end{align*}
\noindent and, therefore,
\begin{equation*}
2\sqrt{\pi(\tau  -s)} \left\langle [1, -1]^\top, \PM^1_{\sigmaVec}(0, \tau | 0, s) \right\rangle =
\begin{bmatrix} 1 \\ -1 \end{bmatrix}
\left[ \frac{1+\Sigma}{\sigma_-}, \frac{1-\Sigma}{\sigma_+}\right]  =
\frac{1+\Sigma}{\sigma_-} - \frac{1-\Sigma}{\sigma_+} = 0.
\end{equation*}
Hence, $\varphi(\tau)$ in \eqref{phiVolterra} becomes an explicit function (since two integrals in the RHS vanish). In a similar way it can be shown that $\fp{\PM^1_{\sigmaVec}}{x}(0, \tau | 0, s) = 0$. Since $\varphi(\tau)$ is already known, the function $\Phi(\tau)$ also becomes known from the second line of \eqref{phiVolterra}. Thus, in this case, we don't need to solve the Volterra equations, and the solution is expressed in a semi-closed form.


Also, note that from computational point of view, it is convenient to re-write \eqref{phiVolterra} in variables $\phi(\tau), \psi(\tau)$ to get
\begin{align} \label{altVolterra}
\varphi(\tau) &= \langle \IndVec_{x_0 | y(0)}, \PM_{\sigmaVec}^{1}(0, \tau, | x_0-y(\tau), 0) \rangle
+ \int_0^\tau \psi(s) {\cal K}_\psi (s,\tau) ds + \int_0^\tau \varphi(s) {\cal K}_\phi (s,\tau) ds, \\
\frac{\psi(\tau) - y'(\tau) \varphi(\tau)}{\sigma_-^2}  &=
\langle \IndVec_{x_0 | y(0)}, \fp{\PM_{\sigmaVec}^{1}}{x}(0, \tau, | x_0-y(\tau), 0) \rangle +
\int_0^\tau \psi(s) {\cal C}_\psi (s,\tau) ds + \int_0^\tau \psi(s) {\cal C}_\phi (s,\tau) ds, \nonumber \\
{\cal K}_\psi (s,\tau) &= \left\langle
\begin{bsmallmatrix}	1 \\ -1 \end{bsmallmatrix}, \PM^1_{\sigmaVec}(0, \tau | y(s)-y(\tau), s) \right\rangle, \quad
{\cal K}_\phi (s,\tau) =  \left\langle
\begin{bsmallmatrix}	1 \\ -1 \end{bsmallmatrix}, \etaM^1_{\sigmaVec}(0, \tau | y(s)-y(\tau), s) \right\rangle, \nonumber\\
{\cal C}_\psi (s,\tau) &= \left\langle
\begin{bsmallmatrix}	1 \\ -1 \end{bsmallmatrix}, \fp{\PM^1_{\sigmaVec}}{x}(0, \tau | y(s)-y(\tau), s) \right\rangle, \quad
{\cal C}_\phi (s,\tau) =  \left\langle
\begin{bsmallmatrix}	1 \\ -1 \end{bsmallmatrix}, \fp{\etaM^1_{\sigmaVec}}{x}(0, \tau | y(s)-y(\tau), s) \right\rangle. \nonumber
\end{align}

When solving this system of equations using quadratures, e.g., the Simpson rule, the matrix in the RHS becomes block-triangular. Therefore, this system can be solved with complexity $O(M^2)$ where $M$ is the number of computational nodes in the time-space $\tau \in [0,T]$, while providing the fourth-order of approximation $O(M^{-4})$.

The representation of the Volterra equations in the form of \eqref{altVolterra} immediately reveals the fact that for the linear boundary $y(\tau) = a + b \tau, \ a,b - const$, \eqref{altVolterra} can be solved by using the Laplace transform. Indeed, taking the Laplace transform of both parts of \eqref{altVolterra} and using the convolution theorem yields a new system of algebraic equations
\begin{align} \label{sysLT}
\bar{\varphi}(p) &= A(p) + \bar{\psi}(p) \bar{{\cal K}}_\psi(p) + \bar{\varphi}(p) \bar{{\cal K}}_\phi(p) \\
\frac{\bar{\psi}(p) - b \bar{\varphi}(p)}{\sigma_-^2}  &=  B(p) + \bar{\psi}(p) \bar{{\cal C}}_\psi(p) + \bar{\varphi}(p) \bar{{\cal C}}_\phi(p), \nonumber
\end{align}
\noindent where the bar over the function $f(\tau)$ means its Laplace transform $\bar{f}(p) = \int_0^\infty f(\tau) e^{- p\tau} d\tau$, and
\begin{align*}
A(p) &= \int_0^\infty e^{- p\tau} \langle \IndVec_{x_0 | y(0)}, \PM_{\sigmaVec}^{1}(0, \tau, | x_0-y(\tau), 0) \rangle d\tau, \,
B(p) = \int_0^\infty e^{- p\tau} \langle \IndVec_{x_0 | y(0)}, \fp{\PM_{\sigmaVec}^{1}}{x}(0, \tau, | x_0-y(\tau), 0) \rangle d\tau,
\end{align*}
\begin{alignat*}{2}
\bar{{\cal K}}_\psi (s,\tau) &= \int_0^\infty e^{- p\tau} \left\langle
\begin{bsmallmatrix}	1 \\ -1 \end{bsmallmatrix}, \PM^1_{\sigmaVec}(0, \tau | -b \tau,0) \right\rangle d\tau,
&\,\, \bar{{\cal K}}_\phi (s,\tau) &=  \int_0^\infty e^{- p\tau}\left\langle
\begin{bsmallmatrix}	1 \\ -1 \end{bsmallmatrix}, \etaM^1_{\sigmaVec}(0, \tau | -b \tau, 0) \right\rangle d\tau, \nonumber\\
\bar{{\cal C}}_\psi (s,\tau) &= \int_0^\infty e^{- p\tau} \left\langle
\begin{bsmallmatrix}	1 \\ -1 \end{bsmallmatrix}, \fp{\PM^1_{\sigmaVec}}{x}(0, \tau | -b \tau, 0) \right\rangle d\tau,
&\,\, \bar{{\cal C}}_\phi (s,\tau) &=  \int_0^\infty e^{- p\tau} \left\langle
\begin{bsmallmatrix}	1 \\ -1 \end{bsmallmatrix}, \fp{\etaM^1_{\sigmaVec}}{x}(0, \tau | -b \tau, 0) \right\rangle d\tau. \nonumber
\end{alignat*}
Solving \eqref{sysLT} we obtain
\begin{align*}	
\bar{\varphi}(p) &= \frac{ \sigma ^2 \left(A(p) \bar{{\cal C}}_\psi(p) - B(p) \bar{{\cal K}}_\psi(p) \right) - A(p)}{Z(p)}, \qquad \bar{\psi}(p) = \frac{B(p) (\bar{{\cal K}}_\varphi(p)-1) \sigma_-^2 - A(p) \left(b + \bar{{\cal C}}_\varphi(p)\sigma_-^2\right)}{Z(p)}, \\
Z(p) &= b \bar{{\cal K}}_\psi(p) + \bar{{\cal K}}_\varphi(p) - 1 + \sigma_-^2 \left(
\bar{{\cal C}}_\psi(p)	+ \bar{{\cal C}}_\varphi(p) \bar{{\cal K}}_\psi(p) -
\bar{{\cal C}}_\psi(p) \bar{{\cal K}}_\varphi(p)\right). \nonumber
\end{align*}
Then application of the inverse Laplace transform (taken numerically) solves the problem.

\subsection{MHE with piecewise constant coefficients and time-dependent interfaces} \label{MHEcurv}

This problem is a generalization of the one considered in Section~\ref{MHEcurv} for the case when the number of layers $N > 2$. When the boundaries between all layers (the interfaces) are constant, this problem was considered, e.g., in \citep{Hickson2011} and solved numerically. As the authors mention, diffusion processes through a multilayered material are of interest for various applications, including industrial, biological, electrical, and environmental areas.  Some industrial applications include annealing steel coils, the performance of semiconductors and electrodes, geological profiles, and measuring greenhouse gas emissions from soil surfaces. Biological applications include determining the effectiveness of drug carriers inserted into living tissue, the probing of biological tissue with infrared light, and analyzing the heat production of muscle, see the corresponding references in \citep{Hickson2011} and the discussion.

When the interfaces are moving, a general semi-analytical method for solving these problems was presented in \citep{ItkinLiptonMuraveyMulti}. Here, as explained in the Introduction, we develop an alternative (but similar) approach.

The problem under consideration is as follows. Given the time $t \in [0,\infty)$, the space coordinate $x \in \mathbb{R}$, and a function $u(t,x): \mathbb{R}^+\times \mathbb{R} \mapsto \mathbb{R}^+, \ u \in {\cal C}^2$. Suppose that the whole domain $\Omega = \mathbb{R}$ could be split into $N+2$ non-overlapping layers $-\infty < y_0(t) < ... < y_{N}(t) < \infty$, $\forall t \geq 0$, i.e. $\Omega = \bigcup_{i=0}^N \Omega_i$, where each layer is a curvilinear strip as in \eqref{Omega_i_def}.  Let us consider a diffusion (thermal conductivity)  process in $\mathbb{R}$ such that the diffusion coefficient $\sigma$ is a piecewise constant function of $x$, i.e.
\begin{align}
\sigma(x) &=
\begin{cases}
 \sigma_i, & \forall x \in [y_{i-1}(t),y_i(t)], \quad i = 1,\dots,N, \\
0,  & x \in (-\infty,y_0(t)], \\
0, &  x \in [y_N(t),\infty),
\end{cases}
\end{align}
\noindent where $\sigma_i > 0, \forall i=1,\dots,N$ are constants. Suppose the evolution of this diffusion process is described by the heat equation
\begin{align} \label{PDE:ex:strip:u_eq}
\fp{u}{\tau} &= \fp{}{x}\left( \langle \sigmaVec^2, \IndVec_{ x | \by(\tau)} \rangle   \frac{\partial u}{dx} \right) + g(\tau, x), \\
u(\tau, y_0(\tau)) &= u(\tau, y_N(\tau)) = 0, \nonumber	\\
u(\tau, x) &= f(x), \qquad \sigmaVec = [0, \sigma_0,\dots,\sigma_N, 0]^\top \nonumber
\end{align}
\noindent
where $f(x), g(\tau,x) \in \mathbb{R}$ are given functions.

To solve this problem we define the integral transform
\begin{equation} \label{invTr2}
\bar u(\tau, \lambda) = \int_{-\infty}^{\infty} u(\tau, x)\langle \Ind_{x | \by(\tau)}, \ThetaVec_{\by(\tau), \lambda}(x)\rangle  dx,
\end{equation}
\noindent where the basis function $\ThetaVec_{\by(\tau), \lambda}(x)$ is defined by \eqref{DS:theta_def} via \eqref{DS:theta_def_CD}. It is clear that
\begin{align} \label{baruStruct}
\bar u(\tau, \lambda) &= \sum_{i=1}^N \int_{y_{i-1}(\tau)}^{y_i(\tau)} u(\tau, x) \left[C_i(\lambda, \by(\tau)) \cos\left( \frac{\lambda x}{\sigma_i}\right) + D_i(\lambda, \by(\tau)) \sin\left( \frac{\lambda x}{\sigma_i}\right) \right]  dx \\
&= \sum_{i= 1}^N \left\{ C_i(\lambda, \by(\tau))\bar u^C_i(\tau, \lambda) + D_i(\lambda, \by(\tau))  u^S_i(\tau, \lambda) \right\}, \nonumber \\
\bar u^C_i(\tau, \lambda) &= \int_{y_{i-1}(\tau)}^{y_i(\tau)} u(\tau, x) \cos\left( \frac{\lambda  x}{\sigma_i}\right)  dx,  \quad \bar u^S_i(\tau, \lambda) = \int_{y_{i-1}(\tau)}^{y_i(\tau)} u(\tau, x) \sin\left( \frac{\lambda x}{\sigma_i}\right)  dx. \nonumber
\end{align}

As shown in Appendix~\ref{app5}, this transform can be found explicitly and the final result reads
\begin{align} \label{PDE:ex:strip:baru_final}
 \nonumber
\bar u(\tau, \lambda) &=e^{-\lambda^2\tau}\int_{-\infty}^{+\infty} \langle \IndVec_{\xi | \by(0)}, \ThetaVec_{\by(\tau), \lambda}(\xi)\rangle   f(\xi) d\xi
+\int_0^\tau \int_{-\infty}^{+\infty} e^{-\lambda^2(\tau-s)} \langle \IndVec_{\xi | \by(s)}, \ThetaVec_{\by(\tau), \lambda}(\xi) \rangle g(s,\xi) ds d\xi
\\
&+ \int_0^\tau e^{-\lambda^2(\tau -s)}\left[ \langle  \PhiVec(s), \OmegaVec_{\by(\tau), \lambda} (s) \rangle
+ \langle  \phiVec(s), \YM'(s)\OmegaVec_{\by(\tau), \lambda} (s)-\omegaVec_{\by(\tau), \lambda} (s) \rangle
\right] ds.
\end{align}
Here the function $\YM$ is defined in \eqref{YM}, and vectors $\OmegaVec, \PhiVec$ - in \eqref{omegaVec}.

Having this representation, the inverse transform can be constructed in a straightforward way. Assuming $\lambda_n = \lambda_n(\tau)$, for each $\tau$ these eigenvalues can be found by solving \eqref{DS:eig_eq}.  Given $\tau$, the functions $\ThetaVec_{\by, \lambda_n(\tau)}(x)$, $ n= 1,\dots,$ form an orthogonal basis.
Therefore, we immediately get an explicit representation for $u(\tau, x)$
\begin{equation} \label{PDE:ex:strip:u_series}
u(\tau, x) = \sum_{n = 1}^\infty \frac{\bar u (\tau, \lambda_n(\tau)) \langle \IndVec_{x | \by(\tau)}, \ThetaVec_{\by(\tau), \lambda_n(\tau)}(x)\rangle } {\int_{-\infty}^{+\infty} \langle \IndVec_{\xi | \by(\tau)}, \ThetaVec_{\by(\tau), \lambda_n(\tau)}(\xi)\rangle ^2 d\xi}.
\end{equation}

Substituting \eqref{PDE:ex:strip:baru_final} into \eqref{PDE:ex:strip:u_series} and introducing the norm $N_n(\tau)$
\begin{equation}
N_n(\tau) = \int_{-\infty}^{\infty} \langle \Ind_{ \xi | \by(\tau)}, \ThetaVec_{\by(\tau), \lambda_n(\tau)}(\xi)\rangle ^2 d\xi,
\end{equation}
\noindent we obtain an explicit solution for $u(\tau, x)$
\begin{align} \label{PDE:ex:strip:u_final}
u(\tau, x) &= \Bigg<\IndVec_{x | \by(\tau)}, \,\,	\sum_{n = 1} ^{\infty} \frac{\ThetaVec_{\by(\tau), \lambda_n(\tau)} (x)}{N_n(\tau)} \Bigg[e^{-\tau \lambda^2_n(\tau)}\int_{-\infty}^{+\infty}
\langle \Ind_{\xi | \by(0)}, \ThetaVec_{\by(\tau), \lambda_n(\tau)}(\xi)\rangle  f(\xi) d\xi 	\\ \nonumber
&+\int_0^\tau \int_{-\infty}^{+\infty} e^{-(\tau-s)\lambda_n^2(\tau)} \langle \IndVec_{\xi | \by(s)}, \ThetaVec_{\by(\tau), \lambda_n(\tau)}(\xi) \rangle g(s,\xi) ds d\xi
\\
&+ \int_0^\tau e^{-(\tau -s)\lambda_n^2(\tau)}\left[ \langle  \PhiVec(s), \OmegaVec_{\by(\tau), \lambda_n(\tau)} (s) \rangle 	+ \langle  \phiVec(s), \YM'(s)\OmegaVec_{\by(\tau), \lambda_n(\tau)} (s)-\omegaVec_{\by(\tau), \lambda_n(\tau)} (s) \rangle  \right] ds. \nonumber
	\Bigg] \Bigg>
\end{align}

The solution in \eqref{PDE:ex:strip:u_final} is expressed in a semi-analytical form, since the functions
$\PhiVec$ and $\phiVec$ are yet unknown. However, they can be found numerically by solving a the system of linear Volterra integral equations of the second kind. To derive these equations we differentiate $u(\tau, x)$ in \eqref{PDE:ex:strip:u_final} with respect to $x$  and set $x \to y_i(\tau)$. This yields
\begin{align} \label{Vol2}
\phiVec(\tau) &= 	\sum_{n = 1} ^{\infty}  \frac{\mathbf{V}_{\by(\tau), \lambda_n(\tau)}(\tau)}{N_n(\tau)} \Bigg\{ e^{-\tau \lambda^2_n(\tau)} \int_{-\infty}^{+\infty} 	\langle \Ind_{\xi | \by(0)}, \ThetaVec_{\by(\tau), \lambda_n(\tau)}(\xi)\rangle  f(\xi) d\xi  \\
&+\int_0^\tau \int_{-\infty}^{+\infty} e^{-(\tau-s)\lambda_n^2(\tau)} \langle \IndVec_{\xi | \by(s)}, \ThetaVec_{\by(\tau), \lambda_n(\tau)}(\xi) \rangle g(s,\xi) ds d\xi
\nonumber \\
&+ \int_0^\tau e^{-(\tau -s)\lambda_n^2(\tau)}\left[ \langle  \PhiVec(s), \OmegaVec_{\by(\tau), \lambda_n(\tau)} (s) \rangle 	+ \langle  \phiVec(s), \YM'(s)\OmegaVec_{\by(\tau), \lambda_n(\tau)} (s)-\omegaVec_{\by(\tau), \lambda_n(\tau)} (s) \rangle \right] ds \Bigg\}, \nonumber \\
\PhiVec(\tau) &= 	\sum_{n = 1} ^{\infty}  \frac{\mathbf{v}_{\by(\tau), \lambda_n(\tau)}(\tau)}{N_n(\tau)} \Bigg\{ e^{-\tau \lambda^2(\tau)} 	\int_{-\infty}^{+\infty} \langle \Ind_{\xi | \by(0)}, \ThetaVec_{\by(\tau), \lambda_n(\tau)}(\xi)\rangle  f(\xi) d\xi \nonumber	\\
&+\int_0^\tau \int_{-\infty}^{+\infty} e^{-(\tau-s)\lambda_n^2(\tau)} \langle \IndVec_{\xi | \by(s)}, \ThetaVec_{\by(\tau), \lambda_n(\tau)}(\xi) \rangle g(s,\xi) ds d\xi
\nonumber \\
&+ \int_0^\tau e^{-(\tau -s)\lambda^2(\tau)}\left[ \langle  \PhiVec(s), \OmegaVec_{\by(\tau), \lambda_n(\tau)} (s) \rangle 	+ \langle  \phiVec(s), \YM'(s)\OmegaVec_{\by(\tau), \lambda_n(\tau)} (s)-\omegaVec_{\by(\tau), \lambda_n(\tau)} (s) \rangle  \right] ds \Bigg\}, \nonumber
\\
\mathbf{V}_{\by(\tau), \lambda_n(\tau)}(\tau) &=
\left(0, \ThetaVec_{\by(\tau), \lambda_n(\tau), 1}(y_1(\tau), ), \ThetaVec_{\by(\tau), \lambda_n(\tau), 2}(y_2(\tau)), \dots, \ThetaVec_{\by(\tau), \lambda_n(\tau), N-1}(y_{N-1}(\tau), 0)\right), \nonumber
\\
\mathbf{v}_{\by(\tau), \lambda_n(\tau)}(\tau) &=
\left(\thetaVec_{\by(\tau), \lambda_n(\tau), 1}(y_0(\tau)), \thetaVec_{\by(\tau), \lambda_n(\tau), 1}(y_1(\tau)), ), \thetaVec_{\by(\tau), \lambda_n(\tau), 2}(y_2(\tau)), \dots, \thetaVec_{\by(\tau), \lambda_n(\tau), N}(y_{N}(\tau))\right). \nonumber
\end{align}

Again, this system can be solved numerically, as discussed in \citep{ItkinLiptonMuraveyBook}. Thus, our method can be conventionally classified as semi-analytical in the sense that the solution in \eqref{PDE:ex:strip:u_final} is explicit but depends on two functions that should be computed numerically by solving the linear Volterra equations in \eqref{Vol2}. The corresponding solution, however, can be found very efficiently; see \citep{ItkinLiptonMuraveyBook}, where the authors also show that their approach favorably compares with the finite-deference method.

\subsection{Freezing and solidification}

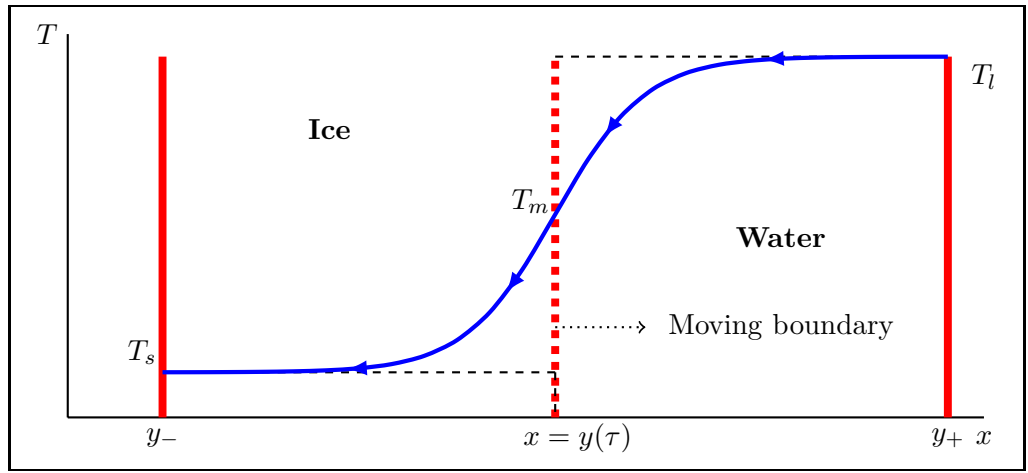
\begin{figure}[!htb]
\begin{center}
\fbox{
\begin{tikzpicture}[thick, scale=0.6]
\def\axisXlength{9}
\def\leftB{-\axisXlength + 0.3}
\def\rightB{\axisXlength - 0.3}
\def\zero{0}
\def\shift{-0.45}
\def\height{8.0}
\def\h1{1.0}
\def\cof{1.2}

\draw (-\axisXlength*\cof, 0) -- (\axisXlength+0.5,0)
         (-\cof*\axisXlength, 0) -- (-\cof*\axisXlength, \height+0.5);

\draw[dashed,red, line width=3pt] (\zero, 0) -- (\zero,\height);
\draw[red, line width=3pt] (\leftB, 0) -- (\leftB,\height);
\draw[red, line width=3pt] (\rightB, 0) -- (\rightB,\height);
\draw[dashed] (\zero, \height) -- (\rightB, \height);
\draw[dashed] (\zero, \h1) -- (\leftB, \h1);
\draw[->, dotted] (\zero, 2*\h1) -- (\zero+2,2*\h1);
\draw[dashed] (\zero, \h1) -- (\zero,0);
\draw[blue,ultra thick,smooth,decoration={ markings,
    mark=at position 0.2 with {\arrowreversed{latex}},
    mark=at position 0.4 with {\arrowreversed{latex}},
    mark=at position 0.6 with {\arrowreversed{latex}},
    mark=at position 0.8 with {\arrowreversed{latex}}
}, postaction={decorate}]
    plot[variable=\x,domain=\leftB:\rightB] (\x,{\height/(1 + exp(-\x)) + \h1/(1 + exp(\x))});

\node at (\zero + 0.5,\shift){$x = y(\tau)$};
\node at (\axisXlength+0.5,\shift){$x$};
\node at (-\cof*\axisXlength + \shift, \height+0.5){$T$};
\node at (\axisXlength+0.5, \height + \shift){$T_l$};
\node at (\leftB + \shift, \h1-\shift){$T_s$};
\node at (\zero + 5, 2*\h1){Moving boundary};
\node at (-\axisXlength + 4, 0.8*\height){{\bf Ice}};
\node at (\axisXlength - 4, 0.5*\height){{\bf Water}};
\node at (\zero + 1.2*\shift, 0.6*\height){$T_m$};
\node at (\leftB, \shift){$y_-$};
\node at (\rightB, \shift){$y_+$};

\end{tikzpicture}
}
\end{center}
\caption{Heat conduction with moving boundary  - a freezing problem: blue - the temperature profile, solid red - the boundaries, dashed red - the interphase (moving) boundary.}
\label{freezing}
\end{figure}

There exist various engineering and scientific problems involving heat conduction with moving boundaries such as freezing or melting for solar storage systems, high-temperature droplet evaporation, ablation, among many others. Let us consider a typical example of the freezing problem, \citep{Han1946, Jiji2009}, which is schematically presented in Fig.~\ref{freezing}.

We consider a rectangular region which is split into two subregions $y(\tau) < x < y_+$ and $y_- < x < y(\tau)$
by an interphase boundary $x=y(\tau)$. For example, suppose that we have liquid water to the right of the boundary and solid ice to the left. The boundary $x=y(\tau)$ is a phase transition boundary which position is unknown yet but could be found based on some physics consideration. The temperature distribution in this two-phase region is governed by two heat equations, one for the solid phase and the other for the liquid phase. Our problem boils down to determining the moving boundary $y(\tau)$ and the temperature distribution in both phases. However, first, we need to describe the physics of the process in more detail. Real geometries are much more complex than the one depicted in Fig.~\ref{freezing}, which is produced based on the following simplifications:
\begin{itemize}
\item properties of each phase are uniform and remain constant;
\item the effect of liquid phase motion due to changes in density is negligible;
\item we consider just one-dimensional heat conduction;
\item we neglect by any energy generation.
\end{itemize}
In addition, we assume that the water temperature at $ x \to y_+$ is constant and equal to $T_l$, and the ice temperature at $x \to y_-$ is also constant and equal to $T_s$. Thus, our system at both ends is connected to thermostats which support those constant temperatures. Also, we assume that water undergoes a phase change at a fixed temperature at the phase interface (the melting or freezing temperature) $T_m$. However, the physical properties of both media are different. Therefore, the system with constant temperatures at both ends and phase transition at the interface can only exist if the interface moves in time.

Based on our assumptions, we can write two heat equations for the temperature $T(\tau,x)$
\begin{alignat}{2} \label{freezing_pde}
\fp{T}{\tau} &= \kappa_W \sop{T}{x}, &\qquad y(\tau) &< x < y_+, \\
\fp{T}{\tau} &= \kappa_I \sop{T}{x}, &\qquad y_- &< x < y(\tau), \nonumber
\end{alignat}
\noindent where $\kappa_W, \kappa_I$ are the corresponding thermal diffusivities. These equations have to be solved subject to the boundary conditions
\begin{align} \label{bcExample}
T(\tau,y_+) &= T_l, \qquad T(\tau,y_-) = T_s, \qquad T(\tau,y(\tau)-0) = T(\tau,y(\tau)+0) = T_m.
\end{align}
The last condition means continuity of temperature at the phase interface. As shown in \citep{Jiji2009}, a conservation of energy argument gives rise to one more boundary condition at the interface
\begin{equation} \label{jump1}
k_I \fp{T}{x}\Big|_{x = y(\tau) -0} - k_W \fp{T}{x}\Big|_{x = y(\tau) + 0} = \rho_I L_h y'(\tau).
\end{equation}
Here $L_h$ is the latent heat of fusion, $\rho_I, \rho_W$ are the corresponding densities in two phases and $k_I,k_W$ - their thermal conductivities. They are connected to $\kappa_I, \kappa_W$ via the relationship $\kappa_j = k_j/(\rho_j C_j), \ j=I,W$ where $C_I,C_W$ are specific heat capacities. The \eqref{jump1} is the interface energy equation and is valid for both solidification and melting. However, for melting $\rho_I$ in the RHS of \eqref{jump} should be replaced with $\rho_W$. Thus, in contrast to temperature, the temperature flux experiences a jump when crossing the phase interface, and the value of this jump is proportional to $y'(\tau)$.

As discussed in more detail in Section~\ref{sec:discussion}, in some papers an alternative form of \eqref{jump} is used, which can be obtained by assuming $\rho_W C_W \approx \rho_I C_I$. Taking the values at temperature $T=0^\circ C$ and atmospheric pressure  $\rho_W$ = 997 kg/m$^3$, $\rho_I$ = 917 kg/m$^3$, $C_W$ = 4.22 kJ/kg/K, $C_I$ = 2.09 kJ/kg/K, \citep{Vargaftik1975,Keenan1979}, one can see that this assumption is not very accurate. However, without loss of generality, below instead of \eqref{jump1} we use its simplified form
\begin{equation} \label{jump}
\kappa_I \fp{T}{x}\Big|_{x = y(\tau) -0} - \kappa_W \fp{T}{x}\Big|_{x = y(\tau) + 0} =  \rho_I L y'(\tau),
\end{equation}
\noindent where $L = L_h/C_a$, and $C_a$ is an average value
\begin{equation} \label{Ca}
C_a = (\rho_W C_W + \rho_I C_I)/2.
\end{equation}

Finally, we need to set the initial state of the system. For instance, one can assume that at $\tau=0$ the interface boundary is at $x = y_-$, and, hence, $T(0,x) = T_l$. Again, it is worth emphasizing that the law of motion $y(\tau)$ is unknown in advance and should be determined together with the temperature profile by solving the described problem. However, this is standard for the so-called Stefan problems, see \citep{Jiji2009} and references therein.

Thus, the stated problem can be solved by using the approach developed in \citep{ItkinMuraveyDBFMF}. Indeed, in the left region (ice), temperature evolution is determined by the heat equation with given temperatures at the edges $T_s, T_m$. Solving this problem and finding the temperature distribution is exactly equivalent (in financial terms) to pricing double barrier option (with barriers at $x = y_-$ and $x = y(\tau)$ and the payoff at $\tau=0$ equal to $T_l$)  which pays rebates at hit equal to $T_s$ and $T_m$, respectively. Given the law $y(\tau)$, this problem can be solved by the method developed in \citep{ItkinMuraveyDBFMF}. Similarly, for the right region same problem is defined at the domain $y(\tau) \le x \le y_+$. Therefore, the whole algorithm could be as follows. We start with some initial guess for $y(\tau) = y_0(\tau)$ and solve the above problems for the left and right regions. As a part of this solution we also obtain its gradients at $x = y_0(\tau)-0$ and $x = y_0(\tau)+0$. Then, substituting them into \eqref{jump} and solving for $y(\tau)$ yields its next approximation $y_1(\tau)$. We proceed in the same way until the solution converges up to a given tolerance. This method is advantageous as compared with sophisticated numerical methods, e.g., described in \citep{Hickson2011}, because it is uniform (no need for a complex construction of the finite-difference scheme at the moving interface) and semi-analytical, so fast.

However, in this paper, for the reasons explained in the Introduction, we solve this problem by using a slight modification of the method described in the previous section. First, here the problem has inhomogeneous boundary conditions, \eqref{bcExample}. Second, the matching condition for the first derivative contains a jump. Fortunately, the problem in \eqref{freezing_pde}, \eqref{bcExample} can be reduced to that in \eqref{PDE:ex:strip:u_eq}, and the latter can be solved in a semi-explicit form via the OIT, see Appendix~\ref{app6}.

We construct the solution of \eqref{freezing_pde} as oscillating Fourier series with the terms
\begin{align*}
\Ind_{x < y(\tau)}\ThetaVec_{\by(\tau), \lambda_n(\tau), 1}(x) &+ \Ind_{x > y(\tau)}\ThetaVec_{\by(\tau)
\lambda_n(\tau), 2}(x) = \Ind_{x < y(\tau)}\sin(x_n^-(\tau,x)) + \Ind_{x > y(\tau)}K_n(\tau) \sin(x_n^+(\tau,x)),
\end{align*}
\vspace{-1.5\baselineskip}
\begin{alignat}{2}
x_n^-(\tau,x) &= \lambda_n(\tau)(x - y_-)/\sqrt{\kappa_I}, &\quad l_-(\tau) &= (y(\tau) - y_-) / \sqrt{\kappa_I},
\quad K_n(\tau) = \frac{\sin\left[\lambda_n(\tau) l_-(\tau)\right]}{\sin\left[\lambda_n(\tau) l_+(\tau)\right]},  \\
x_n^+(\tau,x) &= \lambda_n(\tau) (y_+ - x)/\sqrt{\kappa_W}, &\quad l_+(\tau) &= (y_+ - y(\tau))/\sqrt{\kappa_W}.
\nonumber
\end{alignat}

The spectrum $\lambda_n(\tau), \ n=1,\dots,$ is defined as an ordered sequence of positive roots of the equation
\begin{equation} \label{eigenLam}
\left(\sqrt{\kappa_W} - \sqrt{\kappa_I}\right) \sin\left[ \lambda_n(\tau) \left(l_+(\tau) - l_-(\tau)\right) \right] = \left(\sqrt{\kappa_W} + \sqrt{\kappa_I}\right) \sin \left[ \lambda_n(\tau) \left(l_+(\tau) + l_-(\tau)\right) \right].
\end{equation}
This equation has to be solved numerically. However, in some cases an analytic or asymptotic solution is also possible. For instance, at $\tau = 0$ we have $y(0) = y_-$ and, hence
\begin{equation} \label{eigen0}
\lambda_n(0)  = 2\pi n/l_+(0), \qquad n \in \mathbb{Z}.
\end{equation}
The factor of 2 is required in order to satisfy the boundary conditions.

It can be observed that with this value of $\lambda_n(0)$ the value of $K_n(0)$ becomes an indeterminate form since it is $0/0$. By using L'Hospital's rule and the fact that $\lambda'_n(0) = 0$ (which can be proved by differentiating \eqref{eigenLam} with respect to $\tau$ and taking the limit $\tau \to 0$), we have
\begin{equation} \label{Kn0}
K_n(0) = - \sqrt{\kappa_W}/\sqrt{\kappa_I} \ne 0.
\end{equation}

An asymptotic solution can also be constructed by having in mind that, e.g., at atmospheric pressure and at $T = 0^\circ C$ we have $\kappa_W \approx 0.13 \mathrm{\ mm^2/s}, \kappa_I \approx 1.02 \mathrm{\ mm^2/s}$, \citep{Vargaftik1975,Keenan1979}, so $\kappa_W/\kappa_I \ll 1$. If $l_+(\tau)$ and $l_-(\tau)$ are of the same order of magnitude, then \eqref{eigenLam} can be transformed to
\begin{align*}
0 = \sin \left[\lambda_n(\tau)\left(l_+(\tau) + l_-(\tau)\right) \right] + \sin\left[\lambda_n(\tau)\left(l_+(\tau) - l_-(\tau)\right) \right] = 2 \sin[\lambda_n(\tau)l_+(\tau)]\cos[\lambda_n(\tau)l_-(\tau)],
\end{align*}
\noindent with the obvious solutions\footnote{Again, we take period $2\pi n$ rather then $\pi n$ to correctly obey the boundary conditions.}
\begin{align} \label{eigen1}
\lambda_{n,1}(\tau)  = 2\pi n/l_+(\tau), \quad \lambda_{n,2}(\tau)  = \left(\pi /2 + 2\pi n \right)/l_-(\tau), \qquad
\qquad \tau > 0,  \ n \in \mathbb{Z}.	
\end{align}
Substitution of $\lambda_{n,1}(\tau)$ into the definition of $K_n(\tau)$ makes $K_n(\tau)$ singular, so $\lambda_{n,1}(\tau)$ has to be excluded. Therefore, in this case the asymptotic solution is given by $\lambda_{n,2}(\tau)$. If, however, $l_-(\tau) \ll l_+(\tau)$, then the asymptotic solution is given by $\lambda_{n,1}(\tau)$.

Once the eigenvalues $\lambda_n(\tau)$ are found, the temperature $T(\tau, x)$ can be represented as
\begin{align} \label{T_final}
T(\tau,x) &= \eta(\tau,x) + \sum_{n = 1}^{\infty}\frac{S_n(\tau) + R_n(\tau)}{N_n(\tau)} {\cal M}(\tau,x), \\
{\cal M}(\tau,x) &=
\begin{cases}
\left[\Ind_{x < y(\tau)}\sin[x^-_n(\tau,x)] + \Ind_{x > y(\tau)} K_n(\tau) \sin[x^+_n(\tau,x)] \right], & x \ne y(\tau), \\
\frac{1}{2}\left[s^+(y(\tau))\sin[x^-_n(\tau,y(\tau))] + s^-(y(\tau)) K_n(\tau) \sin[x^+_n(\tau,y(\tau)) \right], & x = y(\tau),
\end{cases}
\nonumber
\end{align}
\noindent where $s^\pm(y(\tau)) = 1 + \Ind_{y_\pm = y(\tau)}$.

After some tedious algebra (which we partly present in Appendix~\ref{app6}, but omit other details) the functions $R_n(\tau)$, $S_n(\tau)$ and $N_n(\tau)$ can be represented in the form (the terms in the below definitions slightly differ from those in \eqref{RSN_def} since for convenience we regrouped them a bit)
\begin{align} \label{RSN_def1}
N_n(\tau) &= \int_{y_-}^{y(\tau)} \ThetaVec_{\by(\tau), \lambda_n(\tau), 1}^2(\xi) d\xi + \int^{y_+}_{y(\tau)} \ThetaVec_{\by(\tau), \lambda_n(\tau), 2}^2(\xi) d\xi = \frac{1}{2}l_-(\tau) \sqrt{\kappa_I}+ \frac{1}{2} K^2_n(\tau) l_+(\tau) \sqrt{\kappa_W}, \nonumber \\
S_n(\tau) &= \frac{e^{-\tau \lambda^2_n(\tau)}}{\lambda_n(\tau)}  \sqrt{\kappa_I} (T_l - T_s) - \frac{1}{\lambda_n^2(\tau)} \left[B_-(\tau) \kappa_I \ThetaVec_{\by(\tau), \lambda_n(\tau), 1}(y(\tau)) - B_+(\tau) \kappa_W \ThetaVec_{\by(\tau), \lambda_n(\tau), 2}(y(\tau))  \right] \nonumber \\
&+ (T_m-T_l)\frac{e^{-\tau \lambda^2_n(\tau)}}{\lambda_n(\tau)} \left[ \sqrt{\kappa_I} \cos\left(x^-(\tau,y(0))\right) +
\sqrt{\kappa_W} K_n(\tau)\cos\left(x^+(\tau,y(0))\right) \right] \nonumber \\
R_n(\tau) &= \int_0^\tau e^{-(\tau -s)\lambda_n^2(\tau)}\Big[ \Phi(s) \Omega_{\by(\tau), \lambda_n(\tau)} (s) + B_-(s)\kappa_I \ThetaVec_{\by(\tau), \lambda_n(\tau), 1}(y(s)) - B_+(s)\kappa_W \ThetaVec_{\by(\tau), \lambda_n(\tau), 2}(y(s)) \Big] ds. \nonumber
\end{align}

Convergence of these series can be an issue.\footnote{Suppose we set $\kappa_W = \kappa_I$, then all basis functions become sine functions and the series in $R_n$ transforms to a sum of Dirac delta functions. Therefore, convergence of this series could be an issue.}
To improve their convergence, we add and subtract $B_-(\tau) \kappa_I \sin [x^-_n(\tau, y(\tau))]$ and $B_+(\tau)\kappa_W K_n(\tau)  \sin[x^+_n(\tau,y(\tau))]$ under the last integral  to obtain
\begin{align}
R_n(\tau) &= \int_0^\tau e^{-(\tau -s)\lambda_n^2(\tau)}\Bigg\{ \Phi(s) \left[\sin[x_n^-(\tau,y(s))] - K_n(\tau) \sin[x_n^+(\tau,y(s))]\right] \\
&\qquad + B_-(s)\kappa_I \sin[x^-_n(\tau,y(s))] -  B_-(\tau)\kappa_I \sin[x^-_n(\tau,y(\tau))] \nonumber \\
&\qquad - B_+(s)\kappa_W K_n(\tau)  \sin[x^+_n(\tau,y(s))] + B_+(\tau)\kappa_W K_n(\tau)  \sin[x^+_n(\tau,y(\tau))] \Bigg\} ds \nonumber \\
&+ \int_0^\tau e^{-(\tau -s)\lambda_n^2(\tau)} \left[ B_-(\tau)\kappa_I \sin[x^-_n(\tau,y(\tau))]  -  B_+(\tau)\kappa_W K_n(\tau)  \sin[x^+_n(\tau,y(\tau))]\right]ds. \nonumber
\end{align}
The last integral can be computed analytically while we move the result to the definition of $S_n$. Hence, finally this yields
\begin{align} \label{RSN_def1}
S_n(\tau) &= \frac{e^{-\tau \lambda^2_n(\tau)}}{\lambda_n(\tau)}  \Bigg\{ \sqrt{\kappa_I} (T_l - T_s) +(T_m-T_l) \left[ \sqrt{\kappa_I} \cos\left(x^-(\tau,y(0))\right) + \sqrt{\kappa_W} K_n(\tau)\cos\left(x^+(\tau,y(0))\right) \right] \\
&- \frac{1}{\lambda_n(\tau)}\left[ B_-(\tau)\kappa_I \sin[x^-_n(\tau,y(\tau))]  -  B_+(\tau)\kappa_W K_n(\tau)  \sin[x^+_n(\tau,y(\tau))]\right] \Bigg\}, \nonumber \\
R_n(\tau) &= \int_0^\tau e^{-(\tau -s)\lambda_n^2(\tau)}\Bigg\{ \Phi(s) \left[\sin[x_n^-(\tau,y(s))] - K_n(\tau) \sin[x_n^+(\tau,y(s))]\right] \nonumber \\
&+ \kappa_I \left[ B_-(s) \sin[x^-_n(\tau,y(s))] -  B_-(\tau)\sin[x^-_n(\tau,y(\tau))] \right] \nonumber \\
&- K_n(\tau) \kappa_W \left[ B_+(s)\sin[x^+_n(\tau,y(s))] - B_+(\tau)\sin[x^+_n(\tau,y(\tau))] \right] \Bigg\}ds. \nonumber
\end{align}

It can be seen, that at $\tau = 0, \ y(0) = y_-$ and from \eqref{RSN_def1} we have $R_n(0) = 0$. It turns out that $S_n(0) = 0$ as well. Indeed, by definition $x_n^-(0,y(0)) = 0, \ x^+_n(0,y(0)) = \lambda_n(0) l_+(0) = 2 \pi n, \ n \in \mathbb{Z}$ (the last equality follows from \eqref{eigen0}). Also, at $\tau=0$ we have $T_s = T_l$, $B_-(0) = B_+(0) = 0$ and $K_n(0)$ is given by \eqref{Kn0}. Then a simple algebra yields the result. However, $N_n(0) \ne 0$. Indeed, \eqref{eigenLam} implies that at $\tau=0$ the value of $\lambda_n(0)$ is given by \eqref{eigen0}. Therefore, we have $l_-(0) = 0, \lambda_{W,n}(0) = 2 \pi n$ and thus,
\begin{equation*}
N_n(0) = \frac{1}{2} K_n^2(0) l_+(0) = \frac{1}{2}\frac{\kappa_W}{\kappa_I} l_+(0)  \ne 0.
\end{equation*}
Thus, from \eqref{T_final} we obtain $T(0,x) = \eta(0,x)$ and, in particular,  $T(0,y(0)) = T_l$. This is the correct value since at $\tau = 0, y(0) = y_-$ we have a uniform media and, hence, $T_s = T_m = T_l$.

Based on \eqref{Volt2} one can see that the free boundary $y(\tau)$ and the function $\Phi(\tau)$ - the flux\footnote{In this section we denote the flux as $\Phi$ while in previous ones that was the gradient. They differ just by a constant multiplier $\kappa_W$ or $\kappa_I$, and hopefully  this change doesn't bring any confusion.} of the modified temperature $\calT$ at the boundary $x=y(\tau)$ (see the definition in \eqref{prCalT}), can be found by solving the system of nonlinear Volterra integral equations of the second kind
\begin{align}  \label{Volt_for_yandPhi}
T_m &= \eta(\tau,y(\tau)) + \frac{1}{2}\sum_{n = 1}^{\infty}\frac{S_n(\tau) + R_n(\tau)}{N_n(\tau)}
\left[s^+(y(\tau))\sin[x^-_n(\tau,y(\tau))] + s^-(y(\tau)) K_n(\tau) \sin[x^+_n(\tau,y(\tau))] \right], \\
\Phi(\tau) &= \frac{1}{2} \sum_{n = 1}^{\infty}\frac{S_n(\tau) + R_n(\tau)}{N_n(\tau)} \lambda_n(\tau) \left[
s^+(y(\tau)) \sqrt{\kappa_I}\cos[x^-_n(\tau,y(\tau))] - s^-(y(\tau))\sqrt{\kappa_W} K_n(\tau) \cos[x^+_n(\tau,y(\tau))] \right]. \nonumber
\end{align}

\paragraph{Behavior of the solution at small $\tau$.} It is worth noting that based on the definitions in \eqref{RSN_def1}, at small $\tau$ we have
\begin{align*}
S_n(\tau) &\approx \frac{e^{-\tau \lambda^2_n(\tau)}}{\lambda_n(\tau)}  \sqrt{\kappa_I} (T_l - T_s),
\qquad N_n(\tau) \approx N_n(0), \qquad R_n(\tau) \approx = 0,
\end{align*}
\noindent while $\lambda_n(\tau)$ is given by \eqref{eigen0}. Assuming $x \ne y(\tau)$, overall, we obtain
\begin{align} \label{PhiShortTau}
\Phi(\tau) &= \frac{\sqrt{\kappa_I} (T_l - T_s)}{2 N_n(0)} \sum_{n = 1}^{\infty} e^{-\tau \lambda^2_n(\tau)} \Big[ s^+(y(\tau)) \sqrt{\kappa_I}\cos[x^-_n(\tau,y(\tau))] \\
&\qquad\qquad - s^-(y(\tau))\sqrt{\kappa_W} K_n(0) \cos[x^+_n(\tau,y(\tau))] \Big], \nonumber \\
x^-_n(\tau,y(\tau)) &\approx \frac{2 \pi n (x - y_-)}{(y(\tau) - y_-)}, \qquad x_n^+(\tau,x) \approx \frac{2 \pi n (y_+ - x)}{(y_+ - y(\tau))}. \nonumber
\end{align}

We proceed with the observation that the sums in \eqref{PhiShortTau} could be expressed via Jacobi theta functions of the third kind, \citep{mumford1983tata}. By definition
\begin{equation}\label{the3}
\theta_3 (z,\omega) = 1 + 2 \sum_{n=1}^{\infty} \omega^{n^2}\cos(2 n z).
\end{equation}
Using these definitions \eqref{PhiShortTau} can be re-written in the form
\begin{align} \label{Ttheta2}
\Phi(\tau) &= \frac{\sqrt{\kappa_I} (T_l - T_s)}{4 N_n(0)} \left[ s^+(y(\tau)) \sqrt{\kappa_I} \theta_3(z_-(x),\omega(\tau)) - s^-(y(\tau))\sqrt{\kappa_W} K_n(0) \theta_3(z_-(x),\omega(\tau)) \right], \\
\omega(\tau) &= e^{- 4 \pi^2 \tau/l_-^2(\tau)}, \qquad z_-(x) = \pi \frac{x - y_-}{y(\tau) - y_-}, \qquad
z_+(x) = \pi \frac{y_+ - x}{y_+ - y(\tau)}. \nonumber
\end{align}
A well-behaved theta function must have parameter $|\omega| < 1$, \citep{mumford1983tata}. In our case this condition holds for any $\tau > 0$.

The RHS of \eqref{Ttheta2} depends on $x$ via functions $z_\pm(x)$. It is easy to check that a similar representation holds  for $T(\tau,x)$ in \eqref{T_final}, but now instead of $\theta_3(z_\pm(x),\omega(\tau))$ we have to use
$\Theta_3(z_\pm(x),\omega(\tau))$, where
\begin{equation*}
\Theta_3(z_\pm(x),\omega(\tau)) = \int_0^x \theta_3(z_\pm(\xi),\omega(\tau)) d\xi.
\end{equation*}
Since function $\Theta_3 (z,\omega)$ vanishes at $z=0$,  the boundary conditions at $x=y_\pm$ are satisfied. The result in \eqref{Ttheta2} to some extent is not a surprise since it is well-known that the Jacobi theta function is the fundamental solution of the one-dimensional heat equation with spatially periodic boundary conditions, \citep{Ohyama95}, and so is its integral in $z$. The representation of the solution via the Jacobi theta function has been used in a series of authors' papers devoted to pricing exotic financial derivatives (the barrier options), see, e.g., \citep{ItkinLiptonMuraveyBook} and references therein. However, when the clock $\tau$ runs ahead from $\tau=0$, the form of the solution in \eqref{T_final} becomes more complex.

\subsubsection{Solution of Volterra equation}

As shown in the previous section, the temperature $T(\tau,x)$ across the whole two-phase domain (see Fig.~\ref{freezing}) is determined by \eqref{T_final} if we know the trajectory $y(\tau)$ of the interphase (moving) boundary as well as the flux $\Phi(\tau)$ at this boundary. These two functions solve \eqref{Volt_for_yandPhi} - the system of two Volterra integral equation of the second kind. This system is linear in $\Phi(\tau)$ and nonlinear in $y(\tau)$.

\citep{ItkinLiptonMuraveyBook} present various methods of solving Volterra integral equations and the corresponding references. Here we suggest solving this system sequentially in time. For doing so
we first choose the upper bound $\bar{\tau}$ for the time $\tau$ and consider all moments of time $\tau \in [0,\bar{\tau}]$. Then we discretize the time by creating a uniform grid in $\tau$ with step $h$, so our discrete time is defined at the points $[0, h, 2 h,\ldots,\bar{\tau})$. We solve \eqref{Volt_for_yandPhi} first for $\tau = h$, the for $\tau = 2 h$ and so on.

To make the structure of the Volterra equations \eqref{Volt_for_yandPhi} more transparent, we re-write it in the form
\begin{align} \label{VoltN}
f_1(\tau) &= \int_0^\tau \Phi(s) \calK_1(\tau, s) ds - \Phi(\tau), \qquad f_2(\tau) = \int_0^\tau \Phi(s) \calK_2(\tau,s) ds,
\end{align}
\noindent where
\begin{align} \label{VoltN2}
f_1(\tau) &=- \sum_{n = 1}^{\infty} \frac{S_n(\tau) + P_n(\tau)}{2 N_n(\tau)} \lambda_n(\tau)
\left[s^+(y(\tau)) \sqrt{\kappa_I}\cos[x^-_n(\tau,y(\tau))] - s^-(y(\tau))\sqrt{\kappa_W} K_n(\tau) \cos[x^+_n(\tau,y(\tau))] \right], \nonumber \\
f_2(\tau) &= T_m - \eta(\tau,y(\tau)) - \sum_{n = 1}^{\infty} \frac{S_n(\tau) + P_n(\tau)}{2 N_n(\tau)} \left[s^+(y(\tau))\sin[x^-_n(\tau,y(\tau))] +  s^-(y(\tau))K_n(\tau) \sin[x^+_n(\tau,y(\tau))] \right], \nonumber \\
\calK_1(\tau,s) &= \sum_{n = 1}^{\infty}\frac{\lambda_n(\tau)}{2N_n(\tau)} e^{-(\tau -s)\lambda_n^2(\tau)}  \left[\sin[x_n^-(\tau,y(s))] - K_n(\tau) \sin[x_n^+(\tau,y(s))]\right] \nonumber \\
&\qquad\qquad \times \Big[ s^+(y(\tau)) \sqrt{\kappa_I} \cos[x_n^-(\tau,y(\tau))] - s^-(y(\tau)) \sqrt{\kappa_W} K_n(\tau) \cos[x_n^+(\tau,y(\tau))] \Big],  \\
\calK_2(\tau, s) &= \sum_{n = 1}^{\infty} \frac{1}{2 N_n(\tau)}  e^{-(\tau -s)\lambda_n^2(\tau)}
\left[\sin[x_n^-(\tau,y(s))] - K_n(\tau) \sin[x_n^+(\tau,y(s))]\right]  \nonumber \\
&\qquad\qquad \times \left[s^+(y(\tau))\sin[x^-_n(\tau,y(\tau))] +  s^-(y(\tau))K_n(\tau) \sin[x^+_n(\tau,y(\tau))] \right], \nonumber \\
P_n(\tau) &=  \int_0^\tau e^{-(\tau -s)\lambda_n^2(\tau)}\Bigg\{ \kappa_I \left[ B_-(s) \sin[x^-_n(\tau,y(s))] -  B_-(\tau)\sin[x^-_n(\tau,y(\tau))] \right] \nonumber \\
&\qquad\qquad - K_n(\tau) \kappa_W \left[ B_+(s)\sin[x^+_n(\tau,y(s))] - B_+(\tau)\sin[x^+_n(\tau,y(\tau))] \right] \Bigg\}ds. \nonumber
\end{align}

As $S_n(\tau)$ and $P_n(\tau)$ are proportional to $e^{-\tau \lambda_n^2(\tau)}$ all sums in \eqref{VoltN2} converge to a finite limit. One can check that $f_1(0) = P_n(0) = \calK_1(0,0) = \calK_2(0,0) = 0, \  f_2(0) = T_m - T_l$, and thus $\Phi(0) = 0$. Also, $\calK_1(\tau,\tau) = \calK_2(\tau,\tau) = 0$.  Therefore, \eqref{VoltN} has an explicit solution for $\Phi(\tau)$
\begin{align} \label{VoltN3}
\Phi(\tau) &= \int_0^\tau \Phi(s) \calK_1(\tau,s) ds - f_1(\tau).
\end{align}
Numerically this solution can be computed recursively. One can observe that $f_1(\tau) \ne 0, f_2(\tau) \ne 0$. Therefore, at the first step $\tau = h$ the second equation in \eqref{VoltN} implies $f_2(h) = 0$. This is a nonlinear algebraic equation with respect to $y(h)$. Solving it  numerically by using some initial guess, e.g., $y(h) = y(0) = y_-$ we obtain $y(h)$, and then from \eqref{VoltN3} get $\Phi(h) = -f_1(h)$. Here the function $f_1(h)$ is computed by using the already found value of $y(h)$.

At the next step $\tau = 2 h$ the values of $y(0), y(h), \Phi(0), \Phi(h)$ are already known. Hence, $y(2h)$ solves the nonlinear equation
\begin{equation*}
f_2(2h) = h \Phi(h) \calK_2(2h,h) = - h f_1(h) \calK_2(2h,h),
\end{equation*}
\noindent while $\Phi(2h)$ immediately follows from \eqref{VoltN3} and reads\footnote{Again, we use trapezoidal quadratures, but at $s=0$ the gradient $\Phi(s)$ vanishes, and at $s=\tau$ so does the kernel $\calK_1(\tau,s)$.}
\begin{equation*}
\Phi(2h) = h \Phi(h) \calK_1(2h,h) - f_1(2h) = - h f_1(h) \calK_1(2h,h) - f_1(2h).
\end{equation*}
And so on.

In our numerical experiments we choose constant values of the model parameters which are given in Tab.~\ref{param}. Note, $y'(\tau)$ is measured in (mm/s), and $\Phi(\tau)$ in (K$\cdot$mm/s).
\begin{table}[!htb]
\begin{center}
\begin{tabular}{|c|c|c|c|c|c|c|c|c|c|c|}
\toprule
$y_-$ & $y_+$ & $T_s$ & $T_m$ & $T_l$ & $\kappa_I$ & $\kappa_W$ & $\rho_I$ & $\rho_W$ & $L$ & $\rho_I L$ \\
\hline
mm &  mm &  K &  K & K & mm$^2$/s & mm$^2$/s & kg/m$^3$ & kg/m$^3$ & K m$^3$/kg & K\\
\bottomrule
1 & 50  & 270 & 273 & 290 & 1.02 & 0.13 & 917 & 997 & 0.054  & 49.86 \\
\bottomrule
\end{tabular}
\caption{Parameters of the model used in the Example. For those that depend on pressure it is assumed to be 1 atm.}
\label{param}
\end{center}
\end{table}

\begin{figure}[!htb]
\centering
\hspace*{-0.15in}\subfloat[]{\includegraphics[width=0.5\textwidth]{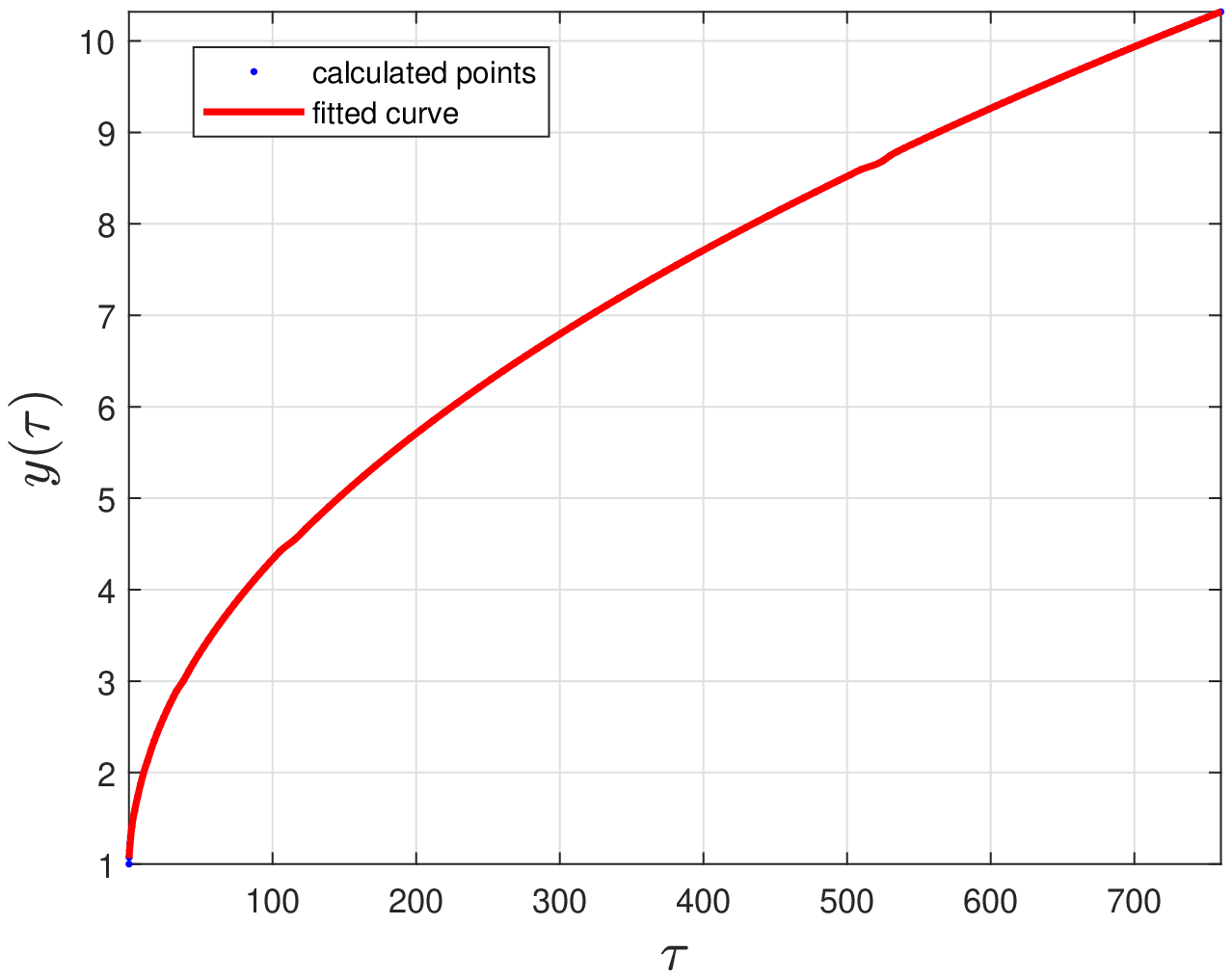}}
\hspace*{-0.15in}\subfloat[]{\includegraphics[width=0.5\textwidth]{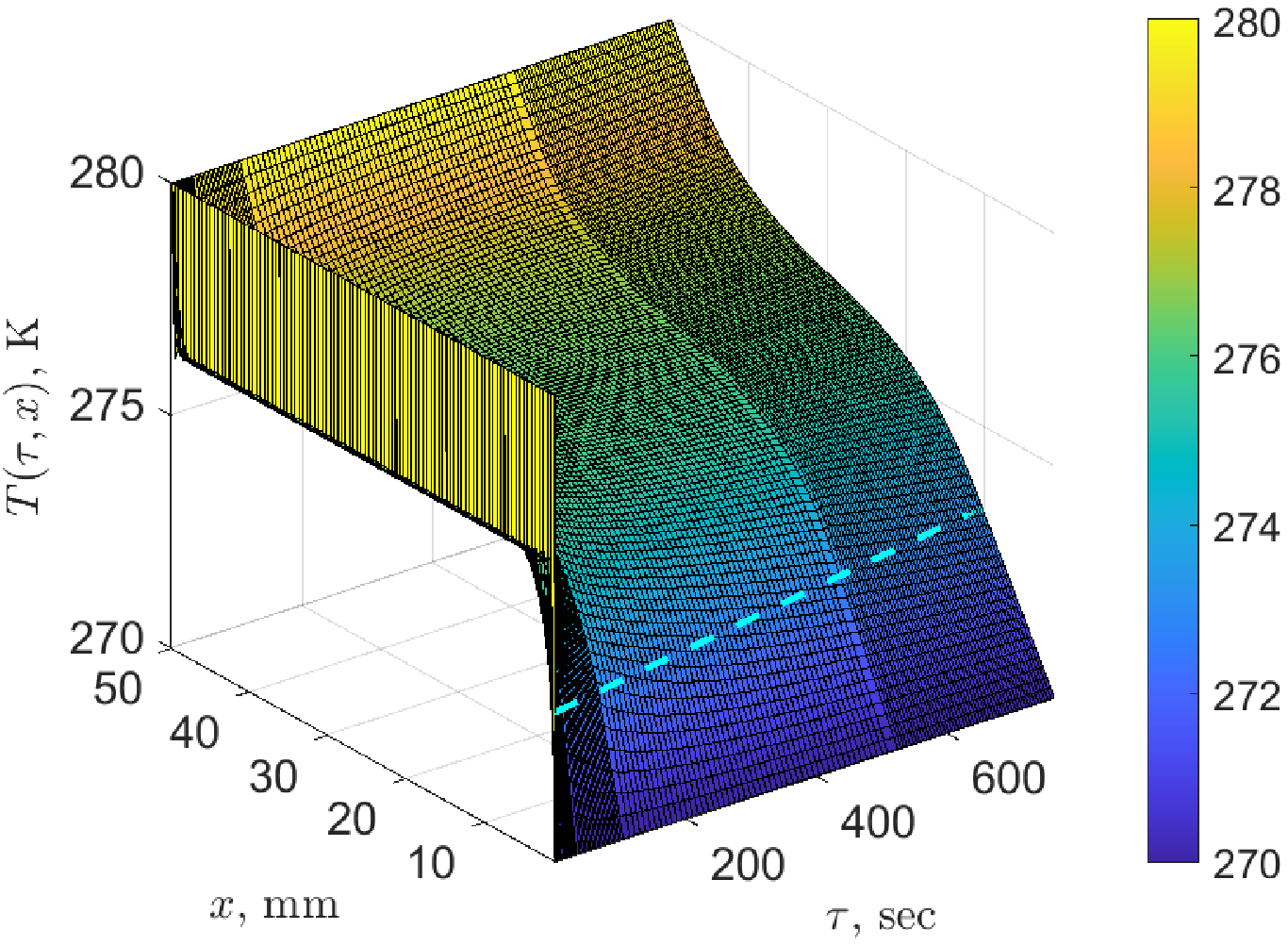}} \\
\hspace*{-0.15in}\subfloat[]{\includegraphics[width=0.5\textwidth]{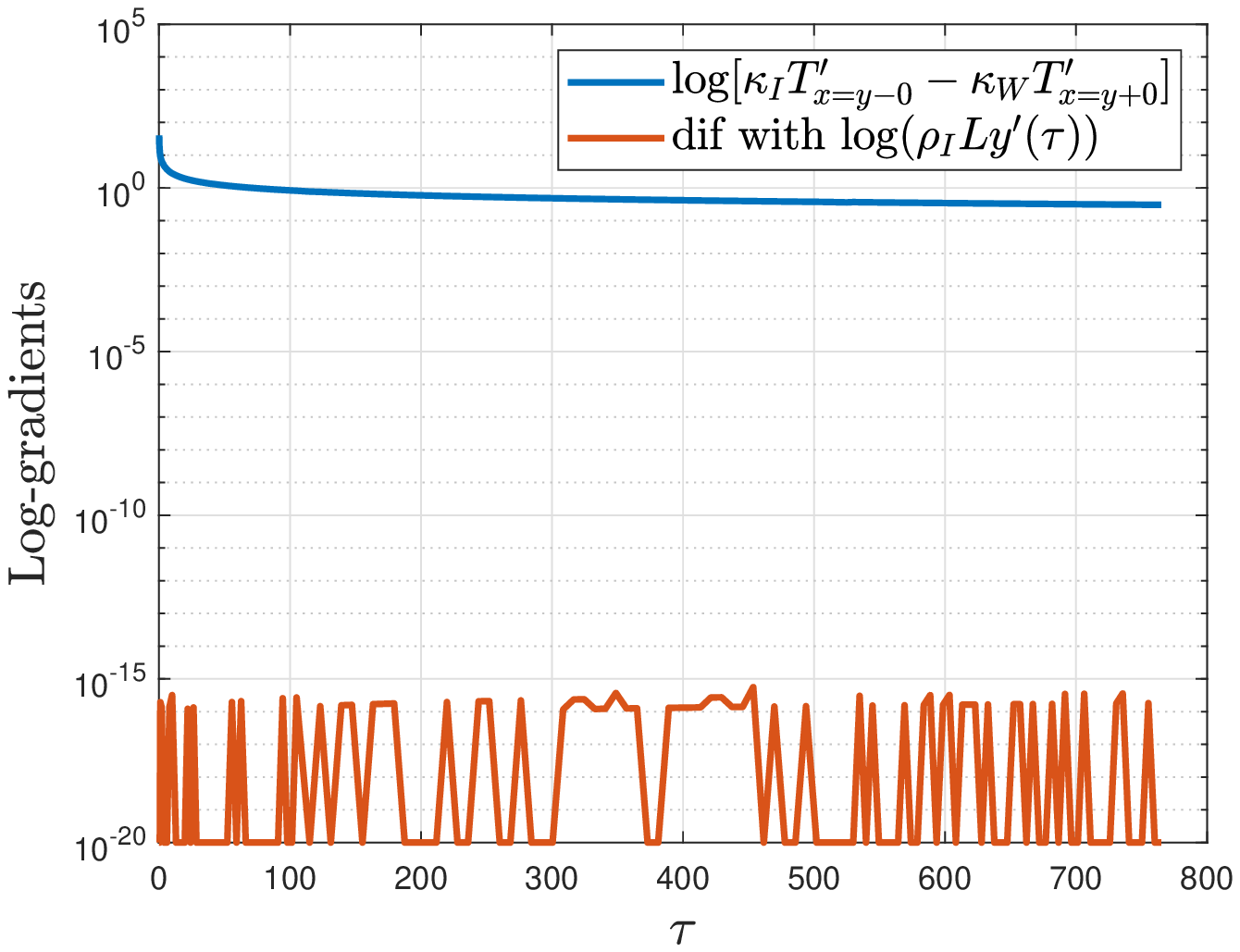}} \\
\caption{ Results on calculations:  a) The temperature profile as a function of the time $\tau$ and distance $x$, the dashed line corresponds to $T(\tau,x) = T_m$; (b) The law of the interphase boundary movement $y(\tau)$; c) Jump in the gradient of temperature at the interphase boundary.}
\label{figTemp}
\end{figure}

Our results are presented in Fig.~\ref{figTemp}: a) the temperature profile; b) the law of the interphase boundary movement $y(\tau)$; the jump in gradients at the interphase boundary; all as a function of time $\tau$. In our calculations we took $n=50$, because further increase of $n$ doesn't affects the results. Since our initial condition is singular (that is because  $y(0) = y_-, T(y_-) = T(y) = T_l$, and right at the first step in time the temperatures jump to their constant values $T(y_-) = T_s, T(y) = T_m$) we construct a nonuniform grid in time starting with small steps $h=0.01$ sec and then increasing the step up to $h=15$ sec. Also, an important point is to compute the correct values of $\lambda_n(\tau)$ by solving \eqref{eigenLam} (see comments after \eqref{eigen1}). We do it by using the \verb|chebfun| package in Matlab, then \verb|lambda = roots(chebfun(fun1,[1.e-6, upLimit]))| and taking only odd roots.  A typical elapsed time for one step in time is 0.14 sec on a standard PC with 3.2 Ghz CPU running under Windows and Matlab 2020.

\section{Conclusions} \label{sec:discussion}

In this paper, we present another approach to solving various MHE problems, which can be viewed as an alternative to the method described in \citep{ItkinLiptonMuraveyMulti}. This approach exploits the series expansion of the Dirac delta function into eigenfunctions of the corresponding Sturm-Liouville problem. We construct some new (oscillating) integral transforms and use them to solve several multilayer problems occurring in physics, finance, and biology. The number of such problems in science and engineering is vast.

The advantage of our method lies in the fact that our solutions are semi-analytical. This means that the answer is represented explicitly (perhaps, via some integrals). These integrals, however, depend on few yet unknown functions that solve a system of linear Volterra integral equations of the second kind. We derive this system in the paper and emphasize that, in general, it has to be solved numerically. However, as is shown in detail in \citep{ItkinLiptonMuraveyBook}, the corresponding numerical procedures are very efficient. In particular, they provide better speed and accuracy than the standard finite-difference (FD) methods.

Our main contribution to the existing literature lies in the fact that we consider moving boundaries, depending on time, while still constructing a semi-analytical solution. All previous approaches to these problems are purely numerical and rely on finite differences, finite elements, and similar methods. But, again, the semi-analytical solution has the advantage that the most challenging part of the problem is solved analytically, and the remaining part is solved numerically but with high efficiency.

As we mentioned earlier in the paper, various time-dependent problems with constant boundaries can be reduced to the MHE (or even to the Multilayer Bessel equations, \citep{Schoutens20061}) by a set of transformations. However, as a result, the boundaries become time-dependent in the new variables. Therefore, our setting covers a broader class of problems than what the title claims.

For financial applications, this semi-analytical form is also very convenient to compute sensitivities of asset prices to various parameters of the model; see a thorough discussion in \citep{ItkinLiptonMuraveyBook}.

For physical and biological problems, the following comment is in order. This paper considers matching conditions at the layer boundaries (internal interfaces), which are the continuity of the solution and its flux over the boundary. As discussed in the literature, see, e.g., in \citep{Hickson2011}, depending on what kind of physics problem is considered, various conditions at the internal interface can be set. It is possible to distinguish at least three different sets of matching conditions of particular interest. The first one assumes the temperature and flux continuity at the interfaces; we use it throughout this paper. The second matching condition deals with the conductivities, $k_{i}$, as opposed to the diffusivities, and reads
\begin{equation*}
\left.k_{i} \frac{\partial u_{i}}{\partial x}\right|_{x_{i}}=\left.k_{i+1} \frac{\partial u_{i+1}}{\partial x}\right|_{x_{i}}
\end{equation*}
\noindent where $k_{i}=\rho_{i} c_{i} \kappa_{i}$. This equation is more general than the continuity of the flux. It is applicable when the density, $\rho_{i}$, and specific heat capacity $c_{i}$ are possibly different in each layer. The continuities of the flux and conductivities highlight the key difference between heat and mass transfer problems, as the former expresses the equality of the mass flux and the latter of the heat flux.

The third matching condition assumes a jump either in the solution or its flux (or both) at the interface (especially if this is a phase interface). Accordingly, it can be written in the form
\begin{align*}
\kappa_{i} \frac{\partial u_{i}}{\partial x} &= H_{i}\left(u_{i+1}-u_{i}\right), \\
\kappa_{i+1} \frac{\partial u_{i+1}}{\partial x} &= H_{i+1}\left(u_{i+1}-u_{i}\right),
\end{align*}
\noindent where $H_{i}$ is the contact transfer coefficient. This matching condition is more general than the previous ones as it models the roughness of the contact between the layers and contact resistance at the interfaces, see \citep{Hickson2011,Illingworth2005}. If $H_{i} \rightarrow \infty$, then the contact is perfect, and hence this limit represents the equivalent matching conditions from the previous cases. In a similar way the jump in the function value $u$ at the interface is often modeled by the condition $u_i(\tau, y_i(\tau)) = P_i u_{i+1}(\tau, y_i(\tau))$ where $P_I$ is the partitioning coefficient which could be set constant or, in a more general case, is a function of time.

As far as our approach is concerned, it can deal with the second and third kinds of interface conditions. The integral transforms are the same, but solutions for the direct and inverse transforms are different. They rely on expanding discontinues functions into the Fourier series, which is possible and works fine. A detailed description of our method will be published elsewhere.

Another extension of the problem under consideration stems from choosing the boundary conditions at the external boundary. Usually, various analytical and semi-analytical solutions are constructed for the infinite or semi-infinite spatial domain. However, in real problems, e.g., that one discussed in \citep{Illingworth2005} and dedicated to finding transient solutions to diffusion problems in two distinct phases separated by a moving boundary,  such an assumption is not valid. Indeed, in addition to ignoring the actual boundary conditions at the boundary walls, analytical methods have to assume that one of the phases has zero initial size. \footnote{We note that, in planar geometries, the exact solution can admittedly be extended to cover the case when both phases are initially non-zero.} Such highly restrictive conditions mean that it is not possible to construct an analytical model for many situations, which are of considerable practical or industrial importance. However, for our method, these conditions are not as restrictive. Again, a detailed consideration will be presented elsewhere.

\section*{Acknowledgments}

We thank Peter Carr who drew our attention to the book  \citep{Antimirov}, as well as for various fruitful discussions. Dmitry Muravey acknowledges support by the Russian Science Foundation under the Grant number 20-68-47030.


\begin{thebibliography}{42}
\providecommand{\natexlab}[1]{#1}
\providecommand{\url}[1]{\texttt{#1}}
\expandafter\ifx\csname urlstyle\endcsname\relax
  \providecommand{\doi}[1]{doi: #1}\else
  \providecommand{\doi}{doi: \begingroup \urlstyle{rm}\Url}\fi

\bibitem[Andreucci et~al.()Andreucci, Cirillo, Colangeli, and
  Gabrielli]{Andreucci2019}
D.~Andreucci, E.N.M. Cirillo, M.~Colangeli, and D.~Gabrielli.
\newblock {Fick and Fokker-Planck Diffusion Law in Inhomogeneous Media}.
\newblock 174:\penalty0 469--493.

\bibitem[Ankirchner et~al.()Ankirchner, Blanchet-Scalliet, and
  Jeanblanc]{Ankirchner2017}
S.~Ankirchner, C~Blanchet-Scalliet, and M.~Jeanblanc.
\newblock Controlling the occupation time of an exponential martingale.
\newblock 70\penalty0 (2):\penalty0 415--428.

\bibitem[Ankirchner et~al.(2021)Ankirchner, Blanchet-Scalliet, Dorobantu, and
  Gay]{ankirchner2021}
S.~Ankirchner, C.~Blanchet-Scalliet, D.~Dorobantu, and L.~Gay.
\newblock {First passage time density of an Ornstein-Uhlenbeck process with
  broken drift}.
\newblock working paper, March 2021.
\newblock URL \url{https://hal.archives-ouvertes.fr/hal-03159498}.

\bibitem[Antimirov et~al.()Antimirov, Kolyshkin, and Vaillancourt]{Antimirov}
M.Ya. Antimirov, A.A. Kolyshkin, and R.~Vaillancourt.
\newblock \emph{Applied integral transforms}, volume~2.
\newblock American Mathematical Society, reprint edition edition.
\newblock ISBN 978-0821843147.

\bibitem[Asvestas et~al.(2014)Asvestas, Sifalakis, Papadopoulou, and
  Saridakis]{Asvestas2014}
M~Asvestas, A.G Sifalakis, E.P Papadopoulou, and Y.G Saridakis.
\newblock Fokas method for a multi-domain linear reaction-diffusion equation
  with discontinuous diffusivity.
\newblock \emph{Journal of Physics: Conference Series}, 490\penalty0 (012143),
  2014.

\bibitem[Carr and March(2018)]{CarrMarch2018}
E.J. Carr and N.G. March.
\newblock Semi-analytical solution of multilayer diffusion problems with
  time-varying boundary conditions and general interface conditions.
\newblock \emph{Applied Mathematics and Computation}, 333\penalty0
  (15):\penalty0 286--303, 2018.

\bibitem[Decamps et~al.()Decamps, Goovaerts, and Schoutens]{Schoutens20061}
M.~Decamps, M.~Goovaerts, and W.~Schoutens.
\newblock Asymmetric skew {B}essel processes and their applications to finance.
\newblock 186\penalty0 (1):\penalty0 130--147.

\bibitem[Deconinck et~al.(2014)Deconinck, Trogdon, and Vasan]{Deconinck2014}
B.~Deconinck, T.~Trogdon, and V.~Vasan.
\newblock The method of {F}okas for solving linear partial differential
  equations.
\newblock \emph{SIAM Review}, 56\penalty0 (1):\penalty0 159--186, 2014.

\bibitem[Deconinck et~al.(2016)Deconinck, Pelloni, N.E., and
  Sheils]{Deconinck2016}
B.~Deconinck, B.~Pelloni, B~N.E., and NE~Sheils.
\newblock Non-steady-state heat conduction in composite walls.
\newblock \emph{Proc. R. Soc. A}, 470\penalty0 (20130605), 2016.

\bibitem[Friz et~al.()Friz, Pigato, and Seibel]{FrizPigatoSeibel2020}
P.~Friz, P.~Pigato, and J.~Seibel.
\newblock The step stochastic volatility model (ssvm).
\newblock URL
  \url{https://papers.ssrn.com/sol3/papers.cfm?abstract_id=3595408}.
\newblock SSRN: 3595408.

\bibitem[Gairat and Shcherbakov()]{Gairat2016}
A.~Gairat and V.~Shcherbakov.
\newblock {Density of Skew Brownian motionand its functionals with application
  in finance}.
\newblock 26\penalty0 (4):\penalty0 1069--1088.

\bibitem[Goovaerts et~al.()Goovaerts, Decamps, and Schoutens]{Schoutens2006}
M.~Goovaerts, M.~Decamps, and W.~Schoutens.
\newblock Self exciting threshold interest rates models.
\newblock 9\penalty0 (7):\penalty0 1093--1122.

\bibitem[Gradshtein and Ryzhik(2007)]{GR2007}
I.S. Gradshtein and I.M. Ryzhik.
\newblock \emph{Table of Integrals, Series, and Products}.
\newblock Elsevier, 2007.

\bibitem[Han()]{Han1946}
Je-Chin Han.
\newblock \emph{Analytical heat transfer}.
\newblock CRC Press, Boca Raton, FL.
\newblock ISBN 978-1-4398-6196-7.

\bibitem[Hickson et~al.()Hickson, Barry, Mercer, and Sidhua]{Hickson2011}
R.I. Hickson, S.I. Barry, G.N. Mercer, and H.S. Sidhua.
\newblock Finite difference schemes for multilayer diffusion.
\newblock 54\penalty0 (1-2):\penalty0 210--220.

\bibitem[Illingworth and Golosnoy()]{Illingworth2005}
T.C. Illingworth and I.O. Golosnoy.
\newblock Numerical solutions of diffusion-controlled moving boundary problems
  which conserve solute.
\newblock 209:\penalty0 207--225.

\bibitem[Itkin and Lipton(2018)]{ItkinLipton2017}
A.~Itkin and A.~Lipton.
\newblock Filling the gaps smoothly.
\newblock \emph{Journal of Computational Sciences}, 24:\penalty0 195--208,
  2018.

\bibitem[Itkin and Muravey(2021)]{ItkinMuraveyDBFMF}
A.~Itkin and D.~Muravey.
\newblock Semi-analytic pricing of double barrier options with time-dependent
  barriers and rebates at hit.
\newblock \emph{Frontiers of Mathematical Finance}, 1:\penalty0 1--36, 2021.

\bibitem[Itkin et~al.(2021{\natexlab{a}})Itkin, Lipton, and
  Muravey]{ItkinLiptonMuraveyBook}
A.~Itkin, A.~Lipton, and D.~Muravey.
\newblock \emph{Generalized Integral Transforms in Mathematical Finance}.
\newblock WSPC, Singapore, 2021{\natexlab{a}}.
\newblock ISBN 978-981-123-173-5.

\bibitem[Itkin et~al.(2021{\natexlab{b}})Itkin, Lipton, and
  Muravey]{ItkinLiptonMuraveyMulti}
A.~Itkin, A.~Lipton, and D.~Muravey.
\newblock Multilayer heat equations: application to finance.
\newblock 1, 2021{\natexlab{b}}.

\bibitem[Ito and McKean()]{ItoMcKean1965}
K.~Ito and {H.P Jr.} McKean.
\newblock \emph{Diffusion Processes and their Sample Paths}.
\newblock Springer-Verlag Berlin Heidelberg.
\newblock ISBN 978-3-642-62025-6.

\bibitem[Jiji()]{Jiji2009}
L.M. Jiji.
\newblock \emph{Heat Conduction}.
\newblock Springer-Verlag, Berlin Heidelberg, 3rd edition.
\newblock ISBN 978-3-642-01266-2.

\bibitem[Keenan()]{Keenan1979}
J.H. Keenan.
\newblock \emph{Steam tables : thermodynamic properties of water including
  vapor, liquid, and solid phases}.
\newblock New York, 2nd edition.
\newblock ISBN 0471042102.

\bibitem[Keilson and Wellner()]{KeilsonWellner1978}
J.~Keilson and J.A. Wellner.
\newblock Oscillating {B}rownian motion.
\newblock 15\penalty0 (2):\penalty0 300--310.

\bibitem[Lee(2002)]{Lee2002}
Roger~W. Lee.
\newblock Implied volatility: Statics, dynamics, and probabilistic
  interpretation,.
\newblock Discussion paper, Department of Mathematics, Stanford University,
  2002.

\bibitem[Lejay(2006)]{Lejay2006}
A.~Lejay.
\newblock {On the constructions of the skew Brownian motion}.
\newblock \emph{Probability Surveys}, 3:\penalty0 413--466, 2006.

\bibitem[Lejay and Pigato()]{LejayPigato2019}
A.~Lejay and P.~Pigato.
\newblock A threshold model for local volatility : evidence of leverage and
  mean reversion effects on historical data.
\newblock 22\penalty0 (4):\penalty0 1--24.

\bibitem[Lejay and Pigato(2018)]{Lejay2018}
A.~Lejay and P.~Pigato.
\newblock Statistical estimation of the oscillating {Brownian Motion}.
\newblock \emph{Bernoulli}, 24\penalty0 (4B):\penalty0 3568 -- 3602, 2018.

\bibitem[Lipton(2018)]{Lipton2018a}
A.~Lipton.
\newblock \emph{{Financial Engineering: Selected Works of Alexander Lipton}}.
\newblock World Scientific, Singapore, 2018.

\bibitem[Lipton and Sepp(2011)]{LiptonSepp2011}
A.~Lipton and A.~Sepp.
\newblock Filling the gaps.
\newblock \emph{Risk Magazine}, pages 66--71, 2011.

\bibitem[Lipton et~al.()Lipton, Gal, and Lasis]{LiptonGal2014}
A.~Lipton, A.~Gal, and A.~Lasis.
\newblock Pricing of vanilla and first-generation exotic options in the local
  stochastic volatility framework: survey and new results.
\newblock 14\penalty0 (11):\penalty0 1899--1922.

\bibitem[Mordecki and Salminen(2019)]{Mordecki2019}
E.~Mordecki and P.~Salminen.
\newblock Optimal stopping of {B}rownian motion with broken drift.
\newblock \emph{High Frequency}, 2\penalty0 (2):\penalty0 113--120, 2019.

\bibitem[Mumford et~al.(1983)Mumford, Nori, Previato, and
  Stillman]{mumford1983tata}
D.~Mumford, C.~Musiliand~M. Nori, E.~Previato, and M.~Stillman.
\newblock \emph{Tata Lectures on Theta}.
\newblock Progress in Mathematics. Birkh{\"a}user Boston, 1983.
\newblock ISBN 9780817631093.

\bibitem[Ohyama(1995)]{Ohyama95}
Y.~Ohyama.
\newblock Differential relations of theta functions.
\newblock \emph{Osaka Journal of Mathematics}, 32\penalty0 (2):\penalty0
  431--450, 1995.

\bibitem[Pigato()]{Pigato2019}
P.~Pigato.
\newblock Extreme at-the-money skew in a local volatility model.
\newblock 23:\penalty0 827--859.

\bibitem[Pontrelli et~al.(2016)Pontrelli, Lauricella, Ferreira, and
  Pena]{Pontrelli2016}
G.~Pontrelli, M.~Lauricella, J.A. Ferreira, and G.~Pena.
\newblock Iontophoretic transdermal drug delivery: A multi-layered approach.
\newblock \emph{Mathematical Medicine and Biology}, 00:\penalty0 1--18, 2016.

\bibitem[Ramirez et~al.(2013)Ramirez, Thomann, and Waymire]{Ramirez2013}
J.M. Ramirez, E.A. Thomann, and E.C. Waymire.
\newblock Advection-dispersion across interfaces.
\newblock \emph{Statistical Science}, 28:\penalty0 487--509, 2013.

\bibitem[Salminen and Stenlund(2021)]{Salminen2021}
P.~Salminen and D.~Stenlund.
\newblock \emph{Journal of Theoretical Probability}, \penalty0 (2):\penalty0
  975--1011, 2021.

\bibitem[Sattin()]{Sattin2008}
F.~Sattin.
\newblock {Fick's law and Fokker-Planck Equation in inhomogeneous
  environments}.
\newblock 172\penalty0 (22):\penalty0 3941--3945.

\bibitem[Stetz()]{Stetza2010}
A.~Stetz.
\newblock A beautiful theory: the relationship between beauty and scientific
  truth.
\newblock URL \url{http://sites.science.oregonstate.edu/~stetza/ph407H/A
  Beautiful Theory.pdf}.

\bibitem[Titchmarsh()]{Titchmarsh1962}
E.C. Titchmarsh.
\newblock \emph{Eigenfunction Expansions Associated with Second-order
  Differential Equations}.
\newblock Oxford University Press, 2 edition.
\newblock ISBN 978-0198533177.

\bibitem[Vargaftik()]{Vargaftik1975}
N.B. Vargaftik.
\newblock \emph{Tables on the Thermophysical Properties of Liquids and Gases}.
\newblock Halsted Press, Division of John Wiley \& Sons, Inc., New York.

\end{thebibliography}

\vspace{0.4in}
\appendixpage
\appendix
\numberwithin{equation}{section}
\setcounter{equation}{0}

\section{The expansion of the Dirac delta function for the continuous spectrum} \label{appDiracCont}

Let us seek for the unknown function $u_\lambda(x | x_0)$ to be in the form
\begin{align}
u_\lambda(x | x_0) &= \IndVec_{x | y}^\top
\begin{bmatrix}
A_- e^{- \tl_- (x- y)} + \frac{1}{2\sigma_- \sqrt{\lambda}}e^{-\tl_- |x- x_0|} \\
A_+ e^{-\tl_+(x- y)} + \frac{1}{2\sigma_+\sqrt{\lambda}}e^{-\tl_+|x- x_0|},
\end{bmatrix}
\end{align}	
\noindent where functions $A_\pm$ are chosen to satisfy the matching conditions \eqref{SL_problem:CONT_SPEC}. In more detail, we need
\begin{align*}
\lim_{x \to y \pm 0} u_\lambda   &= A_\pm + \frac{1}{2\sigma_\pm \sqrt{\lambda}} e^{-\tl_\pm|y- x_0|}, \\
\lim_{x \to y \pm 0} u_{\lambda, x} &= \tl_\pm \left[\mp A_\pm + \frac{1}{2\sigma_\pm \sqrt{\lambda}} e^{-\tl_\pm |y- x_0|} \left(\Ind_{x_0 >y} + \Ind_{x_0 < y}\right)\right].
\end{align*}

Combining these two equations yields a linear system of equations for $A_-$ and $A_+$
\begin{align*}
\begin{bmatrix} 1 & -1 \\ \sigma_- & \sigma_+ \end{bmatrix}
\begin{bmatrix} A_- \\ A_+ 	\end{bmatrix} =
\frac{1}{2\sqrt{\lambda}}
\begin{bmatrix}
\frac{1}{\sigma_+} e^{-\tl_+ |y - x_0|} - \frac{1}{\sigma_-} e^{-\tl_- |y - x_0|} \\
e^{-\tl_+ |y - x_0|} \left(\Ind_{x_0 > y} -\Ind_{x_0 < y}\right) -e^{-\tl_-|y - x_0|} \left(\Ind_{x_0 > y} -\Ind_{x_0 < y} \right),
\end{bmatrix}	
\end{align*}
\noindent with the following solution
\begin{align}
\begin{bmatrix} A_- \\ A_+ \end{bmatrix}
&=
\begin{bmatrix}
\Sigma \frac{e^{-\tl_-|x_0 - y|}}{2\sigma_- \sqrt{\lambda}} 		&
\frac{2\sigma_+}{\sigma_- + \sigma_+}\frac{e^{-\tl_+|x_0 - y|}}{2\sigma_+ \sqrt{\lambda}} - \frac{e^{-\tl_-|x_0 - y|}}{2\sigma_- \sqrt{\lambda}} 		\\
-\frac{e^{-\tl_+|x_0 - y|}}{2\sigma_+ \sqrt{\lambda}} + \frac{2\sigma_-}{\sigma_- + \sigma_+} \frac{e^{-\tl_-|x_0 - y|}}{2\sigma_- \sqrt{\lambda}}		&
-\Sigma	\frac{e^{-\tl_+|x_0 - y|}}{2\sigma_+ \sqrt{\lambda}}
\end{bmatrix}
\IndVec_{x_0 | y}, \\
\Sigma &= \frac{\sigma_- - \sigma_+}{\sigma_- + \sigma_+}. \nonumber
\end{align}

Accordingly, the function $u_\lambda(x|x_0)$ can be represented as
\begin{align}
u_\lambda(x|x_0) =
\begin{bmatrix} \Ind_{x<y} e^{-\tl_-|x-y|} \\ \Ind_{x>y} e^{-\tl_+|x-y|} \end{bmatrix}^\top
\begin{bmatrix} A_-\\ A_+ \end{bmatrix}	
+ 	\IndVec_{x | y}^\top
\begin{bmatrix}
\frac{e^{-\tl_-|x - x_0|}}{2\sigma_- \sqrt{\lambda}} & 	\frac{e^{-\tl_-|x - x_0|}}{2\sigma_- \sqrt{\lambda}} 		\\
\frac{e^{-\tl_+|x - x_0|}}{2\sigma_+ \sqrt{\lambda}} & 	\frac{e^{-\tl_+|x - x_0|}}{2\sigma_+ \sqrt{\lambda}}
\end{bmatrix}
\IndVec_{x_0 | y}.
\end{align}

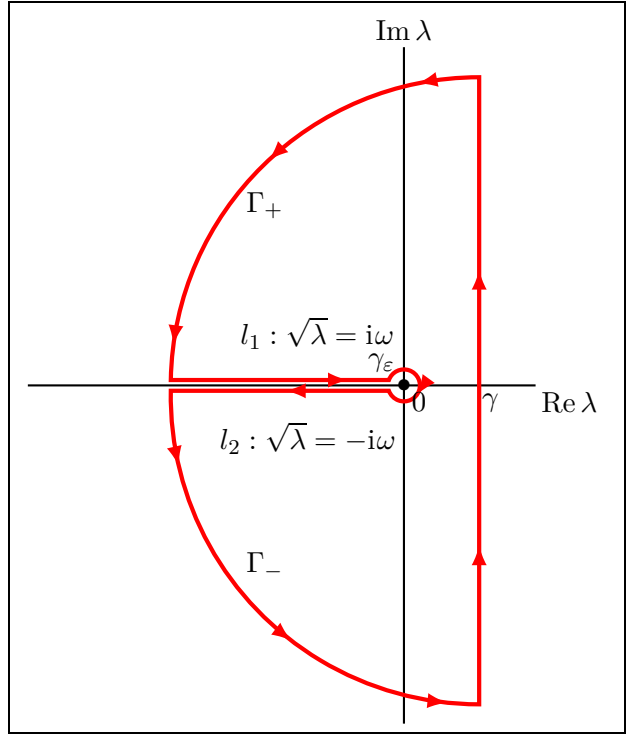
\begin{figure}[!htb]
	\begin{center}
		\fbox{
			\begin{tikzpicture}[thick, scale=0.5]
				\def\axisXlength{9}
				\def\zero{1}
				\def\gammap{3}
				\def\step{0.4}
				\def\alpha{0.75}
				\def\shif{-0.45}
				\def\i{4}
				\def\sr{0.4}
				\def\angshft{1}
				\draw (-\axisXlength, 0) -- (0.5*\axisXlength,0)
				(\zero, -\axisXlength) -- (\zero, \axisXlength);
				
				\draw[red, ultra thick, decoration={ markings,
					mark=at position 0.07 with {\arrow{latex}},
					mark=at position 0.2 with {\arrow{latex}},
					mark=at position 0.32 with {\arrow{latex}},
					mark=at position 0.4 with {\arrow{latex}},
					mark=at position 0.5 with {\arrow{latex}},
					mark=at position 0.6 with {\arrow{latex}},
					mark=at position 0.645 with {\arrow{latex}},
					mark=at position 0.71 with {\arrow{latex}},
					mark=at position 0.8 with {\arrow{latex}},
					mark=at position 0.9 with {\arrow{latex}},
					mark=at position 0.98 with {\arrow{latex}}
				},
				postaction={decorate}]
				let
				\n1 = {\i*\i*\step + \i*\step + 0.5*\step},
				\n2 = {\n1* sin(\angshft)},
				\n3 = {\sr * cos(\angshft)},
				\n4 = {sqrt(\n2 * \n2 + \n3*\n3)},
				\n5= {asin(\n2/ \n4)},
				\n6 = {\gammap - cos(\angshft) * \n1}
				in
				(\gammap,-\n1) -- (\gammap,\n1) arc (90:180-\angshft:\n1)
				--(\zero- \n3, \n2) arc(180 -\n5:-180+\n5:\n4)
				--(\n6, -\n2)  arc (-180-\angshft:-90:\n1)
				--(\gammap,-\n1)
				;
				
				
				\node at (0.6*\axisXlength,\shif){$\operatorname{Re} \lambda$};
				\node at (\zero,\axisXlength-\shif) {$\operatorname{Im} \lambda$};
				\node at (1.1*\gammap,\shif){$\gamma$};
				\node at (1.4*\zero,\shif){$0$};
				\node at (0.4,0.6) {$\gamma_{\varepsilon}$};
				\node at (\zero,0) {$\bullet$};
				\node at (-2.7,4.8) {$\Gamma_+$};
				\node at (-2.7,-4.8) {$\Gamma_-$};
				\node at (-1.3,1.4) {$l_1 : \sqrt{\lambda}= \iu \omega$};
				\node at (-1.55,-1.4) {$l_2 :\sqrt{\lambda} = - \iu \omega$};
			\end{tikzpicture}
		}
	\end{center}
	\caption{Contours of integration in a complex plane in case of the continuous spectrum.}
	\label{osc_contour}
\end{figure}

Now we are can compute the integral in \eqref{SL:Dirac_def}. Let us extend the original contour $(\gamma, -\iu \infty)- (\gamma, \iu \infty)$ to the loop contour $L$ in Fig.~\ref{osc_contour} which can be described as follows. It starts with a parallel line $(\gamma, -\iu R) -(\gamma, \iu R)$,  extending to a big symmetric arcs $\Gamma_-$ and $\Gamma_+$ around the point $(\gamma, 0)$ with the radius $R$ with the two horizontal line segments $l_1, l_2$ connecting to a small circle $\gamma_\varepsilon$ around the origin with the radius $\varepsilon$.

Using a standard technique, we take a limit $\varepsilon \to 0, R \to \infty$, apply the Cauchy Residue theorem and obtain
\begin{equation}
	\oint_L u_{\lambda}(x | x_0) d\lambda = 0.
\end{equation}
Therefore,
\begin{equation}
	\delta (x - x_0) =	\frac{1}{2\pi \iu} \int_{\gamma - \iu \infty}^{\gamma + \iu \infty} u_\lambda (x | x_0) d\lambda =  \int_{0}^{\infty} \frac{u_{e^{-\iu \pi} \omega^2}(x | x_0) - u_{e^{\iu \pi} \omega^2}(x | x_0)}{\pi \iu} \omega d\omega.
\end{equation}

Omitting tedious algebra, we find the following representation of the Dirac delta function
\begin{align} \label{CS:delta_final}
\delta(x - x_0) &= \frac{1}{2\pi}\int_{-\infty}^{+\infty}	
\begin{bmatrix}
\frac{e^{\iu \omega\frac{x -y}{\sigma_-}}}{\sigma_-} \Ind_{x<y}   \\
\frac{e^{\iu \omega\frac{x -y}{\sigma_+}}}{\sigma_+} \Ind_{x>y} 	
\end{bmatrix}^\top
\begin{bmatrix}
1 + \Sigma e^{2\iu \omega\frac{x_0 - y }{\sigma_-}} & 1 + \Sigma  \\ 1 - \Sigma & 1- \Sigma e^{2\iu \omega\frac{x_0 - y }{\sigma_+}}
\end{bmatrix}
\begin{bmatrix}
e^{-\iu \omega\frac{x_0 -y}{\sigma_-}} \Ind_{x_0 < y} \\
e^{-\iu \omega\frac{x_0 -y}{\sigma_+}} \Ind_{x_0 > y}
\end{bmatrix}
d\omega.
\end{align}

We proceed with separation of the variables $x$ and $x_0$ in \eqref{CS:delta_final}. Introducing the matrices (i.e. Forward and Backward)
\begin{align} \label{FBmatrices}
\FM_{z, \sigmaVec}(\omega) &= \StrM_{e^{\frac{\iu \omega z}{\sigma_-}}, e^{\frac{\iu \omega z}{\sigma_+}}} =
\begin{bmatrix} e^{\frac{\iu \omega z}{\sigma_-}} & 0   \\  0 & e^{\frac{\iu \omega z}{\sigma_+}} \end{bmatrix}
, \\
\BM_{z, \sigmaVec}(\omega) &=
\begin{bmatrix}
\frac{1}{\sigma_-}\left(e^{-\frac{\iu \omega z}{\sigma_-}} + \Sigma e^{\frac{\iu \omega z}{\sigma_-}} \right) & 	\frac{1}{\sigma_-} \left(1 + \Sigma \right)e^{-\frac{\iu \omega z}{\sigma_+}} \\
\frac{1}{\sigma_+} \left(1-\Sigma \right)  e^{-\frac{\iu \omega z}{\sigma_-}} &
\frac{1}{\sigma_+}\left(e^{-\frac{\iu \omega z}{\sigma_+}} - \Sigma e^{\frac{\iu \omega z}{\sigma_+}} \right)
\end{bmatrix}
, \nonumber
\end{align}
\noindent the \eqref{CS:delta_final} can be re-written as
\begin{align} \label{CS:delta_FB_repr}
\delta( x - x_0) =\frac{1}{2\pi}\int_{-\infty}^{+\infty}	
\IndVec_{x | y}^\top \FM_{x -y, \sigmaVec}(\omega)\BM_{x_0 - y, \sigmaVec}(\omega)
\IndVec_{x_0 | y} d\omega.
\end{align}

It is worth mentioning that the matrix $\FM_{x, y, \sigmaVec}$ has a semigroup property with respect to the matrix multiplication, i.e.
\begin{equation} \label{CS:semigroup_FM}
\FM_{z, \sigmaVec}(\omega)\FM_{\zeta, \sigmaVec}(\omega) =\FM_{z + \zeta, \sigmaVec}(\omega).
\end{equation}

\section{Contour integration} \label{appContour}

By analogy with computation of the integral in \eqref{SL:Dirac_def}, we extend the contour $(\gamma, -\iu \infty)- (\gamma, \iu \infty)$ to the loop contour $L$ shown in Fig.~\ref{MS_contour}. However, in this case the integration domain contains simple poles which are determined by the equation $\det \LambdaM^*(\lambda) =0$. Again, we proceed in a way similar to that for \eqref{SL:Dirac_def}, i.e., by taking a limit $\varepsilon \to 0, R \to \infty$ and applying the Cauchy Residue theorem. Omitting an intermediate algebra, we provide just the final result which is the representation in \eqref{MS:delta_repr}.

\begin{figure}[!htb]
	\begin{center}
		\fbox{
			\begin{tikzpicture}[thick, scale=0.5]
				\def\axisXlength{9}
				\def\zero{1}
				\def\gammap{3}
				\def\step{0.4}
				\def\alpha{0.75}
				\def\shif{-0.45}
				\def\i{4}
				\def\sr{0.4}
				\def\angshft{1}
				\draw (-\axisXlength, 0) -- (0.5*\axisXlength,0)
				(\zero, -\axisXlength) -- (\zero, \axisXlength);
				\draw[red, ultra thick, decoration={ markings,
					mark=at position 0.07 with {\arrow{latex}},
					mark=at position 0.2 with {\arrow{latex}},
					mark=at position 0.32 with {\arrow{latex}},
					mark=at position 0.4 with {\arrow{latex}},
					mark=at position 0.5 with {\arrow{latex}},
					mark=at position 0.6 with {\arrow{latex}},
					mark=at position 0.645 with {\arrow{latex}},
					mark=at position 0.71 with {\arrow{latex}},
					mark=at position 0.8 with {\arrow{latex}},
					mark=at position 0.9 with {\arrow{latex}},
					mark=at position 0.98 with {\arrow{latex}}
				},
				postaction={decorate}]
				let
				\n1 = {\i*\i*\step + \i*\step + 0.5*\step},
				\n2 = {\n1* sin(\angshft)},
				\n3 = {\sr * cos(\angshft)},
				\n4 = {sqrt(\n2 * \n2 + \n3*\n3)},
				\n5= {asin(\n2/ \n4)},
				\n6 = {\gammap - cos(\angshft) * \n1}
				in
				(\gammap,-\n1) -- (\gammap,\n1) arc (90:180-\angshft:\n1)
				--(\zero- \n3, \n2) arc(180 -\n5:-180+\n5:\n4)
				--(\n6, -\n2)  arc (-180-\angshft:-90:\n1)
				--(\gammap,-\n1)

				;
				\draw[red, ultra thick] (0.25, -2.1) arc(180 :-180: 0.3);
				\draw[red, ultra thick] (0.25, 2.1) arc(180 :-180: 0.3);
				\node at (0.25 + 0.3, -2.1) {$\bullet$};
				\node at (0.25 + 0.3, 2.1) {$\bullet$};
				
				\draw[red, ultra thick] (-0.8, -2.6) arc(180 :-180: 0.3);
				\draw[red, ultra thick] (-0.8, 2.6) arc(180 :-180: 0.3);
				\node at (-0.8 + 0.3, -2.6) {$\bullet$};
				\node at  (-0.8 + 0.3, 2.6) {$\bullet$};
				
				\draw[red, ultra thick] (-1.6, -3) arc(180 :-180: 0.3);
				\draw[red, ultra thick] (-1.6, 3) arc(180 :-180: 0.3);
				\node at  (-1.6 + 0.3, -3) {$\bullet$};
				\node at  (-1.6 + 0.3,  3) {$\bullet$};
				
				\draw[red, ultra thick] (-2.7, -3.6) arc(180 :-180: 0.3);
				\draw[red, ultra thick] (-2.7, 3.6) arc(180 :-180: 0.3);
				\node at  (-2.7 + 0.3, -3.6) {$\bullet$};
				\node at  (-2.7 + 0.3, 3.6) {$\bullet$};
				
				\draw[red, ultra thick] (-3.9, -4.0) arc(180 :-180: 0.3);
				\draw[red, ultra thick] (-3.9, 4.0) arc(180 :-180: 0.3);
				\node at (-3.9 + 0.3, -4.0) {$\bullet$};
				\node at (-3.9 + 0.3, 4.0){$\bullet$};

				
				\node at (0.6*\axisXlength,\shif){$\operatorname{Re} \lambda$};
				\node at (\zero,\axisXlength-\shif) {$\operatorname{Im} \lambda$};
				\node at (1.1*\gammap,\shif){$\gamma$};
				\node at (1.4*\zero,\shif){$0$};
				\node at (0.4,0.6) {$\gamma_{\varepsilon}$};
				\node at (\zero,0) {$\bullet$};
				\node at (-2.7,4.8) {$\Gamma_+$};
				\node at (-2.7,-4.8) {$\Gamma_-$};
				\node at (-1.3,1.4) {$l_1 : \sqrt{\lambda}= \iu \omega$};
				\node at (-1.55,-1.4) {$l_2 :\sqrt{\lambda} = - \iu \omega$};
			\end{tikzpicture}
		}
	\end{center}
	\caption{Contours of integration in a complex plane in case of the mixed spectrum.}
	\label{MS_contour}
\end{figure}

\section{Solution of \eqref{PDE:osc_main_PDE}} \label{app4}

To solve \eqref{PDE:osc_main_PDE} we use the oscillating transform
\begin{align} \label{PDE:osc_ex_transform}
	\boldsymbol{\bar u }(\tau, \omega) =
	\begin{bmatrix} \bar u_- (\tau, \omega) \\ \bar u_+ (\tau, \omega) \end{bmatrix} =
	\int_{-\infty}^{\infty} \FM_{x, \sigmaVec}(\omega) \IndVec_{x | y(\tau)} u(\tau, x)dx,
\end{align}
\noindent where the matrix $\FM_{x, \sigmaVec}(\omega)$ was defined in \eqref{FBmatrices}. Using the semigroup property \eqref{CS:semigroup_FM} in the inversion formula \ref{CS:OSC_FT_inv} yields
\begin{align} \label{u_inv_formula}
	u(\tau, x) = \frac{1}{2 \pi} \int_{-\infty}^{\infty}  \boldsymbol{\bar u}(\tau, \omega)^\top
	\FM_{-y(\tau), \sigmaVec}(\omega)	\BM_{x-y(\tau), \sigmaVec}(\omega) \IndVec_{x | y(\tau)} d\omega.
\end{align}
The images $\bar u_-$ and $\bar u_+$ can be easily found in the explicit form and read
\begin{align}
	\bar u_-(\tau, \omega) = \int_{-\infty}^{y(\tau)}  e^{\frac{\iu \omega x}{\sigma_-}} u(\tau, x) dx,
	\qquad
	\bar u_+(\tau, \omega) = \int^{+\infty}_{y(\tau)}  e^{\frac{\iu \omega x}{\sigma_+}}u(\tau, x) dx
\end{align}
To make these integrals well-behaved we need function $u(\tau,x)$ to vanish at plus and minus infinity faster than the corresponding exponent\footnote{Actually, as shown below integrals on $\omega$ in \eqref{intOmega} are regular, so there is no problem  with convergence of these transforms.}. Now we define new functions $\Phi(\tau), \phi(\tau)$ as
\begin{equation} \label{gradDef}
	\sigma^2_- \frac{\partial u(\tau, x)}{\partial x} \bigg|_{x = y(\tau) -0} =
	\sigma^2_+ \frac{\partial u(\tau, x)}{\partial x} \bigg|_{x = y(\tau) +0} = \Phi(\tau),
	\quad  u(\tau, y(\tau)) = \varphi(\tau),
\end{equation}
\noindent and apply the transform \eqref{PDE:osc_ex_transform} directly to the both sides of the last line in  \eqref{PDE:osc_main_PDE}
\begin{align*}
\int_{-\infty}^{y(\tau)} \sigma_-^2 u_{xx} e^{\frac{\iu \omega x}{\sigma_-}} dx &= e^{\frac{\iu \omega y(\tau)}{\sigma_-}} \Phi(\tau) - \iu \omega \sigma_- \int_{-\infty}^{y(\tau)} u_x e^{\frac{\iu \omega x}{\sigma_-}}  dx \\
&= e^{\frac{\iu \omega y(\tau)}{\sigma_-}}  \Phi(\tau) - \iu \omega \sigma_- e^{\frac{\iu \omega y(\tau)}{\sigma_-}} \varphi(\tau) -\omega^2 \bar u_-(\tau, \omega), \\
\int_{-\infty}^{y(\tau)} u_\tau e^{\frac{\iu \omega x}{\sigma_-}}  dx &= \frac{\partial }{\partial \tau} \int_{-\infty}^{y(\tau)}  e^{\frac{\iu \omega x}{\sigma_-}} u(\tau, x) dx - y'(\tau) e^{\frac{\iu \omega y(\tau)}{\sigma_-}} u(\tau, y(\tau)) = \fp{\bar u_-}{\tau} - y'(\tau) e^{\frac{\iu \omega y(\tau)}{\sigma_-}} \phi(\tau), 	\\
\int^{+\infty}_{y(\tau)} \sigma_+^2 u_{xx} e^{\frac{\iu \omega x}{\sigma_+}}  dx &= -  e^{\frac{\iu \omega y(\tau)}{\sigma_+}} \Phi(\tau) - \iu \omega \sigma_+ \int^{+\infty}_{y(\tau)} u_x e^{\frac{\iu \omega x}{\sigma_+}}  dx \\
&= -  e^{\frac{\iu \omega y(\tau)}{\sigma_+}}  \Phi(\tau) + \iu \omega \sigma_+ e^{\frac{\iu \omega y(\tau)}{\sigma_+}} \varphi(\tau) -\omega^2 \bar u_+(\tau, \omega), \\
\int^{+\infty}_{y(\tau)} u_\tau e^{\frac{\iu \omega x}{\sigma_+}}  dx &= \frac{\partial }{\partial \tau} \int^{+\infty}_{y(\tau)}  e^{\frac{\iu \omega x}{\sigma_+}} u(\tau, x) dx + y'(\tau) e^{\frac{\iu \omega y(\tau)}{\sigma_+}} u(\tau, y(\tau)) = \fp{\bar u_+}{\tau} + y'(\tau) e^{\frac{\iu \omega y(\tau)}{\sigma_+}} \phi(\tau).
\end{align*}

Therefore, the matching conditions in \eqref{PDE:osc_main_PDE} give rise to two independent ordinary differential equations (ODE) (with respect to $\tau$)
\begin{align*}
\fp{\bar u_-}{\tau} + \omega^2 \bar u_- &= e^{\frac{\iu \omega y(\tau)}{\sigma_-}}  \Phi(\tau) - \iu \omega \sigma_- e^{\frac{\iu \omega y(\tau)}{\sigma_-}} \varphi(\tau) +y'(\tau) e^{\frac{\iu \omega y(\tau)}{\sigma_-}}	\phi(\tau), \quad
\bar u_-(0, \omega) = \Ind_{x_0 < y(0)} e^{ \frac{\iu \omega x_0}{\sigma_-}}, \\
\fp{\bar u_+}{\tau} + \omega^2 \bar u_+ &= -e^{\frac{\iu \omega y(\tau)}{\sigma_+}}  \Phi(\tau) + \iu \omega \sigma_+ e^{\frac{\iu \omega y(\tau)}{\sigma_+}} \varphi(\tau) -y'(\tau) e^{\frac{\iu \omega y(\tau)}{\sigma_+}}	\phi(\tau), \quad
\bar u_+(0, \omega) = \Ind_{x_0 > y(0)} e^{ \frac{\iu \omega x_0}{\sigma_+}}.
\end{align*}
These equations can be solved analytically via a standard technique
\begin{align} \label{baruSol}
\bar u_-(\tau, \omega) &=  \Ind_{x_0 < y(0)} e^{-\omega^2 \tau+ \iu \omega\frac{ x_0}{\sigma_-}} \\
&+ \int_0^\tau e^{-\omega^2(\tau -s)} \left[e^{\frac{\iu \omega y(s)}{\sigma_-}}  \Phi(s) - \iu \omega \sigma_- e^{\frac{\iu \omega y(s)}{\sigma_-}} \varphi(s) +y'(s) e^{\frac{\iu \omega y(s)}{\sigma_-}}	\varphi(s) \right] ds, \nonumber \\
\bar u_+(\tau, \omega) &=  \Ind_{x_0 > y(0)} e^{-\omega^2 \tau+ \iu \omega\frac{ x_0}{\sigma_+}} \nonumber \\
&- \int_0^\tau e^{-\omega^2(\tau -s)} \left[e^{\frac{\iu \omega y(s)}{\sigma_+}}  \Phi(s) - \iu \omega \sigma_+ e^{\frac{\iu \omega y(s)}{\sigma_+}} \varphi(s) +y'(s) e^{\frac{\iu \omega y(s)}{\sigma_+}} \varphi(s) \right] ds. \nonumber
\end{align}
This result can be also expressed in the matrix form
\begin{align}
\boldsymbol{\bar u} (\tau, \omega) &= e^{-\omega^2 \tau }\FM_{x_0, \sigmaVec} (	\omega) \IndVec_{x_0 | y(0)}
+ \int_0^\tau  e^{-\omega^2 (\tau - s)} \FM_{y(s), \sigmaVec} (\omega)
\begin{bmatrix}
\Phi(s) - \iu \omega \sigma_- \varphi(s) +y'(s)\varphi(s) \\
-\Phi(s) + \iu \omega \sigma_+ \varphi(s) -y'(s)\varphi(s)
\end{bmatrix}  ds.
\end{align}

Applying the inversion formula \eqref{u_inv_formula} we arrive at the following representation
\begin{align}
u(\tau, x) &= \frac{1}{2 \pi} \int_{-\infty}^{\infty} e^{-\omega^2 \tau} \IndVec_{x_0 | y(0)}^\top
\FM_{x_0-y(\tau), \sigmaVec}(\omega) \BM_{x-y(\tau), \sigmaVec}(\omega) \IndVec_{x | y(\tau)} d\omega \\
&+ \frac{1}{2 \pi} \int_{-\infty}^{\infty} \int_0^\tau e^{-\omega^2 (\tau - s)} \boldsymbol{\Psi}(s,\omega) ^\top \FM_{y(s)-y(\tau), \sigmaVec}(\omega) \BM_{x-y(\tau), \sigmaVec}(\omega) \IndVec_{x | y(\tau)} ds d\omega, \nonumber	
\end{align}
\noindent where $\boldsymbol{\Psi}(s, \omega)$ is defined as
\begin{equation} \label{PsiVec}
\boldsymbol{\Psi}(s,\omega) = \left[\Phi(s) + y'(s) \varphi(s) \right]
\begin{bmatrix} 1 \\ - 1 \end{bmatrix}	
- \iu \omega \varphi(s)
\begin{bmatrix} \sigma_-  & 0 \\ 0 & \sigma_+ \end{bmatrix}	
\begin{bmatrix} 1 \\ - 1 \end{bmatrix}.
\end{equation}

The integrals with respect to $\omega$ can be computed analytically. Indeed, using the identities, \citep{GR2007}
\begin{align} \label{intOmega}
\frac{1}{2\pi} \int_{-\infty}^{+\infty} e^{-\omega^2 (\tau - s) + \iu \alpha \omega} d\omega = \frac{e^{-\frac{\alpha^2}{4(\tau - s)}}}{2\sqrt{\pi (\tau - s)}}, \qquad
\frac{1 }{2\pi} \int_{-\infty}^{+\infty} \iu \omega e^{-\omega^2 (\tau - s) + \iu \alpha \omega} d\omega = - \frac{\alpha e^{-\frac{\alpha^2}{4(\tau - s)}}}{4\sqrt{\pi (\tau - s)^3}},	
\end{align}
\noindent and the explicit formula for the matrix product $	\FM_{\zeta, \sigmaVec}(\omega)\BM_{z , \sigmaVec}(\omega)$
\begin{align*}
\FM_{\zeta, \sigmaVec}(\omega)\BM_{z , \sigmaVec}(\omega) &=
\begin{bmatrix}
\frac{1}{\sigma_-}	e^{\frac{\iu \omega \zeta}{\sigma_-}}  \left(e^{-\frac{\iu \omega z}{\sigma_-}}
+ \Sigma e^{\frac{\iu \omega z}{\sigma_-}} \right) & \frac{1}{\sigma_-} 	e^{\frac{\iu \omega \zeta}{\sigma_-}} \left(1 + \Sigma\right)e^{-\frac{\iu \omega z}{\sigma_+}} \\
\frac{1}{\sigma_+} e^{\frac{\iu \omega \zeta}{\sigma_+}}\left(1-\Sigma \right)  e^{-\frac{\iu \omega z}{\sigma_-}} &	\frac{1}{\sigma_+} e^{\frac{\iu \omega \zeta}{\sigma_+}} \left(e^{-\frac{\iu \omega z}{\sigma_+}} - \Sigma e^{\frac{\iu \omega z}{\sigma_+}} \right) \\
\end{bmatrix}
, \qquad
\Sigma = \frac{\sigma_- - \sigma_+}{\sigma_- + \sigma_+}.
\end{align*}
\noindent yields
\begin{align} \label{etaM}
\frac{1}{2\pi} \int_{-\infty}^{+\infty} & e^{-\omega^2 (\tau - s)\omega} \FM_{\zeta, \sigmaVec}(\omega)\BM_{z, \sigmaVec}(\omega) d\omega = \PM_{\sigmaVec} (z, \tau | \zeta, s), \\
\frac{1}{2\pi} \int_{-\infty}^{+\infty} & \iu \omega e^{-\omega^2 (\tau - s)\omega} \FM_{\zeta, \sigmaVec}(\omega)\BM_{z, \sigmaVec}(\omega) d\omega 	= \etaM_{\sigmaVec}(z, \tau | \zeta, s), \nonumber \\
\PM_{\sigmaVec} (z, \tau | \zeta, s) &= \frac{1}{2\sqrt{\pi(\tau  -s)}}
\begin{bmatrix} \frac{1}{\sigma_-} & 0 \\ 0 &    \frac{1}{\sigma_+} \end{bmatrix} \nonumber \\
&\times
\begin{bmatrix}
e^{-\frac{\left(z - \zeta\right)^2}{4 \sigma_-^2(\tau - s)} } + \Sigma e^{-\frac{\left( z + \zeta \right)^2}{4\sigma_-^2(\tau - s)} } &
\left(1 + \Sigma \right)e^{-\frac{\left( z/\sigma_+ - \zeta / \sigma_-  \right)^2}{4(\tau - s)} } 		\nonumber \\
\left(1-\Sigma  \right)  e^{-\frac{\left( z /\sigma_- - \zeta / \sigma_+ \right)^2}{4(\tau - s)} } 		&
e^{-\frac{\left( z- \zeta\right)^2}{4\sigma_+^2(\tau - s)} } - \Sigma e^{-\frac{\left(z+ \zeta  \right)^2}{4\sigma_+^2(\tau - s)} }
\end{bmatrix}
, \nonumber \\
\etaM_{\sigmaVec}(z, \tau | \zeta, s) &= - \frac{1}{4\sqrt{\pi(\tau  -s)^3}}
\begin{bmatrix} \frac{1}{\sigma_-} & 0 \\ 0 &    \frac{1}{\sigma_+} \end{bmatrix} \nonumber \\
&\times
\begin{bmatrix}
\frac{\zeta - z}{\sigma_-} e^{-\frac{\left( z- \zeta \right)^2}{4\sigma_-^2(\tau - s)} } +\frac{\zeta + z}{\sigma_-} \Sigma e^{-\frac{\left( z + \zeta \right)^2}{4\sigma_-^2(\tau - s)} } 		& 	
\left[\frac{\zeta}{\sigma_-} - \frac{z}{\sigma_+} \right]\left(1 + \Sigma\right)e^{-\frac{\left(z / \sigma_+ - \zeta / \sigma_-  \right)^2}{4(\tau - s)} } 		\\
\left[\frac{\zeta}{\sigma_+} - \frac{z}{\sigma_-} \right] \left(1-\Sigma \right)  e^{-\frac{\left( z / \sigma_- - \zeta / \sigma_+ \right)^2}{4(\tau - s)} } 		&
\frac{\zeta - z}{\sigma_+} e^{-\frac{\left( z- \zeta\right)^2}{4\sigma_+^2(\tau - s)} } - \frac{\zeta + z}{\sigma_+} \Sigma e^{-\frac{\left( z+ \zeta \right)^2}{4 \sigma_+^2(\tau - s)} }
\end{bmatrix}
. \nonumber
\end{align}

Therefore, $u(\tau, x)$ reads
\begin{align} \label{finSol}
u(\tau, x) &=\left\langle \IndVec_{x_0 | y(0)} , \PM_{\sigmaVec}(x - y(\tau), \tau, | x_0 - y(\tau), 0) \IndVec_{x | y(\tau)} \right \rangle \\
&+ \int_0^\tau \left[ \Phi(s) + y'(s) \varphi(s)\right]
\left \langle \begin{bsmallmatrix}  1 \\ -1  \end{bsmallmatrix},
\PM_{\sigmaVec}(x - y(\tau), \tau | y(s) - y(\tau), s) 	\IndVec_{x | y(\tau)}
\right \rangle ds \nonumber \\
&+ \int_0^\tau \varphi(s) \left\langle
\begin{bsmallmatrix}	1 \\ -1 \end{bsmallmatrix},
\etaM_{\sigmaVec}(x - y(\tau), \tau | y(s) - y(\tau), s) \IndVec_{x | y(\tau)}\right \rangle ds. \nonumber
\end{align}

\section{Explicit representation of the inverse transform in \eqref{invTr2}} \label{app5}

An explicit representation of the inverse transform in \eqref{invTr2} can be obtained if we know all its components $\bar u^{C}_i(\tau, \lambda), \ \bar u^{S}_i(\tau, \lambda)$ defined in \eqref{baruStruct}. Our main idea consists in finding them separately and then constructing the whole image $\bar u$ as a weighted sum. For doing so, let us denote the values of  $u(\tau, x)$ and their gradients at all internal interfaces between the layers as
\begin{alignat}{2}
\frac{\partial u(\tau, x)}{\partial x} \Bigg|_{x \to y_{i-1}(\tau) +0} &= \Psi_i^-(\tau),
&\qquad 	\frac{\partial u(\tau, x)}{\partial x} \Bigg|_{x \to y_{i}(\tau) -0} &= \Psi_i^+(\tau), 	\\
u(\tau, y_{i-1}(\tau) +0) &= \phi_i^-(\tau), &\qquad 	u(\tau, y_{i}(\tau) -0) &= \phi_i^+(\tau). \nonumber	
\end{alignat}

Applying the transform in \eqref{invTr2} to \eqref{PDE:ex:strip:u_eq}, integrating by parts and collecting terms gives rise to the initial ODE problems for functions $\bar u^{C}_i(\tau, \lambda), \ \bar u^{S}_i(\tau, \lambda)$
\begin{align*}
\frac{\bar u^C_i(\tau, \lambda)}{d\tau} &= -\lambda^2 \bar u^C_i(\tau, \lambda) +\int_{y_{i-1}(\tau)}^{y_i(\tau)} 	 g(\tau, \xi) \cos\left( \bl_i  \xi \right) d\xi + \sigma^2_i
\left[\cos\left( \bl_i y_i(\tau) \right)  \Psi_i^+(\tau) -   \cos\left( \bl_i y_{i-1} \right) \Psi_i^-(\tau) \right] 	\\
&+\sigma_i \lambda \left[\sin\left( \bl_i y_i(\tau)\right)  \phi_i^+(\tau) -   \sin\left( \bl_i  y_{i-1}(\tau)\right) \phi_i^-(\tau) \right] 	\\
&+ y'_i(\tau)\cos\left( \bl_i y_i(\tau)\right)  \phi_i^+(\tau) -  y'_{i-1}(\tau) \cos\left( \bl_i y_{i-1}(\tau)\right) \phi_i^-(\tau) \\
\bar u^C_i(0, \lambda) &= \int_{y_{i-1}(0)}^{y_i(0)} 	 f(\xi) \cos\left( \bl_i  x\right) d\xi
\end{align*} 	
\noindent and
\begin{align*}
\frac{\bar u^S_i(\tau, \lambda)}{d\tau} &= -\lambda^2 \bar u^S_i(\tau, \lambda) +\int_{y_{i-1}(\tau)}^{y_i(\tau)} 	 g(\tau, \xi) \sin\left( \bl_i  \xi\right) d\xi + \sigma^2_i
\left[\sin\left( \bl_i y_i(\tau)\right)  \Psi_i^+(\tau) -   \sin\left(\bl_i y_{i-1}(\tau) \right) \Psi_i^-(\tau) \right] 	\\
&-\sigma_i \lambda \left[\cos\left(\bl_i y_i(\tau) \right)  \phi_i^+(\tau) -   \cos\left( \bl_i y_{i-1}(\tau)\right) \phi_i^-(\tau) \right], 	\\
&+ y'_i(\tau) \sin\left(\bl_i y_i(\tau)\right)  \phi_i^+(\tau) -  y'_{i-1}(\tau) \sin \left( \bl_i y_{i-1}(\tau)\right) \phi_i^-(\tau) 	\\
\bar u^S_i(0, \lambda) &= \int_{y_{i-1}(0)}^{y_i(0)} f(\xi) \sin\left(\bl_i x \right) d\xi.
\end{align*}

Since each ODE is linear, these problems can be easily solved to get
\begin{align*}
u^C_i(\tau, \lambda) &= e^{-\lambda^2\tau} \int_{y_{i-1}(0)}^{y_i(0)}f(\xi) \cos\left(\bl_i \xi\right) d\xi
+\int_0^\tau e^{-\lambda^2(\tau - s)}\int_{y_{i-1}(s)}^{y_i(s)} 	 g(s, \xi) \cos\left( \bl_i  \xi\right) ds d\xi \\
&+ \sigma^2_i \int_0^\tau e^{-\lambda^2(\tau -s)} \left[\cos\left(\bl_i y_i(s)\right)  \Psi_i^+(s)
-   \cos\left( \bl_i y_{i-1}(s)\right) \Psi_i^-(s) \right] ds 	\\
&+\sigma_i \lambda \int_0^\tau e^{-\lambda^2(\tau -s)} \left[\sin\left(\bl_i y_i(s)\right)  \phi_i^+(s) 	
- \sin\left( \bl_i  y_{i-1}(s)\right) \phi_i^-(s) \right] ds \\
&+\int_0^\tau e^{-\lambda^2(\tau -s)} \left[y'_i(s)\cos\left(\bl_i y_i(s)\right)  \phi_i^+(s)
-  y'_{i-1}(s) \cos\left(\bl_i  y_{i-1}(s)\right) \phi_i^-(s) \right] ds 	\\
u^S_i(\tau, \lambda) &= e^{-\lambda^2\tau} \int_{y_{i-1}(0)}^{y_i(0)}f(\xi) \sin\left( \bl_i \xi\right) d\xi
+\int_0^\tau e^{-\lambda^2(\tau - s)}\int_{y_{i-1}(s)}^{y_i(s)} 	 g(s, \xi) \cos\left( \bl_i  \xi\right) ds d\xi \\
&+ \sigma^2_i \int_0^\tau e^{-\lambda^2(\tau -s)} \left[\sin\left(\bl_i y_i(s)\right)  \Psi_i^+(s)
-   \sin\left(\bl_i y_{i-1}(s)\right) \Psi_i^-(s) \right] ds \\
&-\sigma_i \lambda \int_0^\tau e^{-\lambda^2(\tau -s)} \left[\cos\left( \bl_i y_i(s)\right)  \phi_i^+(s) 	-   \cos\left( \bl_i y_{i-1}(s)\right) \phi_i^-(s) \right] ds 	\\
&+\int_0^\tau e^{-\lambda^2(\tau -s)} \left[y'_i(s)\sin\left( \bl_i  y_i(s)\right)  \phi_i^+(s) -  y'_{i-1}(s) \sin\left( \bl_i y_{i-1}(s)\right) \phi_i^-(s) \right] ds.
\end{align*}

To proceed, it is convenient to define the auxiliary function $\thetaVec_{\by(\tau), \lambda}(x)$
\begin{align} \label{PDE:strip_ex_theta_def}
\thetaVec_{\by(\tau), \lambda}(x) &= 	\left(0,\thetaVec_{\by(\tau), \lambda, 1}(x), \thetaVec_{\by(\tau), \lambda, 2}(x),\dots,\thetaVec_{\by(\tau), \lambda, N}(x), 0 \right)^\top, 	\\ 	
\thetaVec_{\by(\tau), \lambda, k}(x) &= -\sigma_i \lambda \left(C_k(\tau, \lambda) \sin\left( \bl_i x\right) + D_k(\tau, \lambda) \cos\left(\bl_i x\right) \right). \nonumber
\end{align}
With this new notation \eqref{baruStruct} reads
\begin{align} \label{baru2}
\bar u(\tau, \lambda) &= e^{-\lambda^2\tau}\int_{y_0(0)}^{y_N(0)} \langle \IndVec_{\xi |\by(0) }, \ThetaVec_{\by(\tau), \lambda}(\xi)\rangle  f(\xi) d\xi+
\int_0^\tau \int_{y_0(s)}^{y_N(s)}e^{-\lambda^2(\tau - s)} \langle \IndVec_{\xi |\by(s) }, \ThetaVec_{\by(\tau), \lambda}(\xi)\rangle  g(s, \xi)  ds d\xi \nonumber \\
&+\int_0^\tau e^{-\lambda^2(\tau -s)} \sigma^2_N \Theta_{\by(\tau), \lambda, N}(y_N(s))  \Psi_N^+(s)ds
- \int_0^\tau e^{-\lambda^2(\tau -s)} \sigma_1^2\Theta_{\by(\tau), \lambda, 1}(y_{0}(s))\Psi_1^-(s)  ds 	 \\
&+\sum_{i = 1}^{N-1} \int_0^\tau e^{-\lambda^2(\tau -s)} 	\left[\sigma^2_i \Theta_{\by(\tau), \lambda, i}(y_i(s))  \Psi_i^+(s) 	-   \sigma^2_{i+1}\Theta_{\by(\tau), \lambda, i+1}(y_{i}(s))\Psi_{i+1}^-(s) \right] ds \nonumber \\
&-\int_0^\tau e^{-\lambda^2(\tau -s)} 	\theta_{\by(\tau), \lambda, N}(y_N(s))  \phi_N^+(s)ds
+ \int_0^\tau e^{-\lambda^2(\tau -s)}\theta_{\by(\tau), \lambda, 1}(y_{0}(s))\phi_1^-(s)  ds 	\nonumber \\
&-\sum_{i = 1}^{N-1} \int_0^\tau e^{-\lambda^2(\tau -s)} 	\left[\theta_{\by(\tau), \lambda, i}(y_i(s))  \phi_i^+(s)
- \theta_{\by(\tau), \lambda, i+1}(y_{i}(s))\phi_{i+1}^-(s) \right] ds \nonumber \\
&+\int_0^\tau e^{-\lambda^2(\tau -s)} 	 y_N'(s) \Theta_{\by(\tau), \lambda, N}(y_N(s))  \psi_N^+(s)ds
- \int_0^\tau e^{-\lambda^2(\tau -s)} y_0'(s)\Theta_{\by(\tau), \lambda, 1}(y_{0}(s))\psi_1^-(s)  ds 	\nonumber \\
&+ \sum_{i = 1}^{N-1} \int_0^\tau e^{-\lambda^2(\tau -s)} y_i'(s) 	\left[\Theta_{\by(\tau), \lambda, i}(y_i(s))  \psi_i^+(s) - \Theta_{\by(\tau), \lambda, i+1}(y_{i}(s))\psi_{i+1}^-(s) \right] ds. \nonumber
\end{align}

This representation can be further simplified. Taking into account the following identities
\begin{equation*}
\phi_0^-(\tau) = \phi_N^+(\tau ) =0, \quad \sigma_i^2\Psi_i^+(\tau) = \sigma_{i+1}^2\Psi_{i+1}^-(\tau),
\quad \phi_i^+(\tau) = \phi_{i+1}^-(\tau), \qquad i = 1,\dots,N-1,
\end{equation*}
\noindent and making change of variables
\begin{equation*}
\sigma_1^2\Psi_1^-(\tau) = \Phi_0(\tau), \quad \sigma_N^2 \Psi_N^+(\tau) =\Phi_{N}(\tau), \quad \phi_i^+(\tau) =  \phi_{i+1}^-(\tau) =\phi_{i}(\tau)
\end{equation*}
\noindent allows one to re-write \eqref{baru2} in the form
\begin{align} \label{baru3}
\bar u(\tau, \lambda) &=e^{-\lambda^2\tau}\int_{y_0(0)}^{y_N(0)} \langle \IndVec_{\xi |\by(0) }, \ThetaVec_{\by(\tau), \lambda}(\xi)\rangle  f(\xi) d\xi + \int_0^\tau \int_{y_0(s)}^{y_N(s)}e^{-\lambda^2(\tau - s)} \langle \IndVec_{\xi |\by(s) }, \ThetaVec_{\by(\tau), \lambda}(\xi)\rangle  g(s, \xi)  ds d\xi \nonumber \\
&+\int_0^\tau e^{-\lambda^2(\tau -s)} \Phi_{N}(s) 	\Theta_{\by(\tau), \lambda, N}(y_N(s))  ds
- \int_0^\tau e^{-\lambda^2(\tau -s)} \Phi_0(s) \Theta_{\by(\tau), \lambda, 1}(y_{0}(s)) ds  \\
&+\sum_{i = 1}^{N-1} \int_0^\tau e^{-\lambda^2(\tau -s)} \Phi_i(s) \left[\Theta_{\by(\tau), \lambda, i}(y_i(s)) - \Theta_{\by(\tau), \lambda, i+1}(y_{i}(s))\right] ds 	\nonumber\\
&-\sum_{i = 1}^{N-1} \int_0^\tau e^{-\lambda^2(\tau -s)} \phi_i(s) \left[\theta_{\by(\tau), \lambda, i}(y_i(s)) -   \theta_{\by(\tau), \lambda, i+1}(y_{i}(s)) 	\right] ds 	\nonumber\\
&+\sum_{i = 1}^{N-1} \int_0^\tau e^{-\lambda^2(\tau -s)} y'_i(s)\phi_i(s) 	\left[\Theta_{\by(\tau), \lambda, i}(y_i(s)) - \Theta_{\by(\tau), \lambda, i+1}(y_{i}(s))\right] ds. \nonumber
\end{align}

Since by the definition of $\ThetaVec$ and $\phi_i(\tau)$ we have
\begin{equation*}
\ThetaVec_{\by(\tau), \lambda, 0}(x) = 0, \quad \ThetaVec_{\by(\tau), \lambda, N+1}(x) = 0,
\quad  \phi_0(\tau) = 0, \quad \phi_N(\tau) = 0,
\end{equation*}
\noindent the \eqref{baru3} can be transformed to
\begin{align} \label{baru4}
\bar u(\tau, \lambda) &=e^{-\lambda^2\tau}\int_{y_0(0)}^{y_N(0)} \langle \Ind_{\xi | \by(0)}, \ThetaVec_{\by(\tau), \lambda}(\xi)\rangle  f(\xi) d\xi+
\int_0^\tau \int_{y_0(s)}^{y_N(s)} e^{-\lambda^2(\tau - s)}\langle \IndVec_{\xi | \by(s)}, \ThetaVec_{\by(\tau), \lambda}(\xi)\rangle   g(s, \xi) ds d\xi \nonumber
\\
&+\sum_{i = 0}^{N} \int_0^\tau e^{-\lambda^2(\tau -s)} \Phi_i(s) 	\left[\Theta_{\by(\tau), \lambda, i}(y_i(s)) - \Theta_{\by(\tau), \lambda, i+1}(y_{i}(s))\right] ds \\
&-\sum_{i = 0}^{N} \int_0^\tau e^{-\lambda^2(\tau -s)} \phi_i(s) \left[\theta_{\by(\tau), \lambda, i}(y_i(s)) -   \theta_{\by(\tau), \lambda, i+1}(y_{i}(s)) 	\right] ds \nonumber	\\
&+\sum_{i = 0}^{N} \int_0^\tau e^{-\lambda^2(\tau -s)} y_i'(s) \phi_i(s) \left[\Theta_{\by(\tau), \lambda, i}(y_i(s)) - \Theta_{\by(\tau), \lambda, i+1}(y_{i}(s))\right] ds. \nonumber
\end{align}

Finally, we define $N+1$-dimensional vector functions
\begin{align} \label{omegaVec}
\PhiVec(s) &= \left[\Phi_0(s), \Phi_1(s),\dots,\Phi_N(s)\right]^\top, \qquad
\phiVec(s) = \left[0 = \phi_0(s), \phi_1(s), \dots,\phi_N(s) =0\right]^\top, \\
\OmegaVec_{\by(\tau), \lambda}(s) &=  \Bigg[0 - \Theta_{\by(\tau), \lambda, 1}(y_{0}(s)),\dots, \ThetaVec_{\by(\tau), \lambda, i}(y_i(s)) - \ThetaVec_{\by(\tau),\lambda, i+1}(y_{i}(s)),\dots,
\ThetaVec_{\by(\tau), \lambda, N}(y_N(s)) - 0 \Bigg]^\top, \nonumber	\\
\omegaVec_{\by(\tau), \lambda}(s) &=  \Bigg[0 - \thetaVec_{\by(\tau), \lambda, 1}(y_{0}(s)),\dots, 	
\thetaVec_{\by(\tau), \lambda, i}(y_i(s)) - \thetaVec_{\by(\tau),\lambda, i+1}(y_{i}(s)),\dots, 	\thetaVec_{\by(\tau), \lambda, N}(y_N(s)) - 0\Bigg]^\top, \nonumber
\end{align}
\noindent and $N+1 \times N+1$ diagonal matrix $\YM(\tau)$
\begin{equation} \label{YM}
\YM(\tau) = \operatorname{diag} \left[y_0(\tau), y_1(\tau), y_2(\tau), \dots y_N(\tau) \right]^\top, \qquad 	
\YM'(\tau) = \operatorname{diag} \left[ y'_0(\tau), y'_1(\tau), y'_2(\tau), \dots y'_N(\tau) \right]^\top.
\end{equation}
\noindent and arrive at a very compact representation of $\bar u(\tau, \lambda)$
\begin{align} \label{PDE:ex:strip:baru_final1}
\bar u(\tau, \lambda) &=e^{-\lambda^2\tau}\int_{-\infty}^{+\infty} \langle \IndVec_{\xi | \by(0)}, \ThetaVec_{\by(\tau), \lambda}(\xi)\rangle   f(\xi) d\xi +
\int_0^\tau \int_{-\infty}^{+\infty} e^{-\lambda^2(\tau - s)}\langle \IndVec_{\xi | \by(s)}, \ThetaVec_{\by(\tau), \lambda}(\xi)\rangle   g(s, \xi) ds d\xi \nonumber
\\
&+ \int_0^\tau e^{-\lambda^2(\tau -s)}\left[ \langle  \PhiVec(s), \OmegaVec_{\by(\tau), \lambda} (s) \rangle
+\langle  \phiVec(s), \YM'(s)\OmegaVec_{\by(\tau), \lambda} (s)-\omegaVec_{\by(\tau), \lambda} (s) \rangle
\right] ds.
\end{align}

\section{The solution of the problem \eqref{freezing_pde}} \label{app6}

The problem in \eqref{freezing_pde} has inhomogeneous boundary conditions, \eqref{bcExample}. Therefore, first we make a change of variables to transform \eqref{freezing_pde} to the problem with homogeneous boundary conditions by setting
\begin{equation}
T(\tau, x) = \calT(\tau, x) +\eta(\tau, x), \quad \eta(\tau, x) = \theta(y(\tau)-x) F_-(\tau,x) + \theta(x - y(\tau)) F_+(\tau,x),
\end{equation}
\noindent where $\theta(x)$ is the Heaviside theta function with $\theta(0) = 1/2$.
Functions $F_\pm(\tau,x)$ should be such to provide the correct values of the temperature at the lower and upper boundaries, and also a jump of the gradient and continuity of the temperature at the interface boundary. Since this gives four conditions, we proceed by setting
\begin{equation}
F_-(\tau,x) = A_-(\tau) + B_-(\tau) x, \qquad F_+(\tau,x) = A_+(\tau) + B_+(\tau) x.
\end{equation}
Four yet unknown functions $A_-(\tau)$, $B_-(\tau)$, $A_+(\tau)$, $B_+(\tau)$  solve the linear system
\begin{equation} \label{coeffSys}
\begin{bmatrix}
	1 & y_- & 0 & 0 \\
	0 & 0   & 1 & y_+ \\
	1 & y(\tau) & -1 & -y(\tau) \\
	0 & \kappa_I & 0 & -\kappa_W
\end{bmatrix}
\begin{bmatrix}
	A_-(\tau) \\ B_-(\tau) \\ A_+(\tau) \\ B_+(\tau)
\end{bmatrix} =
\begin{bmatrix}
	T_s \\ T_l \\ 0 \\ \rho_I L y'(\tau)
\end{bmatrix}.
\end{equation}

With these definitions the problem in \eqref{freezing_pde} transforms to
\begin{align} \label{prCalT}
\fp{\calT}{\tau} &= \kappa_W \sop{\calT}{x} + g_+(\tau, x), \qquad y(\tau) < x < y_+, \\
\fp{\calT}{\tau} &= \kappa_I \sop{\calT}{x} +  g_-(\tau, x), \qquad y_- < x < y(\tau), \nonumber \\
\calT(\tau,y_-) &= \calT(\tau, y_+) = 0, \qquad \calT(0,x) = T_l - \eta(0, x), \nonumber \\
\calT(\tau,y(\tau)-0) &= \calT(\tau,y(\tau)+0) = T_m - \eta(\tau, y(\tau)) \equiv \phi(\tau), \nonumber \\
\kappa_I \fp{\calT}{x}\Big|_{x = y(\tau) -0} &= \kappa_W \fp{\calT}{x}\Big|_{x = y(\tau) + 0} \equiv \Phi(\tau). \nonumber
\end{align}
Here
\begin{equation*}
g_-(\tau, x) = -A_-'(\tau) - B_-'(\tau) x, \quad 	g_+(\tau, x) = -A_+'(\tau) - B_+'(\tau) x.
\end{equation*}	
Thus, we obtained a similar problem for $\calT(\tau,x)$, but now with homogeneous boundary conditions at both ends, while the heat equations acquire new source terms $g_\pm(x)$.

Solving \eqref{coeffSys} yields
\begin{align} \label{ABpm_def}
A_-(\tau) &=  \frac{-y_-T_l \kappa_W + T_s \left[ y_+ \kappa_I + y(\tau) (\kappa_W - \kappa_I) \right] + y_- \left[ y(\tau) - y_+ \right] \rho_I L y'(\tau) }{D(\tau)}, \\
B_-(\tau) &= \frac{\left[ T_l - T_s \right] \kappa_W + \left[ y_+ - y(\tau) \right]\rho_I L y'(\tau) }{D(\tau)}, \nonumber \\
A_+(\tau) &= \frac{-T_l \left[ y(\tau)(\kappa_I - \kappa_W) + y_- \kappa_W \right] + y_+ \left[ T_s \kappa_I + (y(\tau) - y_-) \rho_I L y'(\tau)\right]}{D(\tau)}, \nonumber \\
B_+(\tau) &= \frac{\left[T_l - T_s \right] \kappa_I + \left[ y_- - y(\tau) \right] \rho_I L y'(\tau)}{D(\tau)}, \nonumber
\end{align}
\noindent where
\begin{equation}
D (\tau) = \det	\begin{bmatrix}
	1 & y_- & 0 & 0 \\
	0 & 0   & 1 & y_+ \\
	1 & y(\tau) & -1 & -y(\tau) \\
	0 & \kappa_I & 0 & -\kappa_W
\end{bmatrix}	= \kappa_I (y_+ - y(\tau)) + \kappa_W(y(\tau) - y_-) > 0.
\end{equation}

Note, that at $\tau=0$ we have $y(0) = y_-$, $T_s = T_m = T_l$ and $y'(0) = 0$.  Then, \eqref{ABpm_def} implies $B_-(0) = B_+(0) = 0, \ A_-(0) = A_+(0) = T_l$, i.e. \eqref{ABpm_def} replicates the correct boundary condition at $x = y_-$.

The problem in \eqref{prCalT} can be solved by directly applying the OIT in \eqref{invTr2}, which immediately gives rise to an explicit representation of $\calT$
\begin{align} \label{calTFin}
\calT(\tau, x) &= \Bigg<\IndVec_{x | \by(\tau)}, \,\,	\sum_{n = 1} ^{\infty} \frac{\ThetaVec_{\by(\tau), \lambda_n(\tau)} (x)}{N_n(\tau)} \Bigg[e^{-\tau \lambda^2_n(\tau)}\int_{-\infty}^{+\infty}
\langle \Ind_{\xi | \by(0)}, \ThetaVec_{\by(\tau), \lambda_n(\tau)}(\xi)\rangle \left[ T_l + \eta(0, \xi) \right] d\xi 	\\
&+\int_0^\tau \int_{-\infty}^{+\infty} e^{-(\tau-s)\lambda_n^2(\tau)} \langle \IndVec_{\xi | \by(s)}, \ThetaVec_{\by(\tau), \lambda_n(\tau)}(\xi) \rangle \left[ \Ind_{\xi< y(s)}g_-(s,\xi) + \Ind_{\xi > y(s)}g_+(s,\xi) \right] ds d\xi \nonumber \\
&+ \int_0^\tau e^{-(\tau -s)\lambda_n^2(\tau)}\left[ \langle  \PhiVec(s), \OmegaVec_{\by(\tau), \lambda_n(\tau)} (s) \rangle 	+ \langle  \phiVec(s), \YM'(s)\OmegaVec_{\by(\tau), \lambda_n(\tau)} (s)-\omegaVec_{\by(\tau), \lambda_n(\tau)} (s) \rangle  \right] ds \nonumber
\Bigg] \Bigg>.
\end{align}
\noindent with
\begin{align*}
	%
\OmegaVec_{\by(\tau), \lambda}(s) &=  \left[0 - \Theta_{\by(\tau), \lambda, 1}(y_-), \ThetaVec_{\by(\tau), \lambda, 1}(y(s)) - \ThetaVec_{\by(\tau),\lambda, 2}(y(s)),	\ThetaVec_{\by(\tau), \lambda, 2}(y_+) - 0 \right]^\top, \nonumber	\\
\omegaVec_{\by(\tau), \lambda}(s) &=  \left[0 - \thetaVec_{\by(\tau), \lambda, 1}(y_-), 	 \thetaVec_{\by(\tau), \lambda, 1}(y(s)) - \thetaVec_{\by(\tau),\lambda, 2}(y(s)),\dots, 	\thetaVec_{\by(\tau), \lambda, 2}(y_+) - 0\right]^\top, \nonumber \\
\thetaVec_{\by(\tau), \lambda_n(\tau), 1}(x) &= \lambda_n(\tau)\sqrt{\kappa_I}\cos\left(\lambda_n(\tau)(x - y_-)/\sqrt{\kappa_I}\right), \nonumber \\
\thetaVec_{\by(\tau), \lambda_n(\tau), 2}(x) &= -\lambda_n(\tau)\sqrt{\kappa_W} K_n(\tau) \cos\left( \lambda(\tau) (y_+ - x)/\sqrt{\kappa_W}\right). \nonumber
\end{align*}

Since $\Theta_{\by(\tau), \lambda, 1}(y_-) =  \Theta_{\by(\tau), \lambda, 2}(y_+) =0$, each scalar product in the final formula contains just one summand. Therefore, in the RHS of \eqref{calTFin} there are only yet two unknown functions $\Phi(s)$ and $\phi(s)$, and we can set
\begin{align*}
\PhiVec(s) &= \left[0, \Phi(s), 0\right]^\top, \qquad
\phiVec(s) = \left[0, \phi(s), 0\right]^\top,
\end{align*}
Also, in our case
\begin{equation*}	
\YM'(\tau) = \operatorname{diag} \left[0, y'(\tau), 0\right]^\top.
\end{equation*}
Together they allow simplification of \eqref{calTFin} to the form
\begin{align} \label{calTfinal1}
\calT(\tau, x) &= \Bigg<\IndVec_{x | \by(\tau)}, \,\,	\sum_{n = 1} ^{\infty} \frac{\ThetaVec_{\by(\tau), \lambda_n(\tau)} (x)}{N_n(\tau)} \Bigg[e^{-\tau \lambda^2_n(\tau)}\int_{-\infty}^{+\infty} \langle \Ind_{\xi | \by(0)}, \ThetaVec_{\by(\tau), \lambda_n(\tau)}(\xi)\rangle \left[ T_l - \eta(0, \xi) \right] d\xi 	\\
&+\int_0^\tau \int_{-\infty}^{+\infty} e^{-(\tau-s)\lambda_n^2(\tau)} \langle \IndVec_{\xi | \by(s)}, \ThetaVec_{\by(\tau), \lambda_n(\tau)}(\xi) \rangle \left[ \Ind_{\xi< y(s)}g_-(s,\xi) + \Ind_{\xi > y(s)}g_+(s,\xi) \right] ds d\xi \nonumber 	\\
&+ \int_0^\tau e^{-(\tau -s)\lambda_n^2(\tau)}\left[ \Phi(s) \Omega_{\by(\tau), \lambda_n(\tau)} (s) + \phiVec(s)
\left( y'(s )\Omega_{\by(\tau), \lambda_n(\tau)} (s)-\omega_{\by(\tau), \lambda_n(\tau)} (s)\right) \right] ds 	\Bigg] \Bigg>, \nonumber
\end{align}
\noindent or
\begin{align} \label{calTfinal2}
\calT(\tau, x) &= \sum_{n = 1} ^{\infty}\frac{F_n(\tau) + G_n(\tau) + H_n(\tau)}{N_n(\tau)} \left[\Ind_{x < y(\tau)}\ThetaVec_{\by(\tau), \lambda_n(\tau), 1}(x) + \Ind_{x > y(\tau)}\ThetaVec_{\by(\tau), \lambda_n(\tau), 2}(x) \right].
\end{align}
\noindent where
\begin{align} \label{FGHN_def}
F_n(\tau) &= e^{-\tau \lambda^2_n(\tau)} \Bigg(\int_{y_-}^{y(0)} \ThetaVec_{\by(\tau), \lambda_n(\tau), 1}(\xi) \left[ T_l - A_-(0) - B_-(0)\xi\right]d\xi 	\\
&+ \int_{y(0)}^{y_+} \ThetaVec_{\by(\tau), \lambda_n(\tau), 2}(\xi) \left[ T_l - A_+(0) - B_+(0)\xi \right] d\xi \Bigg), \nonumber \\
G_n(\tau) &= -\int_0^\tau e^{-(\tau-s)\lambda^2_n(\tau)} \Bigg(\int_{y_-}^{y(s)} 	\ThetaVec_{\by(\tau), \lambda_n(\tau), 1}(\xi) \left[A_-'(s) + B'_-(s)\xi\right]d\xi \nonumber \\
&+\int^{y_+}_{y(s)} \ThetaVec_{\by(\tau), \lambda_n(\tau), 2}(\xi) \left[A_+'(s) + B'_+(s)\xi \right]d\xi \Bigg) ds, \nonumber \\
H_n(\tau) &= \int_0^\tau e^{-(\tau -s)\lambda_n^2(\tau)}\left[ \Phi(s) \Omega_{\by(\tau), \lambda_n(\tau)} (s) + \phiVec(s) \left( y'(s )\Omega_{\by(\tau), \lambda_n(\tau)} (s)-\omega_{\by(\tau), \lambda_n(\tau)} (s)\right) \right] ds, \nonumber \\
N_n(\tau) &= \int_{y_-}^{y(\tau)} \ThetaVec_{\by(\tau), \lambda_n(\tau), 1}^2(\xi) d\xi + \int^{y_+}_{y(\tau)} \ThetaVec_{\by(\tau), \lambda_n(\tau), 2}^2(\xi) d\xi, \nonumber
\end{align}
\noindent and
\vspace{-\baselineskip}
\begin{align}
\Omega_{\by(\tau), \lambda_n(\tau)}(s) &= \ThetaVec_{\by(\tau), \lambda_n(\tau), 1}(y(s)) - \ThetaVec_{\by(\tau),\lambda_n(\tau), 2}(y(s)),	\\
\omega_{\by(\tau), \lambda_n(\tau)}(s) &= \thetaVec_{\by(\tau), \lambda_n(\tau, 1}(y(s)) - \thetaVec_{\by(\tau),\lambda_n(\tau), 2}(y(s)). \nonumber
\end{align}

As now follows from \eqref{calTfinal2} and the boundary conditions, given $y(\tau)$ the unknown functions $\Phi(\tau)$ and $\phi(\tau)$ solve a system of linear Volterra integral equations of the second kind
\begin{align} \label{Volt2}
\phi(\tau) &= \sum_{n = 1} ^{\infty}\frac{F_n(\tau) + G_n(\tau) + H_n(\tau)}{N_n(\tau)} \left[
s^+(y(\tau))\ThetaVec_{\by(\tau), \lambda_n(\tau), 1}(y(\tau)) + s^-(y(\tau))\ThetaVec_{\by(\tau), \lambda_n(\tau), 2}(y(\tau)) \right], \\
\Phi(\tau) &= \frac{1}{2}\sum_{n = 1} ^{\infty}\frac{F_n(\tau) + G_n(\tau) + H_n(\tau)}{N_n(\tau)} \left[
s^+(y(\tau))\thetaVec_{\by(\tau), \lambda_n(\tau), 1}(y(\tau)) + s^-(y(\tau))\thetaVec_{\by(\tau), \lambda_n(\tau), 2}(y(\tau)) \right],\nonumber
\end{align}
\noindent where $s^\pm(y(\tau)) = 1 + \Ind_{y_\pm = y(\tau)}$.

\paragraph{Further simplifications.} The \eqref{calTfinal2} can be further simplified and rewritten in terms of the original function $T(\tau, x)$. Integrating  by parts the formula for $G_n(\tau)$ in \eqref{FGHN_def} yields
\begin{align*}
G_n(\tau) &= e^{-\tau \lambda_n^2(\tau)} \left[ \int_{y_-}^{y(0)} \ThetaVec_{\by(\tau), \lambda_n(\tau), 1}(\xi) \left[A_-(0) + B_-(0)\xi \right] d\xi + \int^{y_+}_{y(0)} \ThetaVec_{\by(\tau), \lambda_n(\tau), 2}(\xi) \left[A_+(0) + B_+(0)\xi \right] d\xi \right] \\
&- \int_{y_-}^{y(\tau)} \ThetaVec_{\by(\tau), \lambda_n(\tau), 1}(\xi) \left[A_-(\tau) + B_-(\tau)\xi\right]d\xi
- \int^{y_+}_{y(\tau)} \ThetaVec_{\by(\tau), \lambda_n(\tau), 2}(\xi) \left[A_+(\tau) + B_+(\tau)\xi \right]d\xi \\
&+ \int_0^\tau  \frac{d}{ds} \left\{e^{-(\tau-s)\lambda^2_n(\tau)} \int_{y_-}^{y(s)} \ThetaVec_{\by(\tau), \lambda_n(\tau), 1}(\xi) \right\} \left[A_-(s) + B_-(s)\xi \right] d\xi ds \\
&+ \int_0^\tau  \frac{d}{ds} \left\{e^{-(\tau-s)\lambda^2_n(\tau)} \int^{y_+}_{y(s)} \ThetaVec_{\by(\tau), \lambda_n(\tau), 2}(\xi) \right\}  \left[A_+(s) + B_+(s)\xi \right]d\xi ds.
\end{align*}

Recall that $B_-(0) = B_+(0) = 0, \ A_-(0) = A_+(0) = T_l$. Using the identities
\begin{alignat*}{2}
T_l\int_{y_-}^{y(0)} \ThetaVec_{\by(\tau), \lambda_n(\tau), 1}(\xi) d\xi &= T_l \frac{\sqrt{\kappa_I}}{\lambda_n(\tau)}
\left\{1 - \cos[x^-_n(\tau,y(0))] \right\}, &\quad x^-_n(\tau,x) &= \lambda_n(\tau) (x - y_-) /\sqrt{\kappa_I}, \\
T_l\int^{y_+}_{y(0)} \ThetaVec_{\by(\tau), \lambda_n(\tau), 2}(\xi) d\xi &= T_l K_n(\tau)  \frac{\sqrt{\kappa_W}}{\lambda_n(\tau)}\left\{1 - \cos[x^+_n(\tau,y(0))] \right\}, &\quad x^+_n(\tau,x) &= \lambda_n(\tau) (y_+ - x) /\sqrt{\kappa_W},
\end{alignat*}
\vspace{-0.5\baselineskip}
\begin{align*}
\int_{y_-}^{y(s)} &\Big[A_-(s) + B_-(s) \xi \Big] \ThetaVec_{\by(\tau), \lambda_n(\tau), 1}(\xi) d\xi =
\frac{\sqrt{\kappa_I} T_s} {\lambda_n(\tau)} +  \frac{\kappa_I B_-(s)}{\lambda_n^2(\tau)} \ThetaVec_{\by(\tau), \lambda_n(\tau), 1}(y(s)) \\
&- \frac{A_-(s) + B_-(s)y(s)}{\lambda_n(\tau)} \thetaVec_{\by(\tau), \lambda_n(\tau), 1}(y(s)), \\
\int^{y_+}_{y(s)} &\Big[A_+(s) + B_+(s) \xi \Big]\ThetaVec_{\by(\tau), \lambda_n(\tau), 2}(\xi) d\xi
= K_n(\tau) \sqrt{\kappa_W} \frac{T_l}{\lambda_n(\tau)} - \frac{B_+(s)\kappa_W}{\lambda_n^2(\tau)} \ThetaVec_{\by(\tau), \lambda_n(\tau), 2}(y(s)) \\
&+ \frac{A_+(s) + B_+(s)y(s)}{\lambda^2_n(\tau)} \thetaVec_{\by(\tau), \lambda_n(\tau), 2}(y(s)), \\
\int_0^\tau  & \frac{d}{ds} \left\{e^{-(\tau-s)\lambda^2_n(\tau)} \int_{y_-}^{y(s)} \ThetaVec_{\by(\tau), \lambda_n(\tau), 1}(\xi) \right\} \left[A_-(s) + B_-(s)\xi \right] d\xi ds = \int_0^\tau  e^{-(\tau-s)\lambda^2_n(\tau)}
\Big\{ \\
& \lambda^2_n(\tau) \int_{y_-}^{y(s)} \ThetaVec_{\by(\tau), \lambda_n(\tau), 1}(\xi) \left[A_-(s) + B_-(s)\xi \right] d\xi + \ThetaVec_{\by(\tau), \lambda_n(\tau), 1}(\xi) \left[A_-(s) + B_-(s)\xi \right]  y'(s)\Big\} ds,
\end{align*}
\noindent taking into account that $A_-(s) + B_-(s) y(s) = A_+(s) + B_+(s) y(s), \ \omega_{\by(\tau), \lambda_n(\tau)}(\tau) = 0$ and collecting all terms, we arrive at the following representations
\begin{align} \label{Volt1}
T(\tau,x) &= \eta(\tau,x) + \sum_{n = 1}^{\infty}\frac{S_n(\tau) + R_n(\tau)}{N_n(\tau)} {\cal M}(\tau,x), \\
{\cal M}(\tau,x) &=
\begin{cases}
\Ind_{x < y(\tau)}\ThetaVec_{\by(\tau), \lambda_n(\tau), 1}(x) + \Ind_{x > y(\tau)} \ThetaVec_{\by(\tau), \lambda_n(\tau), 2}(x), & x \ne y(\tau), \\
\frac{1}{2}\left[s^+(y(\tau))\ThetaVec_{\by(\tau), \lambda_n(\tau), 1}(y(\tau)) + s^-(y(\tau)) \ThetaVec_{\by(\tau), \lambda_n(\tau), 2}(y(\tau)) \right], & x = y(\tau),
\end{cases}
\nonumber \\
\Phi(\tau) &= \frac{1}{2} \sum_{n = 1} ^{\infty}\frac{S_n(\tau) + R_n(\tau)}{N_n(\tau)} \left[s^+(y(\tau))\thetaVec_{\by(\tau), \lambda_n(\tau), 1}(y(\tau)) + s^-(y(\tau))\thetaVec_{\by(\tau), \lambda_n(\tau), 2}(y(\tau)) \right]\Big\}.\nonumber
\end{align}
Here $s^\pm(y(\tau)) = 1 + \Ind_{y_\pm = y(\tau)}$ and
\begin{align} \label{RSN_def}
S_n(\tau) &= \frac{T_l e^{-\tau \lambda^2_n(\tau)}}{\lambda_n(\tau)}  \left[\sqrt{\kappa_I} \left( 1 - \cos[x^-_n(\tau,y(0))]\right) + K_n(\tau)\sqrt{\kappa_W} \left(1 - \cos[x^+_n(\tau,y(0))] \right)
\right]\nonumber \\
&- \frac{1}{\lambda_n(\tau)} \Big[ \sqrt{\kappa_I} T_s + \sqrt{\kappa_W} K_n(\tau) T_l \Big] \\
&- \frac{1}{\lambda_n^2(\tau)} \left[B_-(\tau) \kappa_I \ThetaVec_{\by(\tau), \lambda_n(\tau), 1}(y(\tau)) - B_+(\tau) \kappa_W \ThetaVec_{\by(\tau), \lambda_n(\tau), 2}(y(\tau))  \right], \nonumber \\
R_n(\tau) &= \int_0^\tau e^{-(\tau -s)\lambda_n^2(\tau)}\Bigg\{ \Phi(s) \Omega_{\by(\tau), \lambda_n(\tau)} (s) + [B_-(s)\kappa_I  +  y'(s) T_m] \ThetaVec_{\by(\tau), \lambda_n(\tau), 1}(y(s)) \nonumber \\
&- [B_+(s)\kappa_W +  y'(s) T_m] \ThetaVec_{\by(\tau), \lambda_n(\tau), 2}(y(s))  -  T_m \omega_{\by(\tau), \lambda_n(\tau)}(s) \Bigg\} ds \nonumber \\
&+ \frac{1 - e^{- \tau \lambda_n^2(\tau)}}{\lambda_n(\tau)}\left[\sqrt{\kappa_I} T_s + \sqrt{\kappa_W} K_n(\tau) T_l \right]. \nonumber
\end{align}

The terms containing $y'(\tau)$ can be integrated by parts to yield
\begin{align*}
\int_0^\tau & e^{-(\tau -s)\lambda_n^2(\tau)} y'(s) \ThetaVec_{\by(\tau), \lambda_n(\tau), 1}(y(s)) ds = \int_0^\tau e^{-(\tau-s)\lambda_n^2(\tau)} \sin\left(x^-(\tau,y(s))\right) dy(s)  \\
&=  \frac{\sqrt{\kappa_I}}{\lambda_n(\tau)} e^{-\tau \lambda_n^2(\tau)} \cos\left(x^-(\tau,y(0))\right) - \frac{\sqrt{\kappa_I}}{\lambda_n(\tau)} \cos\left(x^-(\tau, y(\tau))\right) \\
&+ \sqrt{\kappa_I} \lambda_n(\tau) \int_0^\tau e^{-(\tau-s)\lambda_n^2(\tau)} \cos\left(x^-(\tau,y(s)) \right) ds, \\
 \int_0^\tau & e^{-(\tau -s)\lambda_n^2(\tau)} y'(s) \ThetaVec_{\by(\tau), \lambda_n(\tau), 2}(y(s)) ds = K_n(\tau) \int_0^\tau e^{-(\tau-s)\lambda_n^2(\tau)}   \sin\left(x^+(\tau, y(s))\right) dy(s)  \\
&= - K_n(\tau) \frac{\sqrt{\kappa_W}}{\lambda_n(\tau)} e^{-\tau \lambda_n^2(\tau)} \cos\left(x^+(\tau,y(0))\right)
+ K_n(\tau) \frac{\sqrt{\kappa_W}}{\lambda_n(\tau)} \cos\left(x^+(\tau,y(\tau))\right)  \\
&+ K_n(\tau) \sqrt{\kappa_W} \lambda_n(\tau) \int_0^\tau e^{-(\tau-s)\lambda_n^2(\tau)}    \cos\left(x^+(\tau,y(s))\right) ds.
\end{align*}

Taking into account that from \eqref{eigenLam}
\begin{equation} \label{cosIdent}
\sqrt{\kappa_I}\cos[x^-_n(\tau,y(\tau))] + \sqrt{\kappa_W} K_n(\tau) \cos[x^+_n(\tau,y(\tau))] = 0,
\end{equation}
\noindent and slightly regrouping  the terms between $S_n$ and $R_n$ we get another representation of these functions which now doesn't contain $y'(\tau)$
\begin{align} \label{RSN_def}
S_n(\tau) &= \frac{e^{-\tau \lambda^2_n(\tau)}}{\lambda_n(\tau)}  \sqrt{\kappa_I} (T_l - T_s) - \frac{1}{\lambda_n^2(\tau)} \left[B_-(\tau) \kappa_I \ThetaVec_{\by(\tau), \lambda_n(\tau), 1}(y(\tau)) - B_+(\tau) \kappa_W \ThetaVec_{\by(\tau), \lambda_n(\tau), 2}(y(\tau))  \right] \nonumber \\
&+ (T_m-T_l)\frac{e^{-\tau \lambda^2_n(\tau)}}{\lambda_n(\tau)} \left[ \sqrt{\kappa_I} \cos\left(x^-(\tau,y(0))\right) +
\sqrt{\kappa_W} K_n(\tau)\cos\left(x^+(\tau,y(0))\right) \right] \nonumber \\
R_n(\tau) &= \int_0^\tau e^{-(\tau -s)\lambda_n^2(\tau)}\Big[ \Phi(s) \Omega_{\by(\tau), \lambda_n(\tau)} (s) + B_-(s)\kappa_I \ThetaVec_{\by(\tau), \lambda_n(\tau), 1}(y(s)) - B_+(s)\kappa_W \ThetaVec_{\by(\tau), \lambda_n(\tau), 2}(y(s)) \Big] ds. \nonumber
\end{align}

\end{document}